\documentclass[aps,pra,showpacs,twocolumn,amsmath,amssymb,superscriptaddress,nofootinbib]{revtex4}
\usepackage{graphicx}
\usepackage{bm}
\usepackage{amssymb}
\usepackage{amsmath}
\usepackage{latexsym}
\usepackage{color}
\usepackage{palatino}
\usepackage{listings}
\usepackage{siunitx,float}
\usepackage[dvipsnames]{xcolor}  
\usepackage{soul}  
\usepackage{booktabs} 
\usepackage{makecell}
\usepackage{multirow}  
\usepackage{adjustbox}  
\usepackage{array}  
\usepackage[final]{changes}  
\definechangesauthor[name=shvydko,color=red]{yuri}
\newcolumntype{?}{!{\vrule width 1pt}}
\usepackage[flushleft]{threeparttable}  
\usepackage[hidelinks]{hyperref}       

\newcommand{\vc}[1]{\mbox{\boldmath $#1$}} 
\newcommand{\ind}[1]{_{#1}}    
\newcommand{\zr}{Z_{\ind{o}}}   
\newcommand{\zrp}{Z_{\ind{1}}}   
\definecolor{pink}{rgb}{0.99,0.5,0.5}
\begin{document}  
\title{Signatures of misalignment in x-ray cavities of cavity-based x-ray free-electron lasers}
\author{Peng Qi} \thanks{Present address: PSI, CH-5232 Villigen, Switzerland}
\affiliation{Advanced Photon Source, Argonne National Laboratory,  Lemont, Illinois 60439, USA}
\affiliation{University of Saskatchewan, Saskatoon, SK S7N5E5, Canada}
\author{Yuri Shvyd'ko} \email{shvydko@anl.gov}
\affiliation{Advanced Photon Source, Argonne National Laboratory,  Lemont, Illinois 60439, USA}
\begin{abstract} 
  Cavity-based x-ray free-electron lasers (CBXFEL) will allow use of
  optical cavity feedback to support generation of fully coherent
  x-rays of high brilliance and stability by electrons in
  undulators. CBXFEL optical cavities comprise Bragg-reflecting flat
  crystal mirrors, which ensure x-rays circulation on a closed orbit,
  and x-ray refractive lenses, which stabilize the orbit and refocus
  the x-rays back on the electrons in the undulator. Depending on the
  cavity design, there are tens of degrees of freedom of the optical
  elements, which can never be perfectly aligned. Here, we study
  signatures of misalignment of the optical components and of the
  undulator source with the purposes of understanding the effects of
  misalignment on x-ray beam dynamics, understanding misalignment
  tolerances, and developing cavity alignment procedures.  Betatron
  oscillations of the x-ray beam trajectory (both symmetric and
  asymmetric) are one of the characteristic signatures of cavity
  misalignment. The oscillation period is in the general case a
  non-integer number of round-trip passes of x-rays in the
  cavity. This period (unlike the amplitude and offset of the
  oscillations) is independent of the type of misalignment and is
  defined by cavity parameters.  The studies are performed on an
  example of a four-crystal rectangular cavity using analytical and
  numerical wave optics as well as ray-tracing techniques. Both
  confocal and generic stable cavity types are studied.
\end{abstract}
\date{\today}

%
%

\maketitle

\section{Introduction}

The next generation of hard x-ray free-electron lasers (XFELs)
operating in a high-repetition-rate ($\simeq$ MHz) pulse sequence mode
\cite{EXFEL20,Raubenheimer18} will allow for optical cavity--based
feedback, like in classical lasers \cite{Siegman}. Unlike
self-amplified spontaneous emission (SASE) XFELs
\cite{KS80,BPN84,EAA10}, the new cavity-based XFELs (CBXFEL) \cite{HR06,KSR08}
are capable of generating fully coherent x-ray beams of high
brilliance and stability.  Two major CBXFEL schemes are presently
considered: low gain and high gain.

An x-ray free-electron laser oscillator (XFELO)
\cite{KSR08,KS09,LSKF11,KSL12,DDD12} is a low-gain CBXFEL, which
requires a low-loss (high-Q) x-ray cavity.  XFELOs are promising to
generate radiation of unprecedented spectral purity with a meV
bandwidth.

Another possible realization of the CBXFEL is a high-gain regenerative
amplifier free-electron laser (XRAFEL) \cite{HR06,MDD17,FSS19,MHH20},
which can reach saturation after a few round-trip passes. It can
therefore allow for a high (close to 100\%) extraction efficiency.  It
can use a similar or identical cavity as an XFELO; but, with
substantially relaxed tolerances on the efficiency and stability.

Classical laser oscillator cavities in the optical domain are typically
composed of two curved high-reflectance backscattering mirrors, each
ensuring laser beam circulation, refocusing, and stabilization
\cite{Siegman}. In the hard x-ray domain, these functions have to be
assigned to two different types of optical elements \cite{KSR08,KS09}:
high-reflectance flat Bragg-reflecting diamond crystal mirrors
\cite{SSC10,SSB11} and x-ray--transparent beryllium
refractive paraboloidal lenses \cite{LST99,KSG18}. Different cavity designs have
been considered: fixed-photon-energy two-crystal cavity
\cite{SLW02,KSR08,KS09}, tunable-photon-energy four-crystal cavity
\cite{Cotterill68,KS09}, and tunable compact non-planar six-crystal
cavity \cite{Shv13}.

\begin{figure*}[t!]
\includegraphics[width=0.75\textwidth]{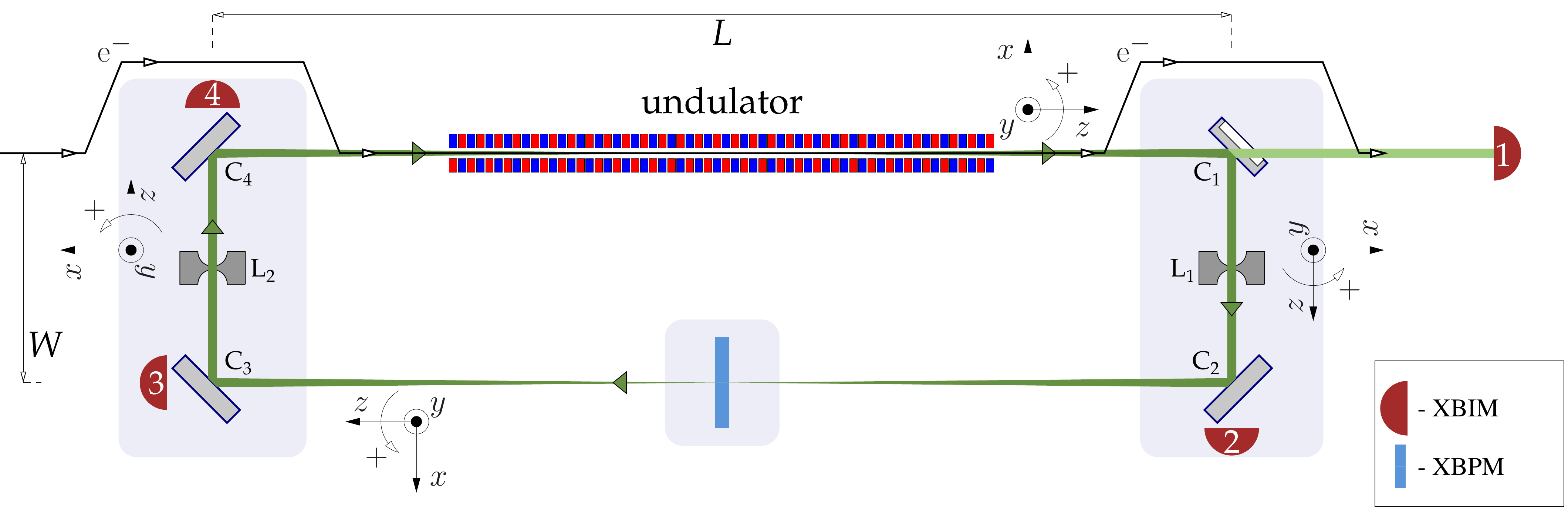}  
\caption{Schematic of a cavity-based XFEL with an undulator x-ray
  source together with a four-crystal (C$_{\ind{1}}$-C$_{\ind{4}}$),
  two-lens (L$_{\ind{1}}$-L$_{\ind{2}}$) rectangular x-ray cavity and
  x-ray diagnostics \cite{MAA19}. Cavity round-trip length
  $2(L+W)\simeq 65$~m. X-ray diagnostics tools include x-ray beam
  intensity monitors (XBIMs) and an x-ray beam position, profile, and
  intensity monitor (XBPM).}
\label{fig000}
\end{figure*}

Here, we are considering as an example a fixed-photon-energy
four-crystal rectangular cavity proposed for a joint Argonne National
Laboratory (ANL) and SLAC National Laboratory (SLAC) CBXFEL R\&D
project \cite{MAA19}. A CBXFEL with a rectangular cavity is presented
in Fig.~\ref{fig000}.  It is composed of an undulator x-ray source and
a cavity comprising four Bragg-reflecting flat-crystal mirrors
C$_{\ind{1}}$, C$_{\ind{2}}$, C$_{\ind{3}}$, and C$_{\ind{4}}$, along
with two refractive lenses (L$_{\ind{1}}$ and L$_{\ind{2}}$) as
focusing elements. The limited space in the Linac Coherent Light
Source (LCLS) undulator tunnel \cite{Raubenheimer18} determines the
choice of the fixed-photon-energy rectangular cavity versus a tunable
bow tie--type four-crystal cavity \cite{KS09}. In the future, the
rectangular cavity could be upgraded to a tunable non-planar
six-crystal cavity \cite{Shv13} taking the same footprint.

There are tens of degrees of freedom of the optical elements (depending
on the cavity design)---many more than in a classical laser
oscillator. All of them have to be aligned with a
precision ensuring that the to-be-amplified x-ray beam returning into
the undulator (after traveling tens to hundreds of meters in the cavity
and being refocused to a spot of a few tens of microns) meets a fresh electron
bunch of a similar size and a few tens of microns in length. Cavity
alignment is one of the major challenges in the realization of CBXFELs.
To this end, several essential questions have to be addressed, in
particular: (i) What are the typical misalignment patterns? (ii) Can
they be identified with x-ray beam monitors? (iii) Can different degrees of
freedom be disentangled and the number of mutually dependent degrees
of freedom reduced? (iv) What are the misalignment tolerances? (v)
How can the cavity be aligned?

In the first step of our analysis to address these questions, we are studying the
spatial, angular, and temporal behavior of x-rays in the rectangular
x-ray cavity using analytical tools developed for classical laser
oscillator cavities \cite{Siegman,Kl66}.  These methods have also been
applied to study statistical misalignment tolerances in CBXFEL
cavities \cite{LSKF11,MAA19,KJK20}.  We are using matrix ray tracing,
paraxial resonator theory, and Gaussian mode theory to derive
stable, self-consistent solutions for the Gaussian beams in the cavity;
to determine cavity parameters; and to analyze beam instabilities in both
a perfectly aligned cavity and a cavity with misaligned
optical components. 

Although in general this analytical approach is quite powerful, it has
limitation, because the Bragg-reflecting crystal mirrors are treated
as 100\%-reflecting mirrors with unrestricted spectral and angular
reflection widths. In the second step of our analysis, we address this
limitation by using numerical simulation tools, which involve
calculations based on the dynamical theory of Bragg diffraction of
x-rays in crystals \cite{Authier}. With these tools, we are able to
take the actual properties of the optical elements into account.

The short-term goal of these studies is to understand the spatial,
angular, and temporal behavior of x-rays and the response of the
cavity x-ray monitors to typical misalignment patterns of x-ray
optical components and of the x-ray source, with a view to developing
human-controlled cavity alignment procedures. The long-term goal is to
develop computer-assisted algorithms (e.g., utilizing machine learning) to guide cavity alignment.

\added{We note, that our studies are limited to empty cavity conditions. 
  XFEL gain effects, in particular, gain guiding effects  typical for XRAFEL are not considered here.}

The paper is organized as follows. Section~\ref{cavity} covers the
function and optical design of the rectangular cavity and its
parameters.  Results of analytical studies of the x-ray cavity are
presented in Section~\ref{analytical}, and results of numerical
simulation studies are given in Section~\ref{numerical}. A strategy
for cavity alignment is presented in Section~\ref{alignment}.
Conclusions and outlook are discussed in Section~\ref{conclusions}.

\section{Four-crystal rectangular cavity optical design and parameters}
\label{cavity}

\begin{table}
\caption{Parameters of the rectangular cavity, x-ray source, and
  optical components \cite{MAA19} used in the calculations presented in the paper. Parameters
  marked with asterisk are derived from other parameters of the
  table.}
\begin{tabular}{|l|l|l|}
  \hline   \hline 
 & &  \\[0pt]    
  & Parameters & Values \\
  \hline
  & & \\[0pt]
   Cavity & Type & rectangular \\[0pt]
          & Length $L$ & 32~m \\[0pt]
          & Width $W$ & 0.65~m \\[0pt]
   & Round-trip length$^{*}$ $\ell=2(L+W)$  & 65.3~m \\[0pt]
             & Round-trip time$^{*}$ $\tau=\ell/c$   & 218~ns\\[0pt]
   & & \\[0pt]
   \hline
    & & \\[0pt]
  Source   & Type & Gaussian beam \\
           & Rayleigh length$^{*}$ $\zr$  & 25.67~m \\   
  & Transverse size$^{*}$  $\sigma_{\ind{x,y}}$ (rms) & 15.9~$\mu$m \\   
  & $\sigma_{\ind{x,y}}=\sqrt{(\lambda/4\pi) \zr}$  & \\   
  & Angular size$^{*}$ $\sigma_{\ind{x,y}}^{\prime}$ (rms) & 0.62~$\mu$rad \\   
  & $\sigma_{\ind{x,y}}^{\prime}=\sqrt{(\lambda/4\pi)/\zr}$  &  \\   
  & Central photon energy$^{*}$ $E(\theta)$         &  9.83102 ~keV \\
  & Spectral bandwidth         &  100~meV \\  
  & (as used in numeric simulations) & \\[0pt]
  \hline
    & & \\[0pt]
   Crystals   & Material & Diamond\\[0pt]
   C$_{\ind{1}}$-C$_{\ind{4}}$       & Bragg reflection & (400) \\[0pt]
          & Central angle of incidence $\theta$ & 45$^{\circ}$ \\[0pt]
          & Thickness & C$_{\ind{1}}$: 20~$\mu$m; \\[0pt]
          &           & C$_{\ind{2}}$-C$_{\ind{4}}$: 500~$\mu$m \\[0pt]
          & Temperature & 300~K \\[0pt]
          & Spectral reflection width $\Delta E$  & 90~meV \\[0pt]
          & Angular reflection width $\Delta \theta$  & 9~$\mu$rad \\[0pt]   
   & & \\[0pt]
   \hline
  & & \\[0pt]
   Lenses   & Material & Beryllium \\[0pt]
   L$_{\ind{1}}$-L$_{\ind{2}}$         &   Lens type & Paraboloidal 2D \\[0pt]
          & Focal length $f^{(1)} = {\ell}/{4}$ & 16.3~m \\[0pt]
          & ~~~~~~~~~~~~~~~~~~~~~~$f^{(2)} = {\ell}/{8}+{2 \zr}^2/{\ell} $ & 28.3~m \\[0pt]
          & Radius of curvature $R^{(1)}$  & 115~$\mu$m \\[0pt]
   & ~~~~~~~~~~~~~~~~~~~~~~~~~~~~~~~~~~~~$R^{(2)}$ & 200~$\mu$m \\[0pt]
   & & \\[0pt]
\hline  \hline  
\end{tabular}
\label{tab1}
\end{table}


Figure~\ref{fig000} shows the schematic of the CBXFEL \cite{MAA19} with the
rectangular x-ray cavity studied here.   Parameters of the x-ray
cavity, x-ray source, and optical components are listed in
Table~\ref{tab1} and explained below. 

X-rays from an undulator source propagate through the x-ray cavity
composed of four Bragg-reflecting crystal mirrors
C$_{\ind{1}}$-C$_{\ind{4}}$ and two focusing lenses
L$_{\ind{1}}$-L$_{\ind{2}}$.  In the perfectly aligned cavity, the
optical axis is a rectangle having sides $L$ and $W$ ($L\gg W$), with
crystals placed in its corners, while the lenses are placed in the
symmetry points at distances $\ell/4$ from the source, where
$\ell=2(L+W)$ is the round-trip length.  Different locations of the
lenses could be considered as well, but only this symmetric
configuration is studied here.

The crystal mirrors compose two backscattering units, which bracket
the undulator, with two crystals C$_{\ind{1}}$ and C$_{\ind{2}}$ on
one end of the undulator and two crystals C$_{\ind{3}}$ and
C$_{\ind{4}}$ on the other end.  In each  unit, two
successive Bragg reflections reverse the direction of the beam, 
which fixes the incidence and reflection angles to
exactly 45$^{\circ}$. This geometry requires an x-ray source
generating photons with a $\sigma$-polarization component (electric
field vector perpendicular to the cavity plane). Diamond crystals are
chosen as Bragg reflecting crystal mirrors because of their high
reflectivity, x-ray transparency, and resilience \cite{SSB11}. In
particular, the 400 Bragg reflection from diamond crystal is used, a choice
dictated by the practical consideration that it is easier to obtain 
high-quality diamond crystal plates in the 100 orientation than in other orientations
\cite{SBT17}. The 400 Bragg reflection with 45$^{\circ}$ incidence
angle predefines a 9.83102-keV central photon energy of the cavity and
of the x-ray source, assuming all crystals operate at the same temperature of
300~K.  Crystals C$_{\ind{2}}$-C$_{\ind{4}}$ are assumed to be
500~$\mu$m thick for high reflectivity and mechanical
stiffness. Crystal C$_{\ind{1}}$ is assumed to be 20~$\mu$m thick for
output coupling of $\simeq 3$~\% of the interactivity power
\cite{KSR08,KS09,Shvydko19}.

The x-ray source is assumed to generate a Gaussian beam with a
Rayleigh length $\zr$ matched to the electron beam beta function at
the undulator center \cite{MAA19}, which is considered to be the
cavity origin with $z=0$.  The Rayleigh length and photon wavelength
$\lambda=hc/E$ define the root mean square (rms) x-ray source
(Gaussian beam waist) transverse size
$\sigma_{\ind{x,y}}=\sqrt{(\lambda/4\pi) \zr}$ and the rms angular
source divergence
$\sigma_{\ind{x,y}}^{\prime}=\sqrt{(\lambda/4\pi)/\zr}$.  The spectral
bandwidth of the source is limited artificially in these calculations
to 100~meV \added{to reduce computation time and} to match the spectral width of the Bragg reflections
$\Delta E=90$~meV.

The focal length $f$ of the focusing elements
L$_{\ind{1}}$-L$_{\ind{2}}$ has to be properly chosen in order for the
Gaussian beam with Rayleigh length $\zr$ to be a stable and
self-consistent (self-reproducing after each round trip) cavity mode,
as discussed in Section~\ref{analytical}, Eq.~\eqref{eq0120}.  The
focal length values -- $f^{(1)}$ and $f^{(2)}$ for confocal and generic
cavities, respectively (Table~\ref{tab1}) -- were chosen based on this
consideration but also for the following practical reason.

We chose beryllium paraboloidal lenses \cite{LST99} as x-ray focusing
elements in the cavity. They are manufactured with discrete values of
the radius of curvature $R$ \cite{RXOPTICS}. For the photon energy in
question, the lens with $R=200~\mu$m has a focal length of
$f^{(2)}=28.3$~m, close to (but not exactly) the value required to
maintain a cavity mode with the desired properties. In the following
calculations, we use this value for $f^{(2)}$ in the generic cavity and adjust appropriately
the values of $\zr$, $\sigma_{\ind{x,y}}$, and
$\sigma_{\ind{x,y}}^{\prime}$, as listed in Table~\ref{tab1}. The same
$\zr$, $\sigma_{\ind{x,y}}$, and $\sigma_{\ind{x,y}}^{\prime}$ values
are used for the confocal cavity with $f^{(1)}=\ell/4=16.3$~m.

In our studies, x-ray beam size, position, angle, and intensity are
calculated and ``monitored'' every round-trip pass in the undulator
center and in the midpoint between crystals C$_{\ind{2}}$ and
C$_{\ind{3}}$. In an experiment, non-invasive x-ray beam position,
profile, and intensity monitors (XBPMs) installed in these locations
can provide all these values except the beam angle. To monitor x-ray
beam angles at different locations in the cavity, x-ray beam intensity
monitors (XBIM) D$_{\ind{1}}$-D$_{\ind{4}}$ can be used to measure the
intensity of x-rays penetrating through the diamond crystals on each
round-trip pass (see Fig.~\ref{fig000}). A high level of transmitted
intensity would signal that the angle of incidence of x-rays to the
(400) reflecting crystal planes is far off the peak reflectivity angle
of the relevant crystal, while a low intensity would indicate
proximity to the peak reflectivity angle. For this reason, we
calculate in our studies the intensity of the x-rays transmitted
through crystals C$_{\ind{1}}$-C$_{\ind{4}}$.

The optical axis of the cavity is defined along the trajectory of
x-rays in the perfectly aligned cavity.  We use local right-handed
reference coordinate systems, which propagate
with the $\vc{z}$-axis always along the optical axis, the $\vc{y}$-axis
perpendicular to the cavity plane, and $\vc{x}\times \vc{y}=\vc{z}$,
as indicated in Fig.~\ref{fig000}. The optical elements inherit the
same coordinate system as the incident rays.  For angular variation of the ray slope in the $(x,z)$ plane and for rotation of the optical elements, the counterclockwise sense is defined as positive.

\section{Analytical studies}
\label{analytical}

Here, we are using the paraxial ray-transfer matrix technique and
paraxial wave optics to propagate x-rays and Gaussian beams through
the optical elements of the cavity shown in Fig.~\ref{fig000}, to
determine parameters of the stable cavity and the self-consistent
Gaussian beam modes and to analyze cavity instabilities due to the
x-ray source and misalignment of optical elements in the cavity plane.

\subsection{Stable and self-consistent solution for a perfect cavity}
\label{self-consistent}

A paraxial ray in any reference plane perpendicular to the optical
axis is given by its distance $x$ from the optical axis and by its
angle $x^{\prime}\ll 1$ with respect to the axis.  A ray vector
$\vc{r}_{\ind{0}}=\left[\begin{array}{c}\! {x}_{\ind{0}}\\ {x}_{\ind{0}}^{\prime} \end{array} \!\right]$
at the source (or input) plane is transformed to
$\vc{r}_{\ind{1}}=\hat{O}\vc{r}_{\ind{0}}$
at the image (or output) plane, where $\hat{O}=\{AB;CD\}$ is a
$2\times 2$ ray-transfer matrix of an optical element placed between
the planes.

The ray-transfer matrix of the cavity is a sequential product of three discrete
ray-transfer matrices: propagation in free space $\hat{P}(l)$ over
distance $l$, reflection from a crystal mirror $\hat{C}$, and 
focusing by lens $\hat{L}(f)$ with focal length $f$, which are given by
\begin{equation}
  \hat{P}(l)=\left[ \begin{array}{cc} 1 & l \\ 0 & 1   \end{array} \right ]\!\! ,\hspace{1mm}
  \hat{C}=\left[\! \begin{array}{cc} -1 & 0 \\ 0 & -1  \end{array} \! \right]\!\! , \hspace{1mm}
  \hat{L}(f)=\left[\!\begin{array}{cc} 1 & 0 \\ -{1}/{f} & 1  \end{array} \! \right]\!\! ,  
\label{eq0010}
\end{equation}
respectively \cite{Kl66,Siegman,SalehTeich}. The only impact
of the crystal mirror is the inversion of the ray vector space and
angular coordinates \cite{Shvydko15}.  The finite angular and spectral acceptance of
the Bragg-reflecting crystal mirrors are not taken into account within
this approach but will be included in the  numerical simulations approach in
Section~\ref{numerical}.

\begin{figure}[t!]
\includegraphics[width=0.5\textwidth]{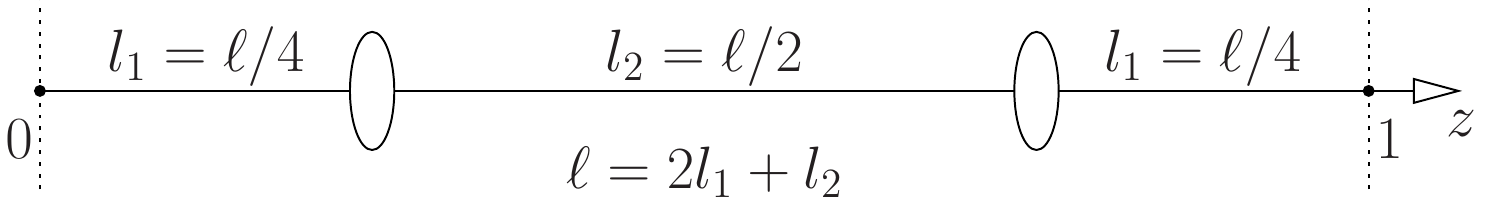}
\caption{Equivalent optical scheme of the unfolded symmetric cavity with two lenses.}
\label{fig003}
\end{figure}

The ray-transfer matrix of the cavity in Fig.~\ref{fig000} with four crystals and two
lenses is calculated as a product
$\hat{R}=\hat{P}(\frac{L}{2})\hat{C}\hat{P}(\frac{W}{2})\hat{L}(f)\hat{P}(\frac{W}{2})\hat{C}\hat{P}(L)\hat{C}\hat{P}(\frac{W}{2})\hat{L}(f)\hat{P}(\frac{W}{2})\hat{C}\hat{P}(\frac{L}{2})$. 
The source ray vector $\vc{r}_{\ind{0}}$ at the undulator center
(considered to be the cavity origin, with $z=0$) is transformed by
this matrix to the image ray vector $\vc{r}_{\ind{1}}$ at the same
location after the round trip.  Because $\hat{C}=-\hat{I}$, where
$\hat{I}=\{10;01\}$ is the unit matrix, the ray-transfer matrix of the
cavity simplifies to
$\hat{R}=\hat{P}({\ell}/{4})\hat{L}(f)\hat{P}({\ell}/{2})\hat{L}(f)\hat{P}({\ell}/{4})$
and becomes
\begin{gather}
\hat{R}\,=\,
\left[ \begin{array}{cc} \frac{8f(f-\ell)+\ell^2}{8f^2} & \frac{\ell\ (32f^2-12f\ell+\ell^2)}{32f^2} \\ \frac{\ell-4f}{2f^2} & \frac{8f(f-\ell)+\ell^2}{8f^2}
  \end{array} \right] ,
\label{eq0020}
\end{gather}
using Eq.~\eqref{eq0010}. The equivalent scheme of the unfolded symmetric cavity is shown in Fig.~\ref{fig003}.

The cavity is stable (meaning a paraxial ray within the cavity does
not escape after numerous passes) when the matrix trace $(A+D)$ obeys
the inequality $-1 \le (A+D)/2\le 1$ \cite{Siegman,Kl66}. In our
particular case, stability requires $f \ge \ell/8$, which is a relaxed
condition.

We now use  the calculated cavity ray-transfer matrix
\eqref{eq0020} to determine self-consistent Gaussian beam modes in the
cavity, i.e., the modes self-reproducing after each round trip.  Knowledge
of the ray-transfer matrix of an optical system is particularly useful
as it also describes the propagation of the Gaussian beams through
the system \cite{Siegman,Kl66}.  Gaussian beams are eigenstates of paraxial
wavefields in free space, and they can be eigenstates of the linear
optical systems presented by ABCD ray-tracing matrices.

The self-consistent solutions can be found by applying the ABCD
propagation law: a Gaussian beam with the complex beam parameter
 $ q_{\ind{0}}(z)=z+{\mathrm i}\zr$
in the source reference plane 0 (here $\zr$ is the Rayleigh range) is
transformed to a Gaussian beam with the beam parameter
\begin{equation}
q_{\ind{1}}\,=\,\frac{A\,q_{\ind{0}}\,+\,B}{C\,q_{\ind{0}}\,+ D}
\label{eq0050}
\end{equation}
in reference plane 1 (image plane).  We assume here that the waist of the
Gaussian beam source is at the undulator center with $z=0$ and
therefore $q_{\ind{0}}={\mathrm i}\zr$.

A mode with the complex beam parameter
$q_{\ind{0}}$ is a self-consistent cavity mode if after the complete
round trip of the beam it returns to its initial value, i.e.,
\begin{equation}
q_{\ind{1}}=q_{\ind{0}}={\mathrm i}\zr.
\label{eq0060}
\end{equation}
Combining Eqs.~\eqref{eq0020}-\eqref{eq0060}, we arrive at a second-order equation in $\zr^2$ for the self-consistent Gaussian mode
containing the focal length $f$ of the lenses and the cavity
round-trip length $\ell$ as parameters. Alternatively, the equation can
be solved for the focal length $f$ required for a specific
self-consistent value of $\zr^2$.  There are two possible solutions:
\begin{equation} 
f^{(1)}\,=\,\frac{\ell}{4}, \hspace{1cm} f^{(2)}\,=\,\frac{\ell}{8}\,+\,\frac{2\zr^2}{\ell}. 
\label{eq0120}
\end{equation}
In both cases $f \ge \ell/8$, and therefore both represent stable
solutions.  We note that $f^{(2)}=f^{(1)}$, provided $\zr=\ell/4$. We
next analyze the beam dynamics for each solution.

\subsubsection{Self-consistent confocal cavity solution}
\label{confocal}

A Gaussian beam with any Rayleigh range $\zr$ is
a self-consistent solution if the relationship $f^{(1)}={\ell}/{4}$ of
Eq.~\eqref{eq0120} is fulfilled.  This relationship corresponds to a
confocal symmetric stable laser cavity \cite{Kl66,Siegman}, in which
the focal points of both lenses coincide with each other, with the
cavity origin (at the undulator center), and with the midpoint on the
return path.

\begin{figure}[t!]
\includegraphics[width=0.5\textwidth]{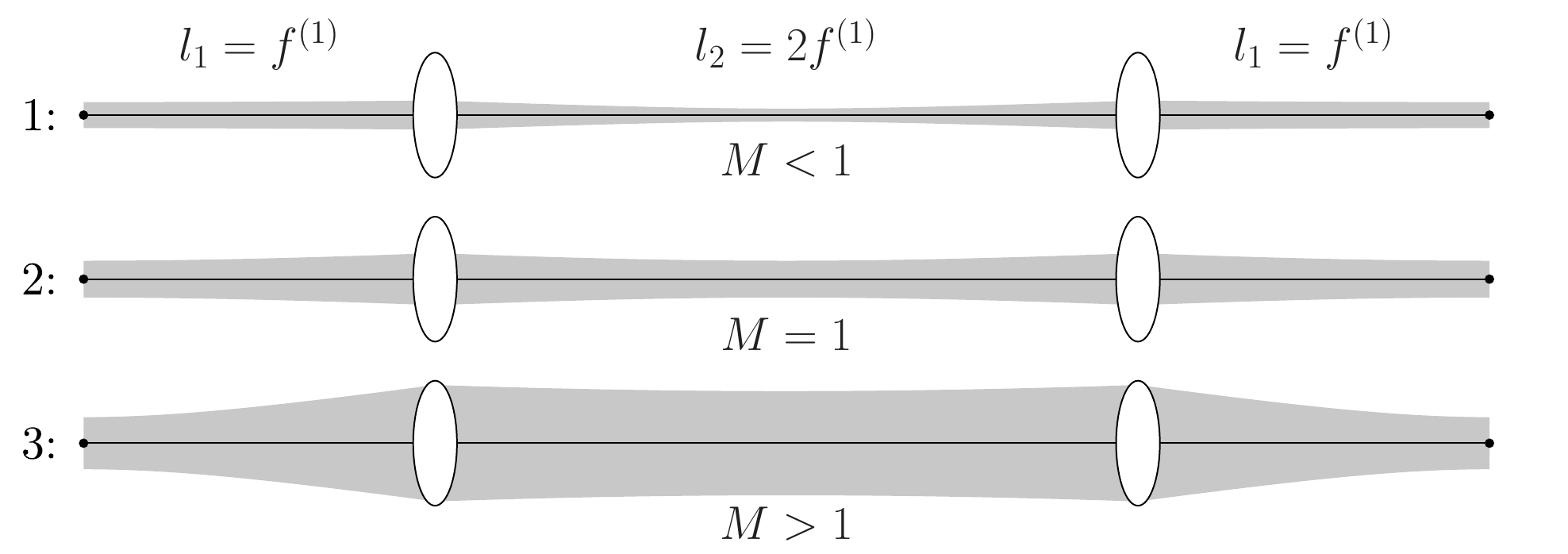}
\caption{Gaussian beam profiles for self-consistent solutions in the
  symmetric confocal cavity, shown for different initial Rayleigh range
  values $\zr=f^{(1)}/M$. The second Gaussian beam waist is in the
  midpoint of the round-trip path, with the size of the beam waist (relative to the original waist size) scaling with the
  magnification factor $M$.}
\label{fig002}
\end{figure}

Any Gaussian beam of Rayleigh range $\zr$ and beam waist of rms width
$\sigma_{\ind{0}}=\sqrt{\lambda \zr/4\pi}$ located at the cavity
origin (which is also a focal point of lens L$_{\ind{1}}$) is
refocused by lens L$_{\ind{1}}$ into a Gaussian beam of Rayleigh
length $\zrp=\zr M^2$ with beam waist of rms width $\sigma_{\ind{1}}=M
\sigma_{\ind{0}}$ located at the second focal point of lens
L$_{\ind{1}}$ (which is also midpoint of the return path).  Here
\begin{equation}
  M=f^{(1)}/\zr=\ell/4\zr
\label{eq0125}
\end{equation}
is the absolute value of the magnification factor (see Section~3.2 of
\cite{SalehTeich}). The location of the second Gaussian beam waist is
in the focal point of the second lens L$_{\ind{2}}$. As a result, this
Gaussian beam is refocused with magnification $1/M$ by the second lens
L$_{\ind{2}}$ to a Gaussian beam with the waist located at the cavity
origin (the second focal point of lens L$_{\ind{2}}$), and with
exactly the same Rayleigh length and rms width as the original beam.
In the particular case of $\zr=f_{\ind{0}}=\ell/4$, magnification
$M=1$ results in the same Rayleigh length $\zrp=\zr$ and beam width
$\sigma_{\ind{1}}=\sigma_{\ind{0}}$ at the cavity midpoint as well as
at the cavity origin. Figure~\ref{fig002} shows appropriate beam
profiles with different magnification $M$.

The ABCD cavity matrix given by Eq.~\eqref{eq0020} reduces in the
confocal cavity case to a very simple
\begin{gather}
  \hat{R}^{(1)}\,=\,\left[ \begin{array}{cc} -1 & 0 \\ 0 & -1  \end{array} \right]\,=\, -\hat{I}.
\label{eq0122}
\end{gather}
The $n$-pass confocal cavity matrix can be immediately calculated as
$[\hat{R}^{(1)}]^n = (-1)^n \hat{I}$, which in turn allows us to
calculate the transformation of an arbitrary initial source ray
$\vc{r}_{\ind{0}}$ at the cavity origin after $n$ passes:
\begin{equation}
\vc{r}_{\ind{n}}\,=\,[\hat{R}^{(1)}]^n\vc{r}_{\ind{0}}\,=\,(-1)^n\vc{r}_{\ind{0}}.
\label{eq0126}
\end{equation}
For an x-ray beam with a distribution of source rays characterized
by an initial average ray vector $\bar{\vc{r}}_{\ind{0}}$ that
represents the offset of the beam from the optical axis, 
Eq.~\eqref{eq0126} also describes the pass-to-pass variation of the beam
position $\bar{\vc{r}}_{\ind{n}}=\,(-1)^n \bar{\vc{r}}_{\ind{0}}$ at
the source location.  The beam position changes periodically but is
reproduced every second pass. If there is no initial beam offset, i.e., 
$\bar{\vc{r}}_{\ind{n}}=0$, the beam position is reproduced (is self-consistent)
every pass. In contrast, the beam size
\begin{equation}
\sigma_{\ind{n}}= \sqrt{
  \overline{ \left( x_{\ind{n}}-\bar{x}_{\ind{n}}\right)^2 }} = \sqrt{
  \overline{x_{\ind{0}}^2}-(\bar{x}_{\ind{0}})^2 } = \sigma_{\ind{0}}
  \label{eq0340}
\end{equation}  
never changes at the source origin in the confocal cavity.

\subsubsection{Self-consistent generic cavity solution}
\label{selfconsistent-generic}

The second solution to Eq.~\eqref{eq0120} represents a generic cavity
configuration. A Gaussian beam with any Rayleigh length $\zr$ is a
valid self-consistent stable solution if focal length $f^{(2)}$ obeys
Eq.~\eqref{eq0120}.  For this case, the magnification factor is always
$M=1$ for any $\zr$ (see Section~3.2 of \cite{SalehTeich}). The second
waist is located at the midpoint of the return path and has the same
beam waist size as the original Gaussian beam, as shown schematically
in Fig.~\ref{fig002} for $M=1$.


The cavity ray-transfer matrix Eq.~\eqref{eq0020} in this case is 
\begin{gather}
  \hat{R}^{(2)}\,=\,\left[ \begin{array}{cc} \frac{1-96r^2+256r^4}{(16r^2+1)^2} & \frac{\ell\ (256r^4-16r^2)}{(16r^2+1)^2} \\\frac{16-256r^2}{\ell\ (16r^2+1)^2} & \frac{1-96r^2+256r^4}{(16r^2+1)^2}   \end{array} \right],
  \hspace{0.15cm} r=\zr/\ell,
\label{eq0130}
\end{gather}
which transforms to
\begin{gather}
  \hat{R}^{(2)}\,=\,\left[ \begin{array}{cc} \cos\phi & \zr \sin\phi \\ -(1/\zr)\sin\phi & \cos\phi   \end{array} \right],
\label{eq0132}
\end{gather}
using a  new parameter $\phi$ defined as 
\begin{gather}
\cos\phi= A= \frac{1-96r^2+256r^4}{(16r^2+1)^2},\\ \sin\phi= B/\zr = -C \zr = \frac{16 (16r^2-1)}{(16r^2+1)^2},\\
\phi=4\arctan(1/4r).   
\label{eq0135}
\end{gather}
Phase $\phi$ is the accumulated within one round trip Gouy phase
$\arctan(z/\zr)$ of the Gaussian beams \cite{PeifanLiu21}.
With the cavity parameters provided in Table~\ref{tab1},
\begin{equation}
\phi=2\pi/2.76. 
\label{eq0137}
\end{equation}

Using Eq.~\eqref{eq0132}, a ray-transfer matrix of a multi-pass cavity can be calculated as
\begin{gather}
  [\hat{R}^{(2)}(\phi)]^n\,=\,\left[ \begin{array}{cc} \cos\phi n & \zr \sin\phi n \\ -(1/\zr)\sin\phi n & \cos\phi n   \end{array} \right],
\label{eq0200}
\end{gather}
where $n$ is a number of passes.
This expression in turn allows us to calculate how an arbitrary initial source
ray $\vc{r}_{\ind{0}}$ at the cavity origin  changes after $n$ passes:
\begin{gather}
\vc{r}_{\ind{n}}\!=\![\hat{R}^{(2)}(\phi)]^n \vc{r}_{\ind{0}}\!=\!  
\left[
\begin{array}{c} {x}_{\ind{0}}\cos\phi n+ {x}_{\ind{0}}^{\prime} \zr \sin\phi n \\ {x}_{\ind{0}}^{\prime}\cos\phi n - ({x}_{\ind{0}}/\zr) \sin\phi n
\end{array}  
\right].
\label{eq0210}
\end{gather}

We now consider  a beam of rays with non-zero average
distribution of the initial spatial $\bar{x}_{\ind{0}}$ and initial
angular $\bar{x}_{\ind{0}}^{\prime}$ offset  from the optical
axis. From Eq.~\eqref{eq0210} it follows that the average spatial and
angular beam positions at the cavity origin 
oscillate after each round-trip pass as
\begin{equation}
\bar{\vc{r}}_{\ind{n}}\!=\!\left[
\begin{array}{c} \bar{x}_{\ind{n}} \\ \bar{x}_{\ind{n}}^{\prime}
\end{array}  
\right]
=
\left[
\begin{array}{c} \bar{x}_{\ind{0}}\cos\phi n+ \bar{x}_{\ind{0}}^{\prime} \zr \sin\phi n \\ \bar{x}_{\ind{0}}^{\prime}\cos\phi n - (\bar{x}_{\ind{0}}/ \zr) \sin\phi n
\end{array}  
\right],
\label{eq0220}
\end{equation}
with a period $2\pi/\phi$, which is a non-integer number of round
trips in the general case.  We will refer to this effect as betatron
oscillations of x-rays in a cavity, by analogy to similar effects in
periodic accelerator structures \cite{ES92}. This effect is of course typical
of laser cavities as well \cite{Siegman}. 
The importance of the betatron oscillations for a closed optical cavity was pointed out recently \cite{KJK20}.
The phase $\phi$ corresponds to the phase advance of the betatron oscillations with each round-trip pass.

Note that, from Eq.~\eqref{eq0220}, betatron oscillations are absent if the initial beam (source) is
perfectly aligned with the optical axis (i.e., $\bar{x}_{\ind{0}}=0$ and
$\bar{x}_{\ind{0}}^{\prime}=0$).  The presence of the betatron
oscillations thus indicates a misaligned source. Table~\ref{tab2} shows
an example of non-zero betatron oscillation for particular source
alignment errors in the cavity whose parameters are provided in
Table~\ref{tab1}.  As shown in the following sections, the presence of betatron
oscillations with a period of $2\pi/\phi$ is a signature of 
misalignment (of any kind) in the cavity.

Betatron oscillations are also characteristic of the confocal
cavity case, but there the period of the oscillations is always two cycles; compare Eq.~\eqref{eq0126} and Eq.~\eqref{eq0220}.

\begin{table}
\caption{Initial x-ray beam (source) alignment errors and subsequent
  spatial and angular beam betatron oscillation amplitudes calculated with
  Eq.~\eqref{eq0220}.}
\begin{tabular}{|l||l|l|}
  \hline
  Initial alignment errors & $\bar{x}_n$ amplitude & $\bar{x}_n^{\prime}$ amplitude \\
  \hline\hline
  $\bar{x}_0=0$ &  $\bar{x}^{\prime}_0 \zr= 5 \mu$m  & $\bar{x}^{\prime}_0 = 200$~nrad \\
  $\bar{x}^{\prime}_0=200$~nrad & & \\
  \hline
  $\bar{x}_0=5 \mu$m &  $\bar{x}^{\prime}_0 = 5 \mu$m  & $\bar{x}_0/\zr = 200$~nrad \\
  $\bar{x}^{\prime}_0=0$ & & \\
  \hline
\end{tabular}
\label{tab2}
\end{table}

Similarly, Eqs.~\eqref{eq0210}-\eqref{eq0220} can be used to calculate the variation of
the rms spatial and angular spread of the x-ray beam (beam size) after each round
trip:
\begin{gather}
  \label{eq0301}
  \sigma_{\ind{x_n}}^2 = 
  \overline{ \left( x_{\ind{n}}-\bar{x}_{\ind{n}}\right)^2 } = \sigma_{\ind{x_0}}^2\left(\cos^2\phi n+ S^2 \sin^2\phi n\right) ,\\
\label{eq0303}
\sigma_{\ind{x_n'}}^2 =
  \overline{ \left( x_{\ind{n}}^{\prime}-\bar{x}_{\ind{n}}^{\prime}\right)^2 } = \sigma_{\ind{x_0'}}^2\left(\cos^2\phi n+ S^{-2} \sin^2\phi n\right),\\
  {\mathrm {where}~~}S=\sigma_{\ind{x_0'}} \zr/\sigma_{\ind{x_0}},
\label{eq0300}
\end{gather}
and $\sigma_{\ind{x_0}}=\sqrt{
  \overline{ \left( x_{\ind{0}}-\bar{x}_{\ind{0}}\right)^2 }}$ and $\sigma_{\ind{x_0'}}=\sqrt{
  \overline{ \left( x_{\ind{0}}^{\prime}-\bar{x}_{\ind{0}}^{\prime}\right)^2 }}$ are initial rms beam
size parameters.  Here we assume that
the spatial and
angular distributions are statistically independent: $\left< (x_{\ind{0}}-\bar{x}_{\ind{0}})(x_{\ind{0}}^{\prime}-\bar{x}_{\ind{0}}^{\prime}) \right>=0$.

If $S=1$, or equivalently
\begin{equation}
  \sigma_{\ind{x_0}}=\sigma_{\ind{x_0'}} \zr,
  \label{eq0310}
\end{equation}
we arrive at a stable, self-reproducible beam solution  
\begin{equation}
  \sigma_{\ind{x_n}} = \sigma_{\ind{x_0}}, \hspace{1cm} \sigma_{\ind{x_n'}} =
  \sigma_{\ind{x_0'}},
\label{eq0330}
\end{equation}
valid also for a  misaligned source.
It is similar to the case of the self-consistent solution for Gaussian beams
given by Eq.~\eqref{eq0060}. 

Remarkably, according to Eq.~\eqref{eq0310}, this stable solution is
valid only if the x-ray spatial and angular source sizes
$\sigma_{\ind{x_0}}$ and $\sigma_{\ind{x_0'}}$, respectively, obey the
same relationship as those of the Gaussian beam with Rayleigh range
$\zr$ at its waist location.

In all other cases, when $S\not = 1$, the beam size and angular spread
oscillate periodically from pass to pass with a twice smaller period
around average values that can be smaller or larger than the initial
beam size values, depending on the $|S|^2$ value.
  
This result implies that the ABCD ray-transfer matrix propagates a
collection of rays with a Gaussian distribution of position and angles
given by Eq.~\eqref{eq0310}, exactly in the same way as it propagates
the Gaussian beam with Rayleigh range $\zr$. The equivalence of the
paraxial Gaussian ray and Gaussian wave optics, which is well
established \cite{Kl66,Siegman,KJK20}, justifies using ray-transfer
techniques to study other aspects of cavity performance, in
particular for a cavity with misaligned optical components (Section \ref{numerical}).

\subsection{Ray optics with misaligned elements}
\label{misaligned}

The ray-transfer matrix approach can also be used to propagate rays
through optical systems with misaligned optical elements
\cite{Siegman}.  Here, we calculate how the rays are transformed in
such cases. These results will be applied in Section~\ref{numerical}
to calculate x-ray trajectories in a cavity with misaligned
elements. Those trajectories will be compared with results obtained
with a more advanced numerical simulations approach, which uses the
x-ray optics modeling package Shadow3 \cite{SHADOW3}.

\begin{figure}
\includegraphics[width=0.325\textwidth]{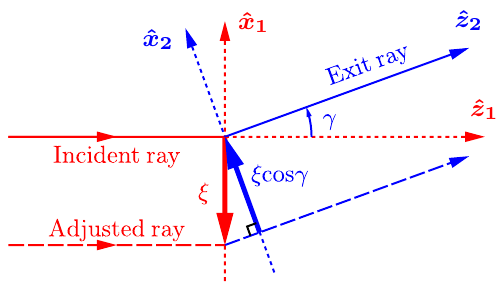}
  \caption{Schematic of incident and adjusted rays clarifying calculation of the x-ray trajectory distorted by a misaligned optical element.}
  \label{fig:axis-transform}
\end{figure}

Figure\(~\)\ref{fig:axis-transform} illustrates the relationships used
in calculating misalignment. Suppose an optical element (a crystal or a
lens) is displaced from the perfect configuration by a distance $\xi$
and an angle $\delta$. The distorted propagation of the x-rays can be
calculated by (i) displacing the incident x-rays by the same amount as
the optical element but in the opposite direction ($-\xi$ and
$-\delta$), (ii) transferring the ray with a standard matrix
[$\hat{C}$ or $\hat{L}(f)$], and (iii) moving the transferred ray back
by the adjusted amount. As a result, the coordinates of the distorted
ray are given by
\begin{align}
    \begin{bmatrix}
      x_n\\
      x^\prime_n
    \end{bmatrix}
    = M   
    \left(
      \begin{bmatrix}
      x_{n-1}\\
      x^\prime_{n-1}
      \end{bmatrix}
      -
      \begin{bmatrix}
        \xi\\
        \delta
      \end{bmatrix}
    \right)
    +
    \begin{bmatrix}
      \xi\cos\gamma\\
      \delta
    \end{bmatrix} ,
\end{align} 
where \(M\) is the standard ray-transfer matrix corresponding to the optical element [$\hat{C}$ or $\hat{L}(f)$] and \(\gamma\) is the angle between the exit ray and the incident ray at the location of the optical element, as shown in
Fig.\(~\)\ref{fig:axis-transform}. 
Note that the first adjustment of the  ray position (\(-\xi\)) is relative to the incident ray.
To restore the the exit ray to its correct position, it needs to
be moved by the amount \(\xi\cos\gamma\). 
For a crystal, the exit ray angle \(\gamma=2\theta\) after crystal reflection with Bragg angle $\theta$.  For a lens, $\gamma=0$.

Thus, the coordinates of the ray vector reflected by a misaligned crystal with a  Bragg angle $\theta=45^{\circ}$ (corresponding to the case of our cavity) are
\begin{align}
    \begin{bmatrix}
      x_n\\
      x^\prime_n
    \end{bmatrix}
    &= \hat{C}     
      \begin{bmatrix}
        x_{n-1}\\
        x_{n-1}^\prime
      \end{bmatrix}
      +
      \begin{bmatrix}
        \xi\\
        2\delta
      \end{bmatrix}
\label{eq:crystal-misalign}
    , 
\end{align} and the coordinates after a misaligned lens are
\begin{align}
    \begin{bmatrix}
      x_{n}\\
      x_{n}^\prime
    \end{bmatrix}
   &= L(f)   
      \begin{bmatrix}
        x_{n-1}\\
        x_{n-1}^\prime
      \end{bmatrix}
      +
      \begin{bmatrix}
        0\\
        \xi/f
      \end{bmatrix}
     \label{eq:crl-misalign}.
\end{align}
Eq.\(~\)\eqref{eq:crystal-misalign} shows that crystal misalignment
displaces the ray by an amount equal to the crystal's positional
displacement and deflects the ray by twice the crystal's angular
misalignment.  Eq.\(~\)\eqref{eq:crl-misalign} indicates that the ray
propagation is not sensitive to angular misalignment of the lens and
the angular deflection of the ray is proportional to positional
displacement of the lens.
Given the initial status of a ray and the misalignment status of the
optical elements, the ray position and angle at locations of interest
in the cavity can be traced, as will discussed in the next section.

\section{Numerical studies}
\label{numerical}

The paraxial ray-transfer matrix and wave-optics Gaussian beam
approaches used in the previous section are limited because the cavity
optical components---the crystals and lenses---are treated in  
simplified way.  The Bragg-reflecting crystals are considered as
100\%-reflecting flat mirrors with unrestricted spectral and angular
reflection widths. Dynamical diffraction effects \cite{Authier} are
not taken into account. X-ray absorption in lenses, which limits their
effective aperture \cite{LST99}, is also not considered. As a result,
the round-trip losses in the cavity cannot be studied accurately.

We are using numerical simulation tools to provide for a more realistic
account of the actual properties of the optical elements and to
acquire more detailed insight into the spatial, angular, and temporal
behavior of x-rays (x-ray beam dynamics) and the response of 
x-ray beam monitors to typical alignment errors.

The numerical simulations are performed in three dimensions, unlike
the two-dimensional analytical studies performed in
Section~\ref{analytical}.

In most cases, we consider the generic cavity, but we do examine the confocal
cavity case in Section~\ref{confocal-numeric} for comparison.

\begin{figure}[t!]
    \centering
    \includegraphics[width=0.99\linewidth]{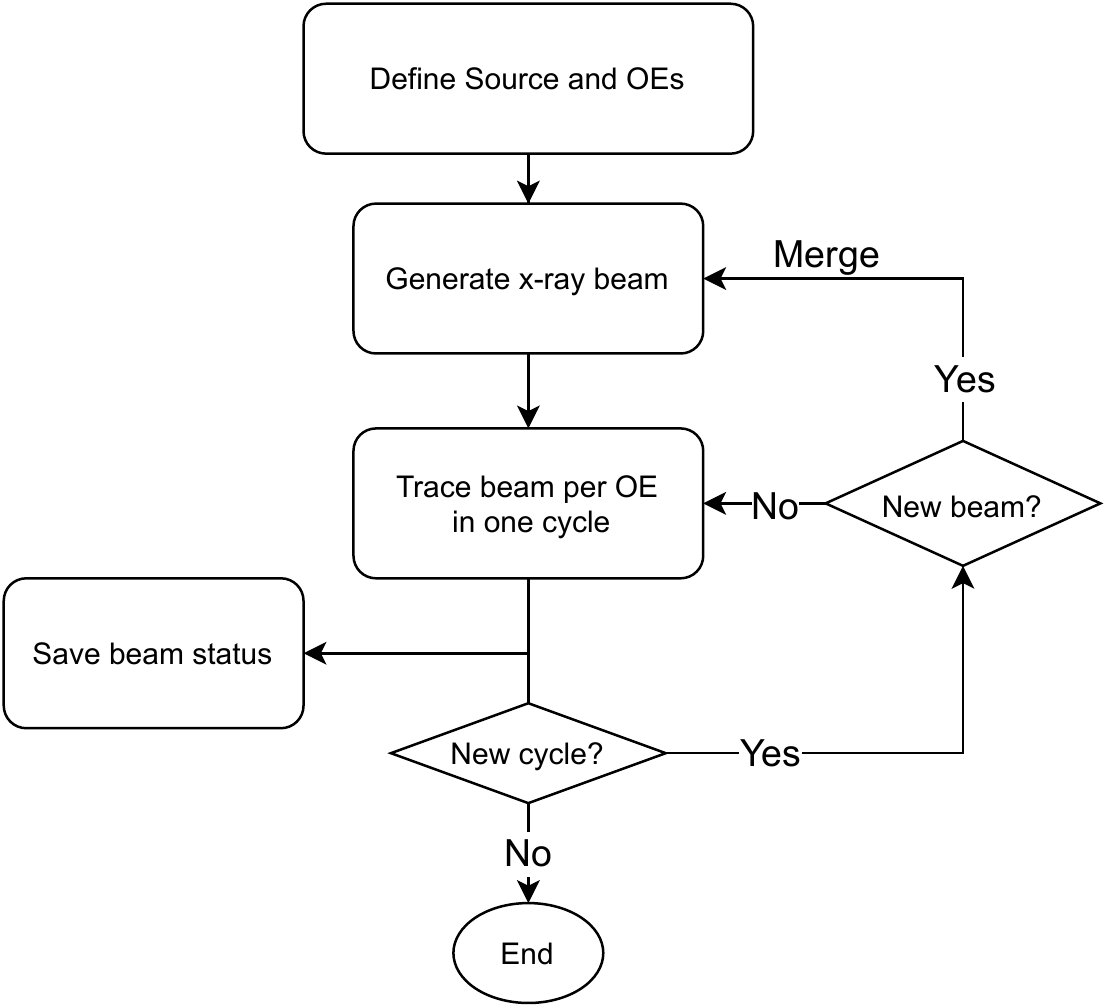}
    \caption{A diagram of the simulation work flow. (OE stands for optical elements.)}

    \label{fig:diagram}
\end{figure}

\subsection{Method}

Numerical simulations in this work are performed using the x-ray optics
modeling package Shadow3 \cite{SHADOW3} in the Oasys
environment \cite{RS16,RS17} and custom Python
scripts.

As shown in Fig.~\ref{fig:diagram}, the simulation program starts with
defining the x-ray source and all optical elements (crystals and
refractive lenses) that exist in the cavity.  The source generates
x-rays with randomized positions, angles, and energies with a given
Gaussian distribution. The rays propagate and interact with each optical
element (OE) in the cavity and return to the source. When
multiple-cycle simulation is desired, the x-ray beam keeps propagating
and repeats similar simulations as in the first cycle. The status of
all x-rays after each OE is recorded for the analysis.

In cases that are not affected by the limited
reflection width of the crystals or by photoabsorption in lenses (e.g., for small spatial and angular beam deflections from the optical axes),
the x-ray beam dynamics are also calculated, using the ray-tracing matrix approach (Section~\ref{misaligned}), and compared with the numerical simulation results
obtained with the Shadow3 package.

The x-ray source and optical elements parameters of the simulated CBXFEL system are listed in Table~\ref{tab1}.

\subsection{Perfectly aligned system}
\label{perfect}

We commence with a perfectly aligned system and a generic cavity in
which the optical elements and the x-rays source are in the spatial
and angular positions shown in Fig.~\ref{fig000}.

\begin{figure}[t!]
    \centering
        \includegraphics[width=1.0\linewidth]{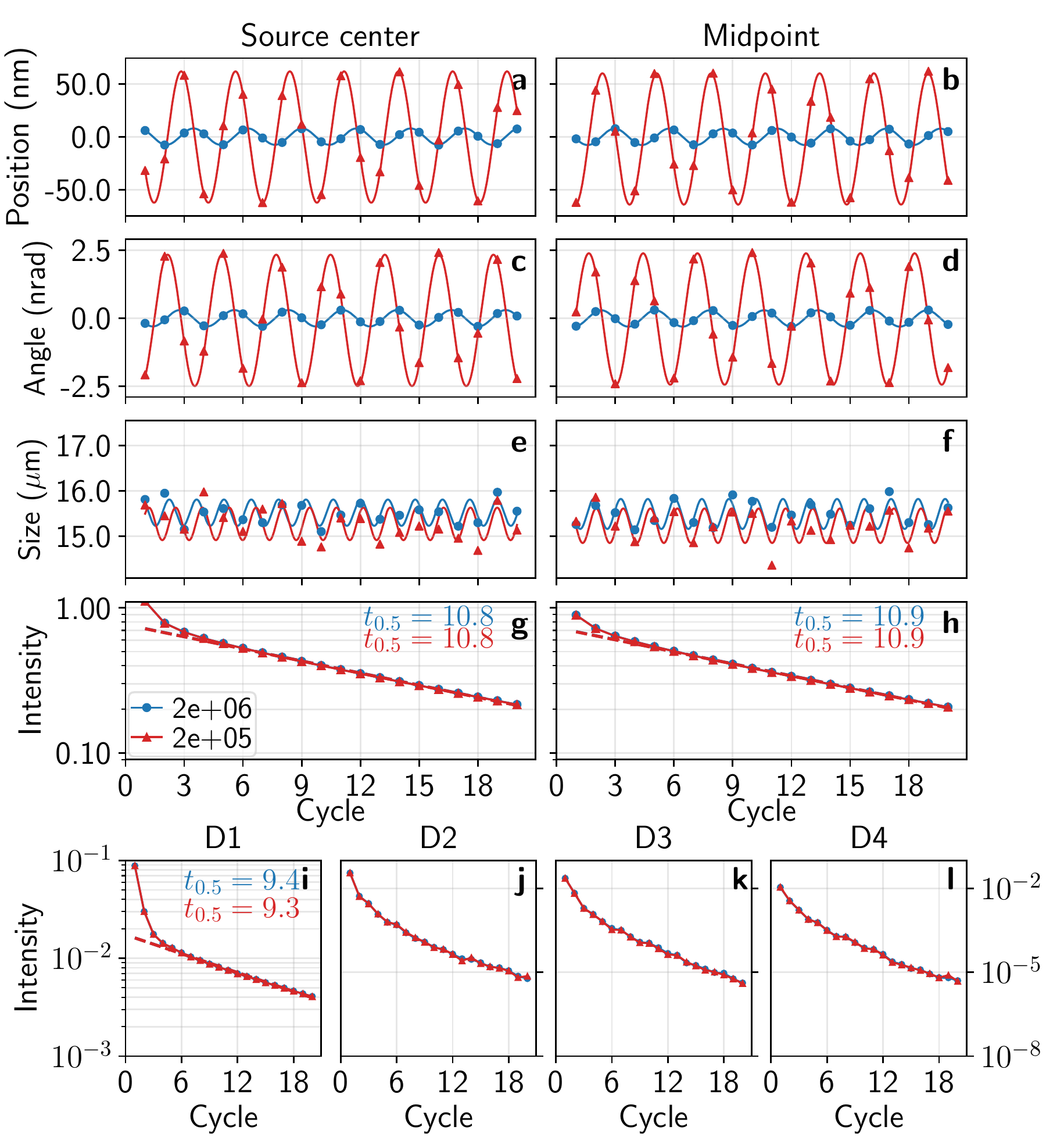}
    \caption{Numerical simulations of x-ray beam dynamics in the
      perfectly aligned generic four-crystal rectangular cavity shown
      in Fig.~\ref{fig000}. Values are presented simulated at cavity
      midpoint (top, left column), source center (top, right column),
      and beam intensity monitor positions (bottom row). The cavity
      and source parameters are provided in Table~\ref{tab1}. Two
      resolutions are shown: $2\times 10^{5}$ rays (red) and $2\times
      10^{6}$ rays (blue). The positional and angular oscillations are
      due to numerical simulation artifacts, as discussed in the
      text. See text for details.}
    \label{fig:perfect}
\end{figure}

Figure~\ref{fig:perfect} presents results of the numerical
simulations, which will serve as a baseline to identify signatures of
misalignment in subsequent studies.  The graphs in the top four rows
show variations after each round-trip cycle of the x-ray beam spatial
position (a)-(b), angle (c)-(d), transverse beam size (e)-(f), and beam intensity
(g)-(h) calculated either at the midpoint of the return path (left
column) or at the x-ray source location (right column). The beam
positions and angles are measured as a relative deviation of the
average ray distribution from the optical axis. The beam size is
calculated as the rms of the ray distribution.

Graphs (i)-(l) in the bottom row of Fig.~\ref{fig:perfect} show the
intensities (calculated for each cycle) of x-rays
transmitted through crystals C$_{\ind{1}}$-C$_{\ind{4}}$,
respectively.  In an experiment, these intensities would  be measured by
the corresponding XBIMs D$_{\ind{1}}$-D$_{\ind{4}}$ shown in
Fig.~\ref{fig000}.

The analytical studies performed in Section~\ref{analytical} (for a
perfectly aligned cavity with an x-ray source generating a
self-consistent Gaussian beam) predict a perfectly reproducible beam
after each round trip. In contrast, the numerical simulations show
pass-to-pass periodic non-zero variation of the x-ray beam position,
angle, and size in the perfectly aligned cavity at both source and
midpoint.  The x-ray beam trajectory oscillates with a period of 2.77
cavity cycles.  This value corresponds exactly to the period of the
betatron oscillations derived for a system with the same parameters
featuring a perfectly aligned cavity but a misaligned source [see
  Eqs.~\eqref{eq0137}, \eqref{eq0210}, and \eqref{eq0220}].  

As we noted in Section~\ref{analytical}, betatron oscillations are a
signature of misalignment, in particular, of the x-ray source. We do
in fact have a misaligned source in this case, but here the
``misalignment" is an artifact arising from the use of a finite number
of rays. Such a ``beam" consisting of discrete rays is never perfectly
aligned with the optical axis. Figure~\ref{fig:perfect} provides a
baseline for the scale of beam variations due to this artifact.  In
fact, the spatial and angular variations in
Figs.~\ref{fig:perfect}(a)-(f) are very small, about 1/1000 of the
beam size and angular beam divergence, and therefore can be neglected
in this case. To explore the scale of this artifact, we use
simulations at two resolutions: $2\times 10^5$ rays (red) and $2\times
10^6$ rays (blue).  Increasing the number of rays decreases the
amplitudes of the oscillations, albeit at a cost of increased
computation time, but they never reach zero (that would occur only
with an infinite number of rays). The simulations of a misaligned
cavity presented in the following sections use the lower resolution of
$2\times 10^5$ rays.

The beam size in Figs.~\ref{fig:perfect}(e)-(f) oscillates with a
twice smaller period and an amplitude of $\simeq 0.3~\mu$m around an
average value of $\simeq 15.2~\mu$m. The oscillations are independent
of the number of rays and therefore this effect is not the artifact
discussed in the previous paragraph. The average beam size is smaller
than the source size value of $15.9~\mu$m [See
  Table~\ref{tab1}]. According to Eqs.~\eqref{eq0301}-\eqref{eq0300},
this indicated that the numerical value of parameter $|S|^2 \not =1$
[See Eq.~\eqref{eq0300}], i.e.  x-ray source parameters are chosen
slightly off those of a Gaussian beam with Rayleigh length $\zr$ at
its waist.


Figure~\ref{fig:perfect} also shows the evolution of the beam
intensity. The intra-cavity intensity of x-rays decays exponentially,
with a half lifetime of $t_{\ind{1/2}}=10.8(1)$ cycles [see
  Figs.~\ref{fig:perfect}(g)-(h)]. We note that the faster decay
during the first cycles occurs because the x-ray source bandwidth used
in the calculations ($\simeq 100$~meV) is slightly larger than the
$\simeq 90$~meV bandwidth of the 400 Bragg reflection of the diamond
crystals.  The round-trip intensity losses are about 6.4\%. Of this,
$\simeq 3$~\% is coupled out through 20-$\mu$m-thick crystal
C$_{\ind{1}}$. Figure~\ref{fig:perfect}(i) shows the relevant
out-coupling time dependence, with a decay constant of 9.3 cycles.
The remaining losses are due to photoabsorption in the lenses and
crystals.

Figures~\ref{fig:perfect}(j)-(l) show the time dependence of leakage
through crystals C$_{\ind{2}}$-C$_{\ind{4}}$, respectively, of x-rays
belonging to the tails of the intra-cavity spectral distribution. The
magnitudes of the signals are small, and they vanish rapidly as the
tails become sharper after each Bragg reflection. If the crystal
C$_{\ind{1}}$ thickness is increased to match that of crystals
C$_{\ind{2}}$-C$_{\ind{4}}$, the decay constant in
Fig.~\ref{fig:perfect}(i) becomes as fast as those in
Fig.~\ref{fig:perfect}(j)-(l).

\begin{figure}[t!]
    \centering
   \includegraphics[width=1.0\linewidth]{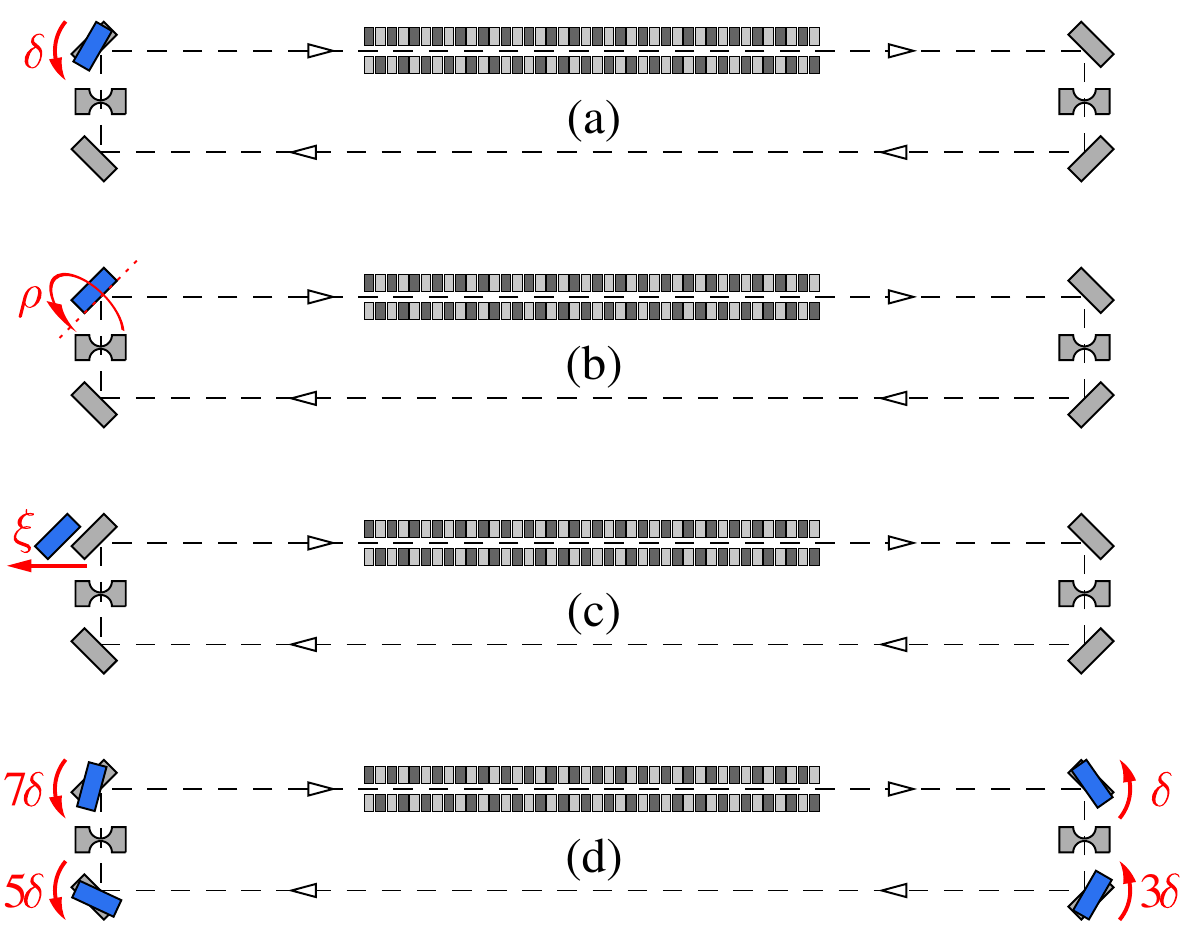}
    \caption{Schematics of crystal misalignment scenarios.
      One-crystal misalignment: (a) yaw angle $\delta$,
    (b) roll angle $\rho$, and (c) position $\xi$ for crystal C$_{\ind{4}}$.
      Systematic multi-crystal misalignment: (d) yaw angles $\delta$, $3\delta$, $5\delta$, and
      $7\delta$ for crystals C$_{\ind{1}}$-C$_{\ind{4}}$, respectively.}
    \label{fig:schematic-crystal-errors}
\end{figure}

\subsection{Alignment errors}

Having established the baseline performance for a perfectly aligned cavity,
we now model the effects of typical
angular and spatial alignment errors on the x-ray beam trajectory in
the cavity and how these in turn are reflected in the temporal response
of the beam monitors.  In most cases, we consider alignment error of a single
degree of freedom of a single optical component or of the x-ray source.

\begin{figure}[b]
    \centering
   \includegraphics[width=1.0\linewidth]{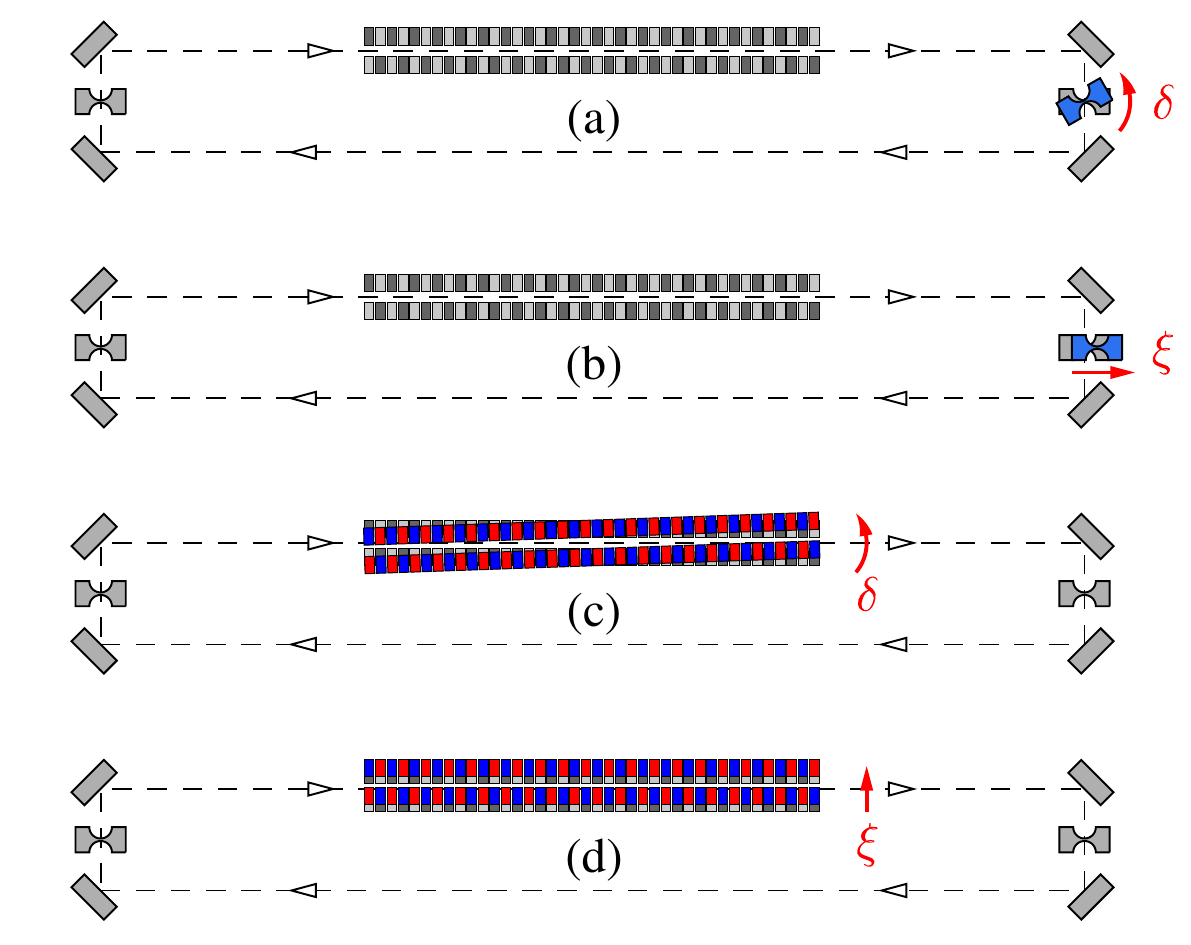}  
   \caption{Additional misalignment scenarios. 
     (a) Lens L$_{\ind{1}}$ with yaw angle error $\delta$ or (b) spatial error $\xi$;  (c) undulator source with yaw angle $\delta$ or (d) position misalignment $\xi$.}
    \label{fig:schematic-other-errors}
\end{figure}

For example, one crystal in the cavity can be misaligned by yaw
angle $\delta$ [crystal C$_{\ind{4}}$ in
Fig.~\ref{fig:schematic-crystal-errors}(a)], which corresponds to the rotation around the
$\hat{y}$-axis perpendicular to the crystal cavity, while the other
crystals are aligned perfectly at the designed \SI{45}{\degree} to the
optical axis and perpendicular to the cavity plane. Alternatively,
the roll angle $\rho$ of one crystal can be misaligned [crystal C$_{\ind{4}}$ in
Fig.~\ref{fig:schematic-crystal-errors}(b)], corresponding
to rotation around the axis formed by the intersection of the
crystal surface and the cavity
plane. A typical spatial
alignment error $\xi$ of crystal C$_{\ind{4}}$ along the $\hat{x}$-axis is shown in Figure~\ref{fig:schematic-crystal-errors}(c).
The arrows on the schematics show in all cases the positive direction of
the alignment error.

\begin{figure*}[t!]
\centering
\begin{minipage}{0.49\textwidth}
\centering
        \includegraphics[width=0.63\linewidth]{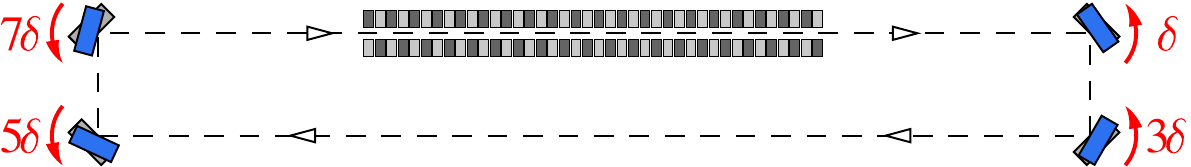}
\includegraphics[width=0.99\linewidth]{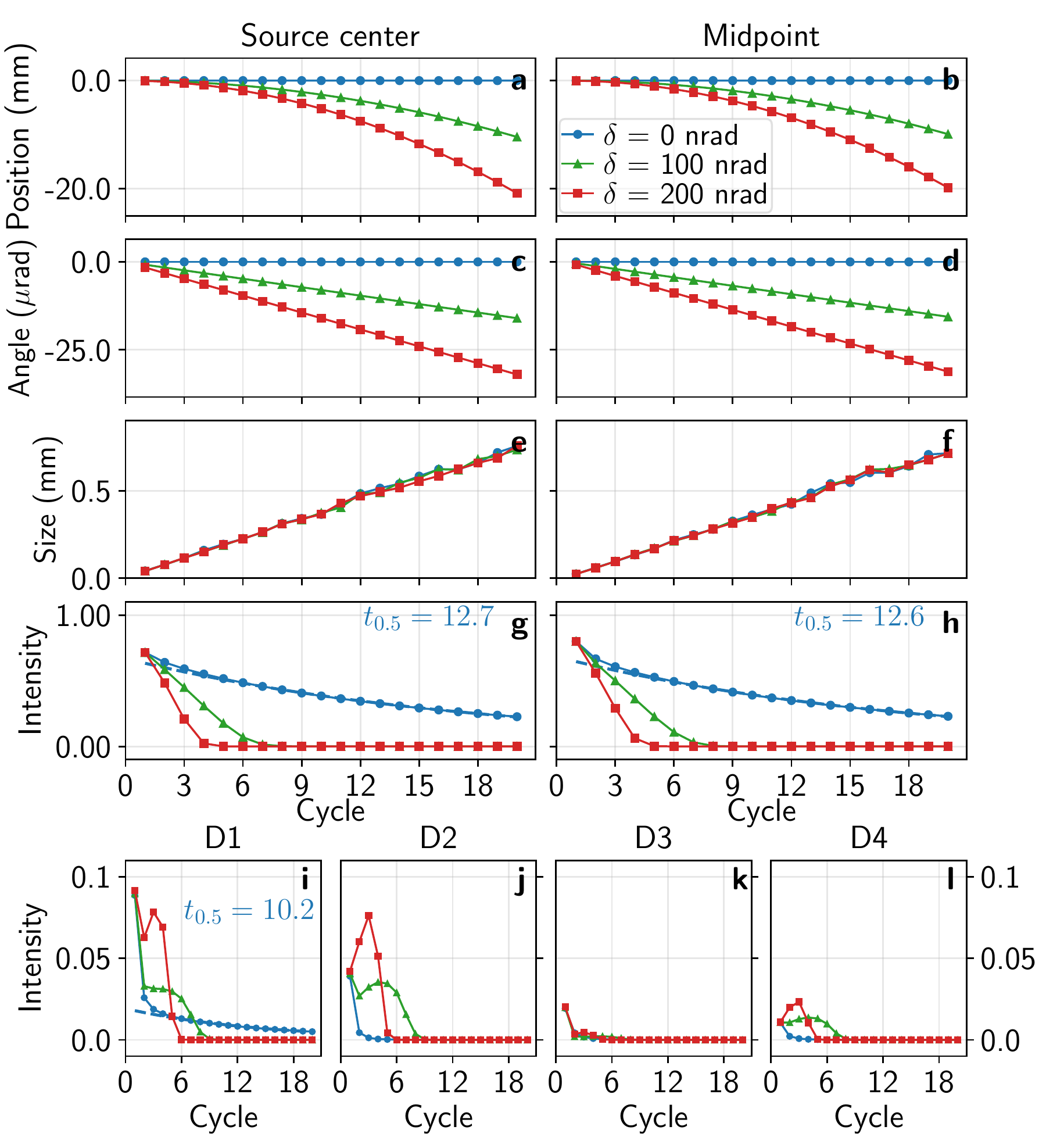}
\caption{Simulation results of the beam dynamics in the rectangular four-crystal
  cavity without lenses with systematically misaligned crystal yaw angles.}
\label{fig:systematic-nolenses}
\end{minipage}
\begin{minipage}{0.49\textwidth}
  \centering
          \includegraphics[width=0.63\linewidth]{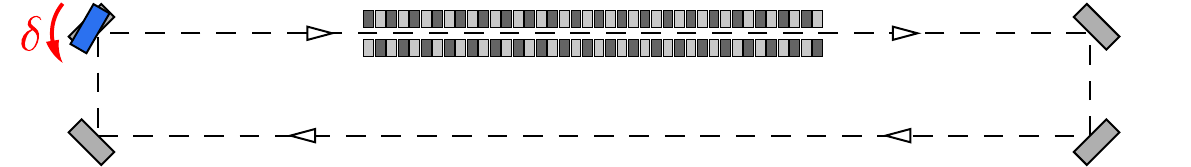}
\includegraphics[width=0.99\linewidth]{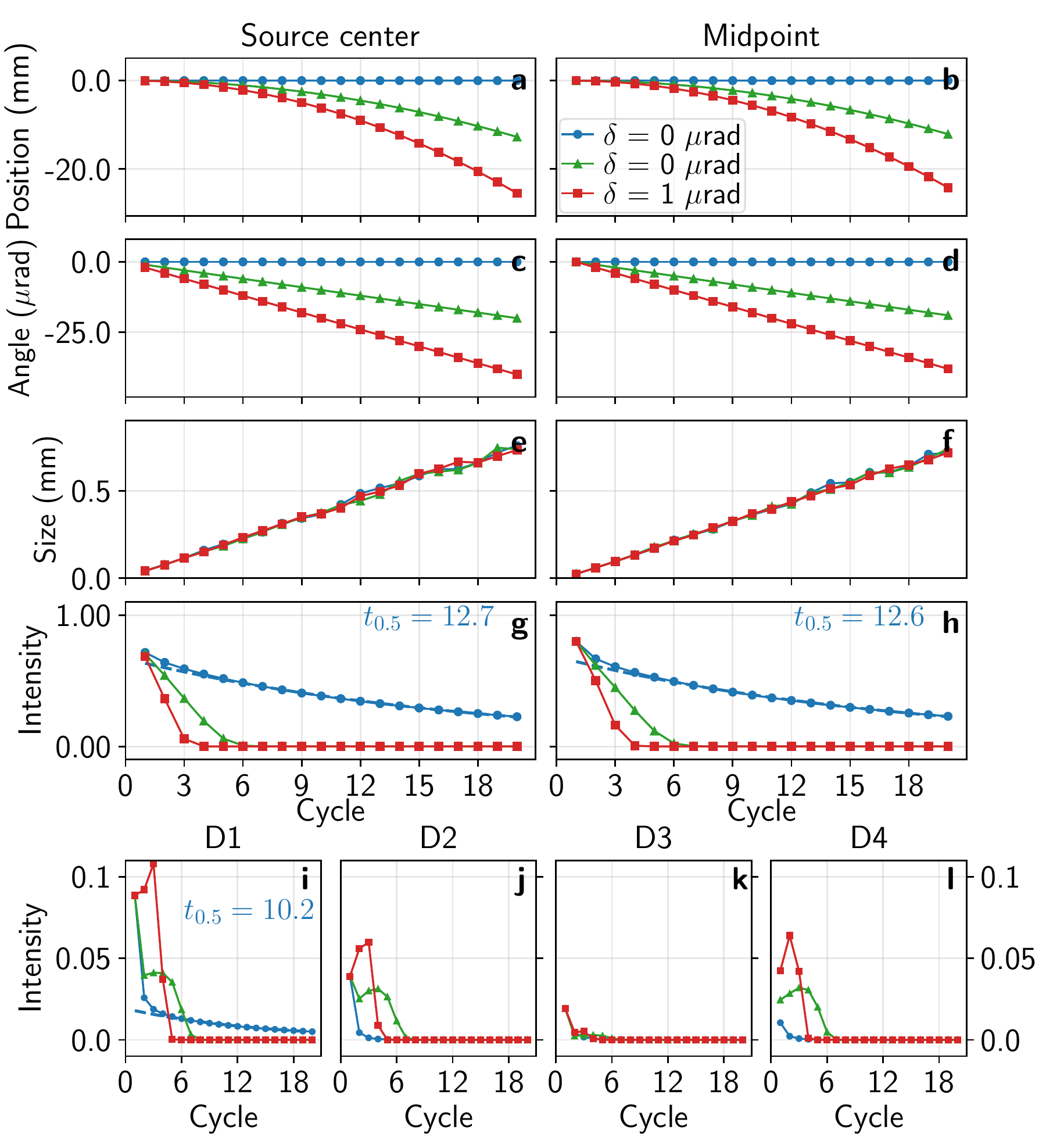}
\caption{Similar to Fig.~\ref{fig:systematic-nolenses}, but with yaw angle  error      $\delta$ in crystal C$_{\ind{4}}$.}
\label{fig:c4-angle-nolenses}
\end{minipage}
\centering
\begin{minipage}{0.49\textwidth}
  \centering
          \includegraphics[width=0.63\linewidth]{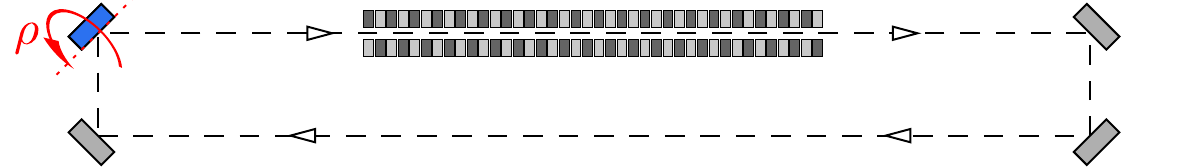}  
\includegraphics[width=0.99\linewidth]{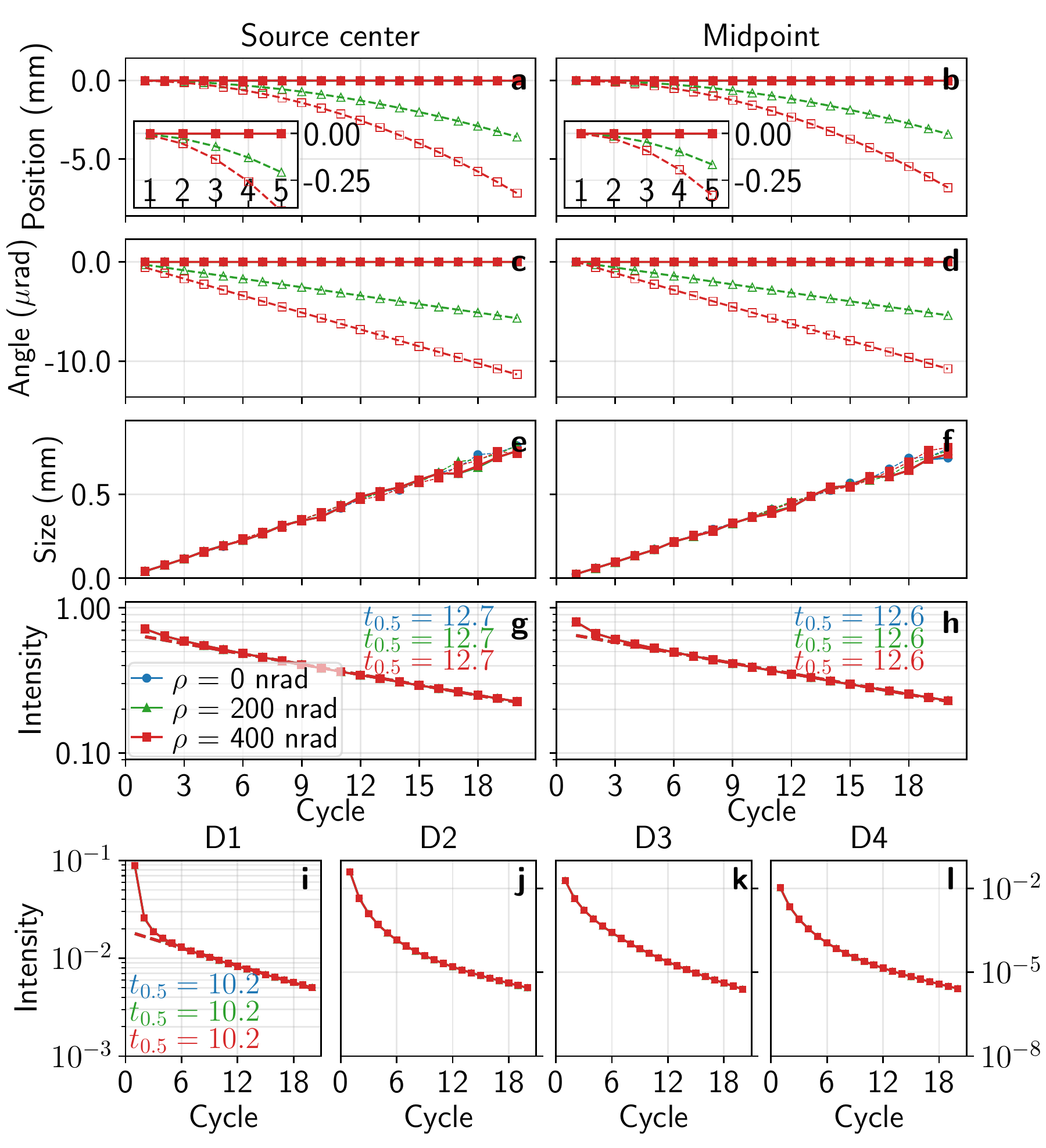}
\caption{Similar to Fig.~\ref{fig:systematic-nolenses}, but with roll angle error $\rho$ in crystal C$_{\ind{4}}$. The dashed
  lines show out-of-cavity-plane deflections.}
\label{fig:c4-roll-angle-nolenses}
\end{minipage}
\begin{minipage}{0.49\textwidth}
  \centering
            \includegraphics[width=0.63\linewidth]{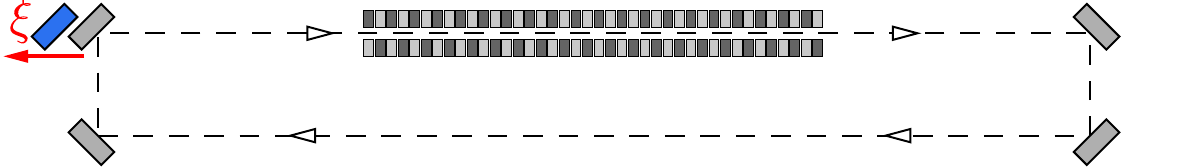}
\includegraphics[width=0.99\linewidth]{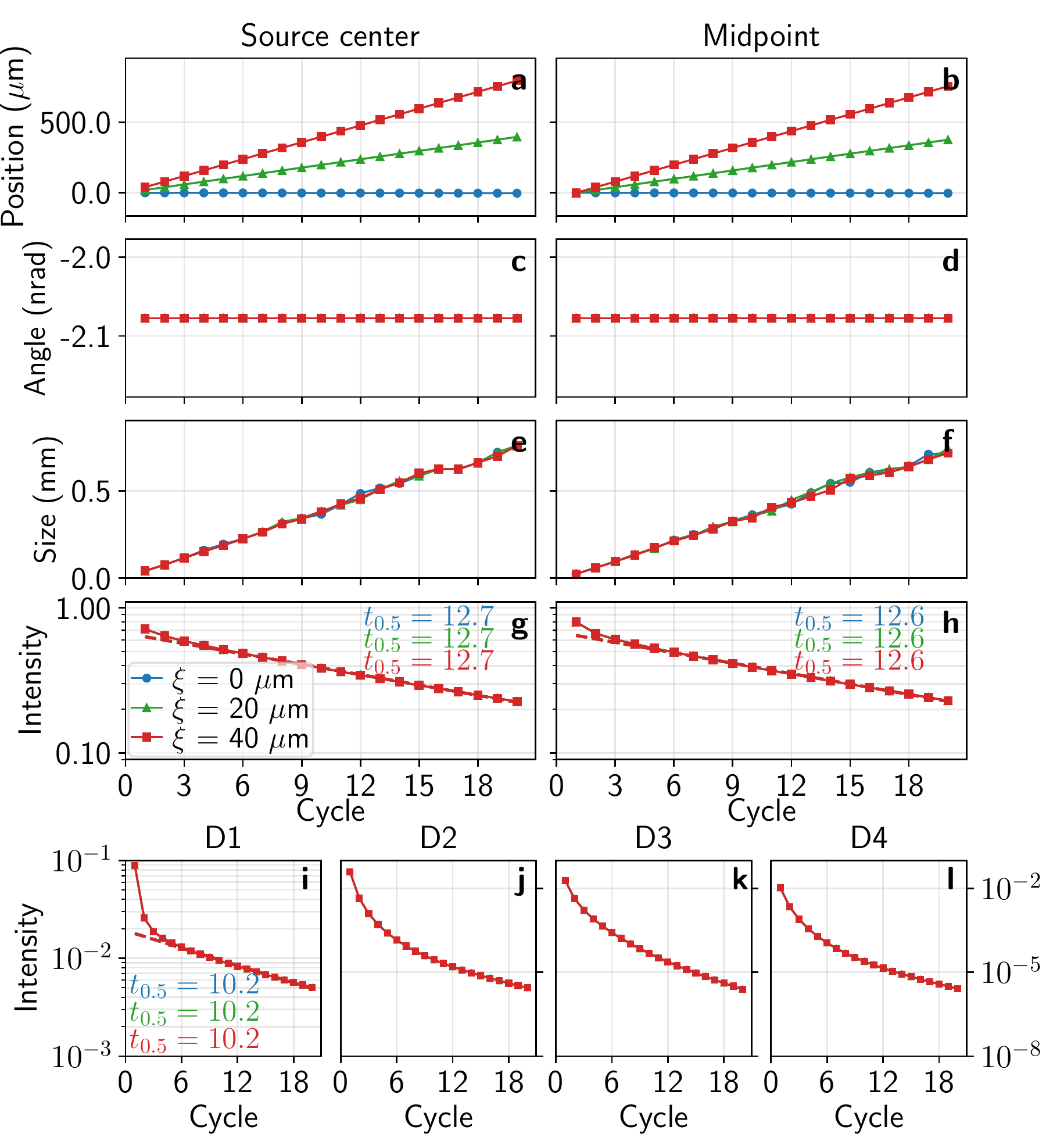}
\caption{Similar to Fig.~\ref{fig:systematic-nolenses}, but 
      with spatial  error $\xi$ in crystal C$_{\ind{4}}$.}
\label{fig:c4-position-nolenses}
\end{minipage}
\end{figure*}

Along with single-element alignment errors, we also consider a special
multi-crystal error case, which we refer to as systematic
angular misalignment.  This kind of error is typical for
fixed-photon-energy cavities, such as the four-crystal cavity we are
considering here, which require the yaw angles (Bragg angles) of all
crystals to be fixed. Such multi-crystal error occurs naturally as a
result of the cavity alignment procedure. Cavity alignment starts
with installation of crystal C$_{\ind{1}}$ as close as possible to
\SI{45}{\degree}to the incident beam; however, this placement can never be
exactly \SI{45}{\degree}.  If the first crystal is misaligned by yaw
angle $\delta$, then the reflected beam is $2\delta$ off the optical
axis. To achieve the highest reflectivity at every next crystal, the
second, third, and fourth crystals have to be misaligned by angles
$3\delta$, $5\delta$, and $7\delta$, respectively, as shown in
Fig.~\ref{fig:schematic-crystal-errors}(d). The beams reflected from
these crystals are $4\delta$, $6\delta$, and $8\delta$ off the optical
axis, respectively. In this scenario, we assume that no lenses
are installed.

In later sections, we study the effects of such crystal alignment
errors on x-ray beam dynamics in the cavity; however, we first consider
a pure crystal cavity without lenses, in which the effect of the
angular and spatial errors can be disentangled.

Other possible misalignment errors discussed later are
presented in Fig.~\ref{fig:schematic-other-errors}.

\subsection{Lensless cavity with crystal alignment errors}

Figure~\ref{fig:systematic-nolenses} shows simulation results of x-ray
beam dynamics in the four-crystal cavity without lenses with
systematically misaligned crystal yaw angles, corresponding to
Fig.~\ref{fig:schematic-crystal-errors}(d).
Figures~\ref{fig:c4-angle-nolenses}, \ref{fig:c4-roll-angle-nolenses},
and \ref{fig:c4-position-nolenses} show numerical simulation results
when crystal C$_{\ind{4}}$ has yaw angle alignment error $\delta$,
roll angle error $\rho$, or spatial alignment error $\xi$,
respectively, corresponding to
Fig.~\ref{fig:schematic-crystal-errors}(a)-(c).  In all these figures,
the results are organized like those for the perfectly aligned cavity
in Fig.~\ref{fig:perfect}, with one exception: in
Figs.~\ref{fig:systematic-nolenses}(i)-(l) and
~\ref{fig:c4-angle-nolenses}(i)-(l), the time response in XBIMs
D$_{\ind{1}}$-D$_{\ind{4}}$ is shown on a linear scale. Each figure
includes data for a perfectly aligned lensless cavity.

The intra-cavity intensity half lifetime increases from 10.6 cycles
for a perfect cavity with lenses (Fig.~\ref{fig:perfect}) to 12.7
cycles for a perfect lensless cavity, because the lenses and their
associated losses are absent.

For all instances of the lensless cavity (including perfectly aligned)
the beam size rapidly increases by $\simeq 40~\mu$m/cycle (rms),
because there is no refocusing.

For cavities with yaw angle errors, each cycle increases both the
angular and spatial deviation of the beam from the optical axis at
both source and midpoint. The angular deviation increases linearly,
and the spatial deviation increases quadratically. The systematic
multi-crystal yaw error produces an effect 4 to 5 times larger than
the one-crystal yaw angle error.  Already after a few cycles, the
angular deviations become larger than the half width (4.5~$\mu$rad) of
the angular width of the 400 Bragg reflection from diamond. As a
result, leakage through the crystals increases, causing increased
signals at XBIMs D$_{\ind{1}}$-D$_{\ind{4}}$ compared to the perfectly
aligned lensless cavity [see
  Figs.~\ref{fig:systematic-nolenses}(i)-(l) and
  Figs.~\ref{fig:c4-angle-nolenses}(i)-(l)].  The intra-cavity
intensity also decreases rapidly with yaw angle error $\delta$ [see
  Fig.~\ref{fig:systematic-nolenses}(g)-(h) and
  Fig.~\ref{fig:c4-angle-nolenses}(g)-(h)].

In contrast, the effect of the roll angle error $\rho$ of crystal
C$_{\ind{4}}$ is much smaller, both on the in-cavity-plane beam
position and on the in-cavity-plane angular deviation from the optical
axis, as shown by solid lines in
Figs~\ref{fig:c4-roll-angle-nolenses}(a)-(d). The effect is small
because there is only a weak coupling between variation in roll angle
and variation in the angle of incidence of x-rays to the reflecting
atomic planes in the crystals.  As a result, the intra-cavity decay
curves in Figs.~\ref{fig:c4-roll-angle-nolenses}(g)-(h) and the signal
variations in XBIMs D$_{\ind{1}}$-D$_{\ind{4}}$ in
Figs.~\ref{fig:c4-roll-angle-nolenses}(i)-(l) do not differ from those
for the perfectly aligned lensless cavity.  The main effect of the
roll angle error is in the out-of-plane deviation of the beam position
and angle at both source and midpoint, as shown by the dashed lines in
Figs~\ref{fig:c4-roll-angle-nolenses}(a)-(d).  The roll angle
misalignment can be corrected in the lensless crystal cavity by
observing the beam profile at the midpoint location.

For all types of crystal angular errors, a positive angular error
results in negative angular and negative spatial deviation of the
x-ray beam from the optical axis.

The spatial alignment error $\xi$ of crystal C$_{\ind{4}}$ results
only in a rapid linear increase of the beam position deviation from
the optical axis (positive for positive error) at both source and
midpoint, while the beam angle does not change. As a result, the decay
curves in Figs.~\ref{fig:c4-position-nolenses}(g)-(l) are the same as
for the perfectly aligned lensless cavity.

In the lensless cavity, the spatial crystal error does not affect the
angular beam deviations. This property can be used for cavity alignment, as
discussed in Section~\ref{alignment}.

\subsection{Complete generic cavity with crystal alignment errors}

\begin{figure*}[t!]
\centering
\begin{minipage}{0.49\textwidth}
\centering
        \includegraphics[width=0.63\linewidth]{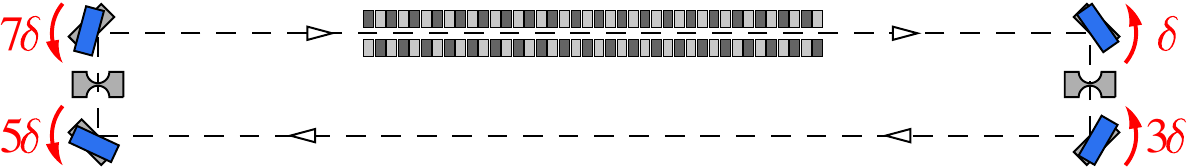}
      \includegraphics[width=0.99\linewidth]{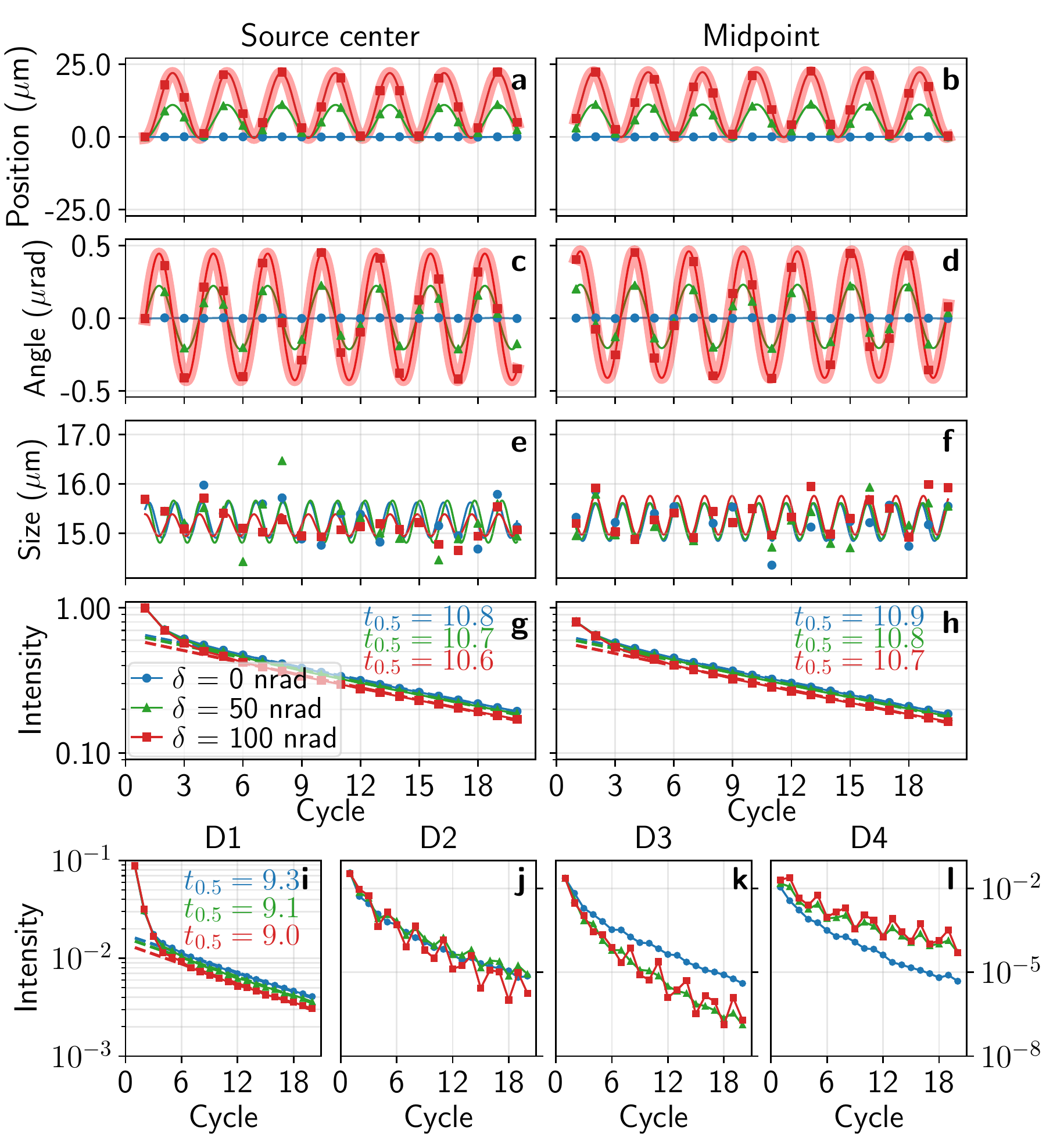}
    \caption{Numerical simulation results  of x-ray beam dynamics in the rectangular four-crystal generic  cavity with systematically misaligned crystal yaw angles.}
    \label{fig:full-simulation-crystal-angle-sys}
\end{minipage}
\begin{minipage}{0.02\textwidth}
\end{minipage}  
\begin{minipage}{0.49\textwidth}
\centering
        \includegraphics[width=0.63\linewidth]{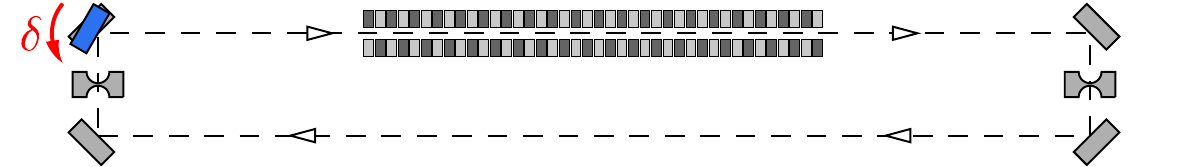}
      \includegraphics[width=0.99\linewidth]{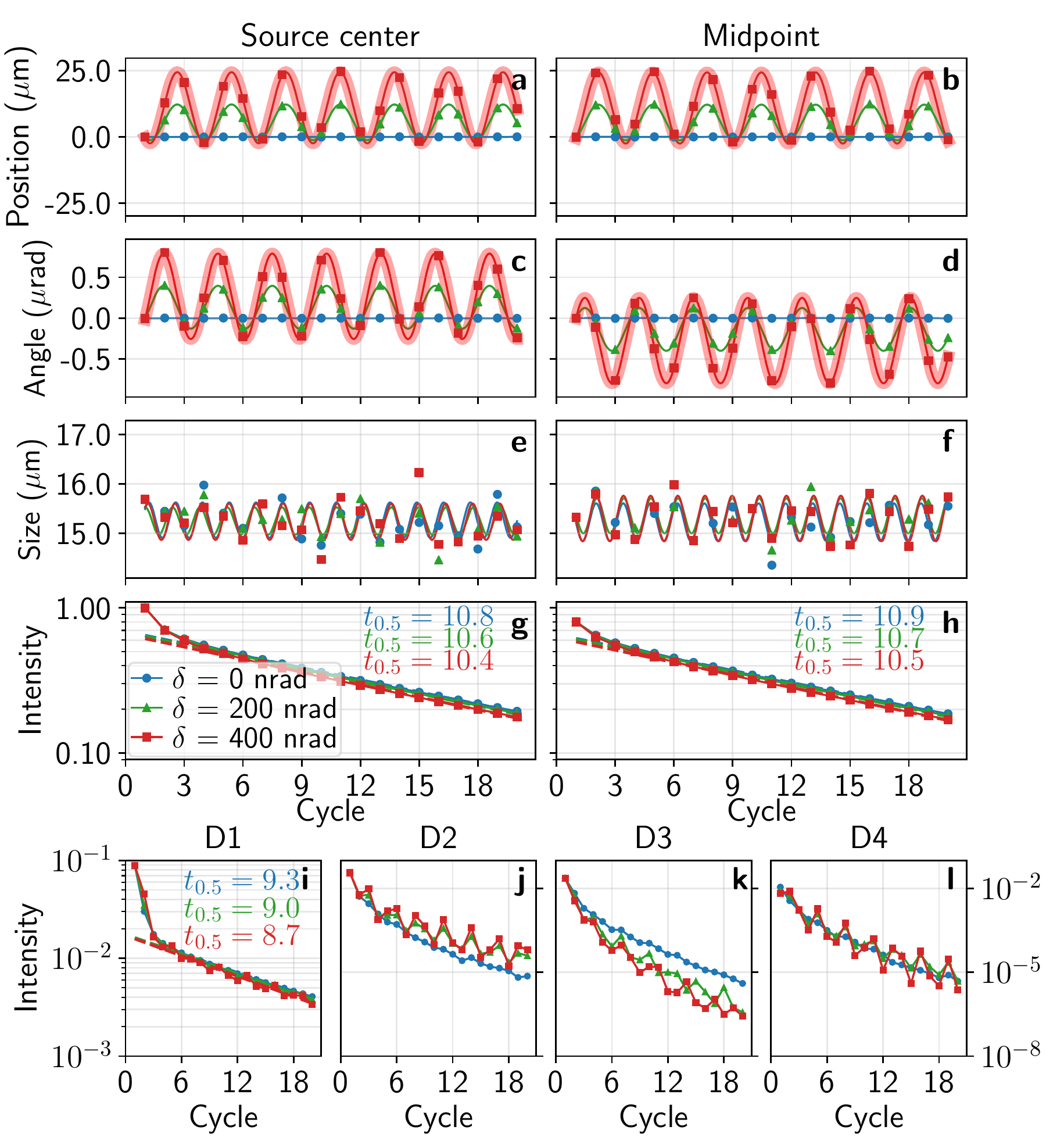}
    \caption{Similar to Fig.~\ref{fig:full-simulation-crystal-angle-sys}, but with yaw angle error $\delta$ in crystal~C$_{\ind{4}}$.}
    \label{fig:full-simulation-crystal-angle-one}
\end{minipage}
\centering
\begin{minipage}{0.49\textwidth}
  \centering
        \includegraphics[width=0.63\linewidth]{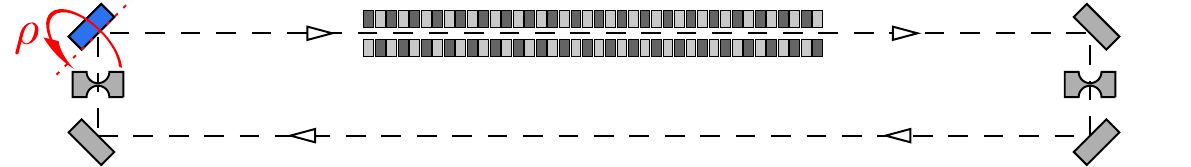}
          \includegraphics[width=0.99\linewidth]{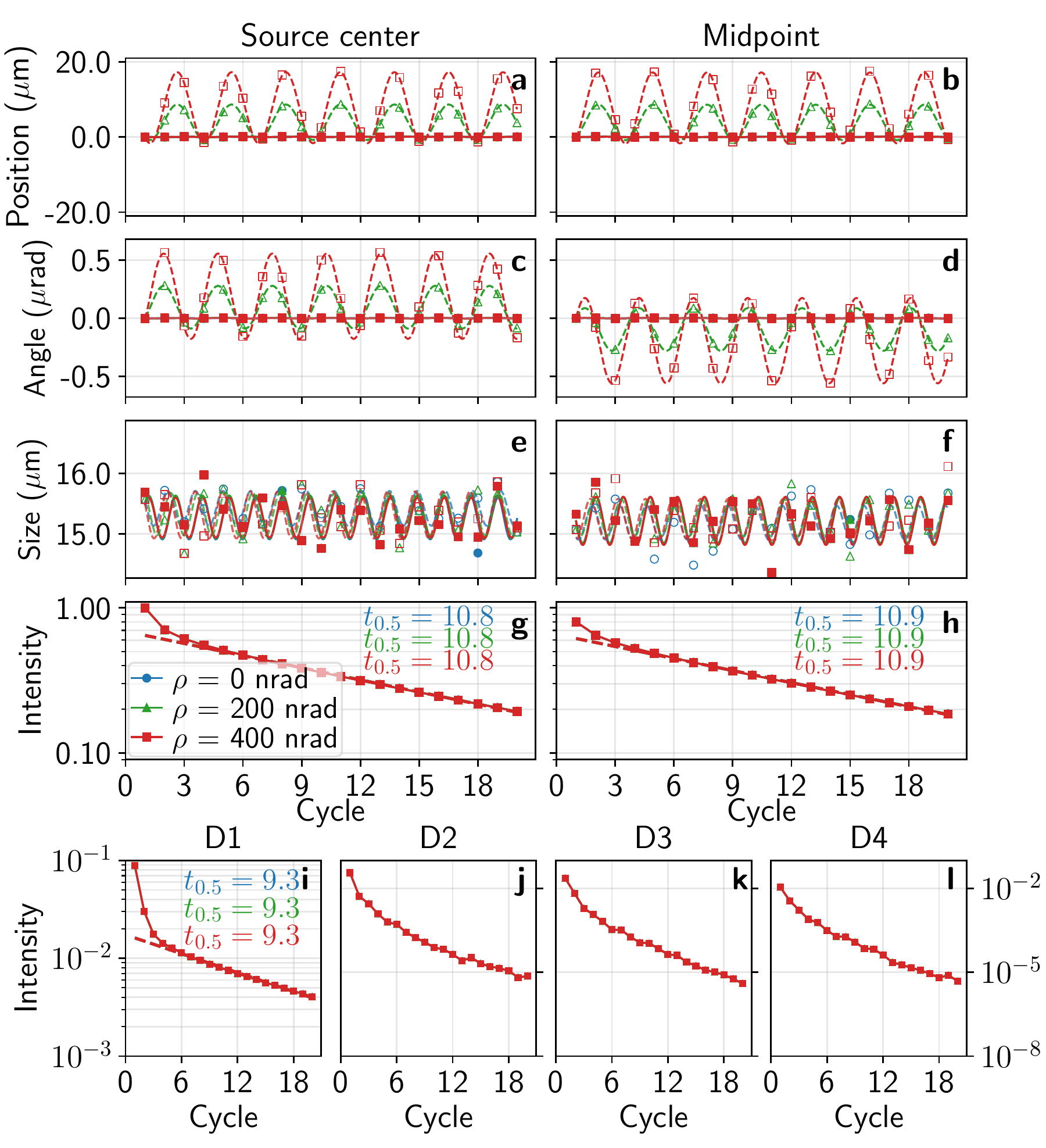}
    \caption{Similar to Fig.~\ref{fig:full-simulation-crystal-angle-one}, but with roll angle error $\rho$ in C$_{\ind{4}}$.\newline The dashed lines represent out-of-cavity-plane deviations.}
    \label{fig:full-simulation-crystal-angle-roll}
\end{minipage}
\begin{minipage}{0.02\textwidth}
\end{minipage}  
\begin{minipage}{0.49\textwidth}
\centering
        \includegraphics[width=0.63\linewidth]{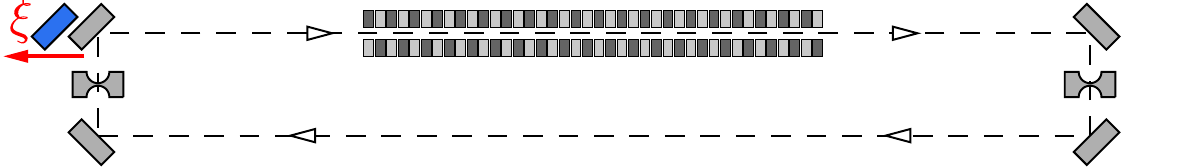}
\includegraphics[width=0.99\linewidth]{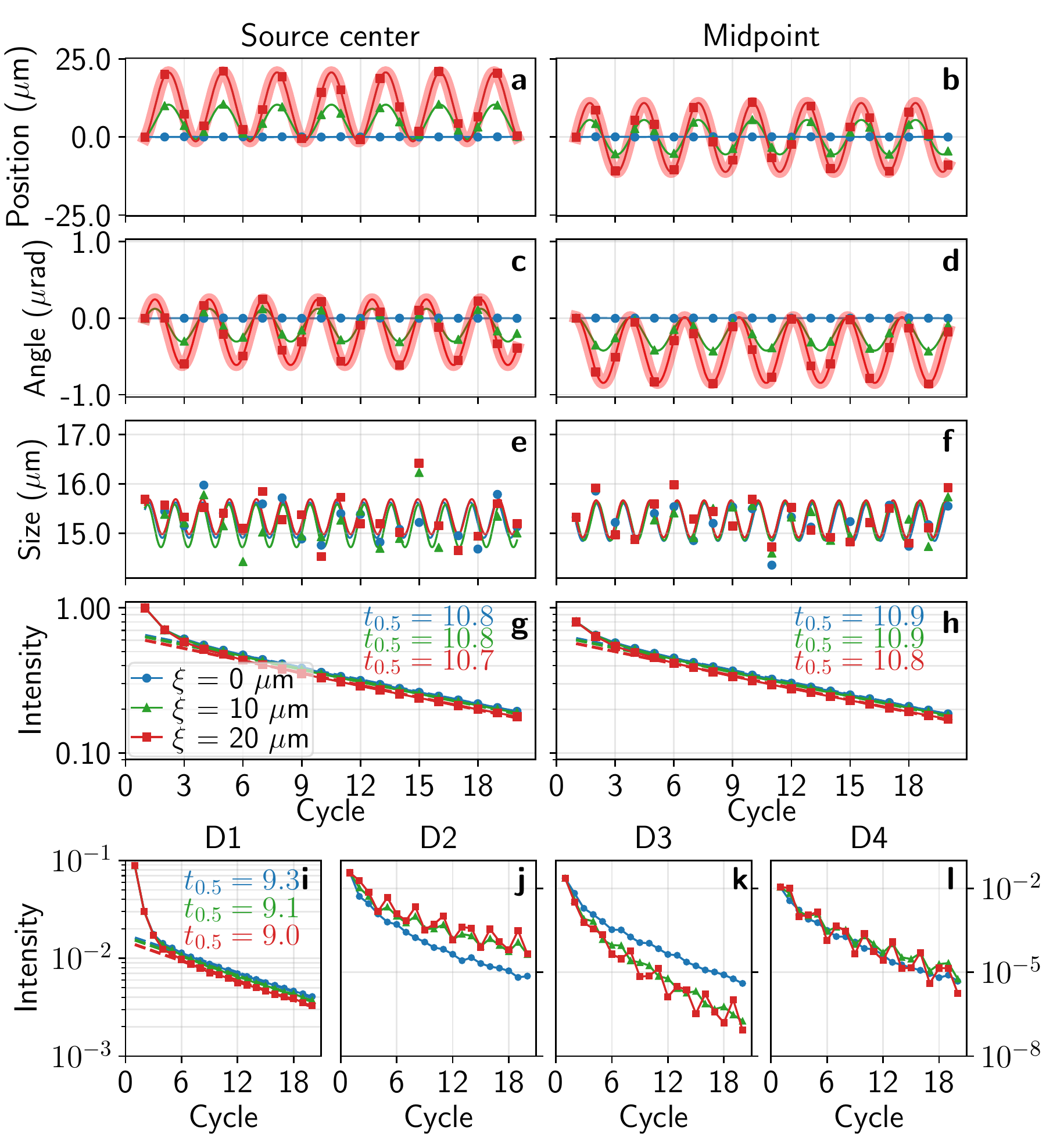}
      \caption{Similar to Fig.~\ref{fig:full-simulation-crystal-angle-one}, but with position error $\xi$ in crystal~C$_{\ind{4}}$.}
    \label{fig:full-simulation-crystal-position}
\end{minipage}
\end{figure*}

We now consider a complete four-crystal rectangular cavity with
lenses, as in Fig.~\ref{fig000}, featuring various crystal alignment
errors.
Figures~\ref{fig:full-simulation-crystal-angle-sys}-\ref{fig:full-simulation-crystal-position}
show numerical simulation results of the x-ray beam dynamics for the
following cases: systematically misaligned yaw angles of crystals
C$_{\ind{1}}$-C$_{\ind{4}}$
(Fig.~\ref{fig:full-simulation-crystal-angle-sys}), one-crystal yaw
angle error in C$_{\ind{4}}$
(Fig.~\ref{fig:full-simulation-crystal-angle-one}), roll angle error
in crystal C$_{\ind{4}}$
(Fig.~\ref{fig:full-simulation-crystal-angle-roll}), or position error
in crystal C$_{\ind{4}}$
(Fig.~\ref{fig:full-simulation-crystal-position}).  The simulation
results in
Figs.~\ref{fig:full-simulation-crystal-angle-sys}-\ref{fig:full-simulation-crystal-position}
are arranged in the same way as those for the perfect cavity in
Fig.~\ref{fig:perfect}, with one exception.  In panels (a)-(d) of
Figs.~\ref{fig:full-simulation-crystal-angle-sys},
\ref{fig:full-simulation-crystal-angle-one}, and
\ref{fig:full-simulation-crystal-position}, we show with thick pink
lines the results of calculations using ray-transfer matrices (see
Section~\ref{misaligned}), for comparison with the results obtained
with the Shadow3 simulation package \cite{SHADOW3}.  Remarkably, they
are practically identical, as long as the angular deviations are much
smaller than the angular width of the Bragg reflections.

\subsubsection{Betatron oscillations}
  
Adding the focusing lenses dramatically changes the x-ray beam
dynamics in the cavity. For the same crystal
alignment errors, the beam size, beam trajectory, and beam
monitor response are completely different in the complete cavity.

Unlike the lensless cavity case, any alignment error produces both
spatial and angular deflections of the beam from the optical axis.

In all presented cases, the pass-to-pass
variations of the spatial or angular deviations from the optical axis
as well as of the beam size are much smaller, compared to the lensless
cavity analogs shown in
Figs.~\ref{fig:systematic-nolenses}-\ref{fig:c4-position-nolenses}. The
beam size variation and the beam position deflection from the optical
axis per cycle are now about 100 times smaller, while the angular
deflection per cycle is 10 times smaller, compared to the analogous
conditions for the lensless cavity.  These changes are clear evidence
that the beam size and beam trajectory are stabilized by the
lenses. This stabilization takes place only if the parameters and
locations of the lenses are chosen properly, as discussed in
Section~\ref{self-consistent}.  We note that we are considering here
the generic cavity case; the confocal cavity case is considered later,
in Section~\ref{confocal-numeric}.

\newcolumntype{g}{>{\columncolor{lightgray}}c}
\begin{table}[t!]
    \centering
    \caption{Comparison of the beam characteristics at the source
      location and at the midpoint of the return path for the cavity with
      the systematically misaligned crystal yaw angles
      ($\delta$ = \SI{0.10}{\micro\radian}).}
    \begin{tabular}{|c||c|c|c|c|c|}
        \hline      
          & \makecell{Beam\\ position \\ (\si[]{\micro\metre})}&
        \makecell{Beam\\ angle \\ (\si[]{\nano\radian})}& 
        \makecell{Beam\\ size\\ (\si[]{\micro\metre})}& 
        \makecell{Half\\ lifetime \\ (cycles)}&
        \makecell{Oscillation\\ period\\(cycles)}\\
        \hline\hline
         Source & 10.82 & 8.67 & 15.17 & 10.8 & 2.77 \\
        \hline
        Midpoint & 11.10 & 8.91 & 15.17 & 10.9 & 2.77 \\
        \hline
    \end{tabular}
    \label{tab:comparison}
\end{table}

\begin{table*}
    \centering
    \caption{X-ray beam characteristics at the source position
      calculated for various crystal misalignment types from the data in
      Figs.~\ref{fig:full-simulation-crystal-angle-sys}-\ref{fig:full-simulation-crystal-position}. A second sign for offset
      values is provided if the sign differs between the source and the midpoint. Offset values for both
      locations are given for the crystal position error
      case.}
    \begin{threeparttable}[t]
      \begin{tabular}{l|c||c|c||c|c||c|c||c}
        \hline
        \Xhline{1pt}
            \multicolumn{2}{c||}{\makecell{\makecell{Alignment\\error}}} &
            \multicolumn{6}{c|}{Betatron oscillation type and parameters} &
            \makecell{ \\ Half\\ lifetime\\(cycles)}\\
            \cline{3-8}
            \multicolumn{2}{c||}{} &
            \multicolumn{2}{c||}{\makecell{\makecell{Position \\(\si{\micro\metre})}}} &
            \multicolumn{2}{c||}{\makecell{\makecell{Angle \\(nrad)}}} &
            \multicolumn{2}{c||}{\makecell{\makecell{Size \\(\si{\micro\metre})}}} & \\
            \cline{1-8}
            Type & \makecell{Value} & Offset & \makecell{Amplitude \\(rms)} & Offset & \makecell{Amplitude \\(rms)} & Average & \makecell{Amplitude \\(rms)} \\
            \hline 
            \Xhline{1pt}
             None &0 & 0 & 0.02 & 0 & 0 & 15.47 & 0.26 & 10.8  \\ 
            \Xhline{1pt}
            \multirow{ 2}{*}{\makecell{C$_{\ind{1}}$-C$_{\ind{4}}$ systematic\\ yaw angle $\delta$ (nrad) }} 
            &50 & 5.41 &4.03& 4 & 151 &  15.24 & 0.45 &  10.7  \\
            \cline{2-9}
            &100 & 10.82 & 8.03 & 9 & 303 &  15.17 & 0.27 &  10.6  \\ 
            \Xhline{1pt}
            \multirow{ 2}{*}{\makecell{C$_{\ind{4}}$  yaw angle $\delta$\\ (nrad)}} 
            &200 & 5.49 & 4.69 & $\pm$ 134 & 188 & 15.25 & 0.33 &  10.6  \\
            \cline{2-9}
            &400 & 10.99 & 9.34 & $\pm$ 268 & 375 & 15.24 & 0.39 &  10.4  \\ 
            \Xhline{1pt}
            \multirow{2}{*}{\makecell{C$_{\ind{4}}$  roll angle $\rho$\tnote{1} \\ (nrad)}} 
            &200 & 3.88& 3.29 & $\pm$ 95 & 132 &15.33 &0.33  & 10.8\\
            \cline{2-9}
            &400 &7.77 &6.57 &$\pm$ 189 & 264 &15.31 &0.32  & 10.8\\
            \Xhline{1pt}
            \multirow{2}{*}{\makecell{C$_{\ind{4}}$ position $\xi$\\ ($\mu$m)}} 
            &10 &  \makecell{4.81\\(-0.09)} & 4.03 &  \makecell{-0.09\\ (-0.21)} & 0.15 & 15.15 & 0.42   & 10.8\\
            \cline{2-9}
            &20 & \makecell{9.63\\(-0.17)} & 8.04 & \makecell{-0.18\\(-0.42)} & 0.29  &15.34  & 0.41   & 10.7\\
            \hline
            \Xhline{1pt}
        \end{tabular}
        \begin{tablenotes}
            \item[1] \footnotesize{The roll-angle alignment errors
              result primarily in off-cavity-plane x-ray beam
              deflections, with the corresponding values shown in the
              table. The effect on the in-cavity-plane beam
              deflections is minor and not shown.}
        \end{tablenotes}
    \end{threeparttable}
\label{tab:angle_misalignment}
\end{table*}

Unlike the steadily increasing beam size and increasing spatial and
angular beam deviations from the optical axis in the lensless cavity,
the beam size and beam deviations in the complete cavity experience
periodic variations having a constant amplitude and constant offset
from the optical axis. This effect is found at both source and
midpoint for all simulated cases, as shown in
Figs.~\ref{fig:full-simulation-crystal-angle-sys}-\ref{fig:full-simulation-crystal-position}.
Remarkably, the oscillation period of the beam deviations is always
2.77 cycles, independent of the type of misalignment, while the beam
size oscillation period is half that value.  These values correspond
exactly to the period of the betatron oscillations in a perfectly
aligned generic cavity with a misaligned source, as derived
analytically in Section~\ref{selfconsistent-generic} [see
  Eqs.~\eqref{eq0137}, \eqref{eq0220}, and
  \eqref{eq0301}-\eqref{eq0303}] for a CBXFEL system with the same
cavity and source parameters.  These betatron oscillations are not
specific to this particular cavity. Rather, they are a generic
signature of misalignments in periodic focusing systems composed of
lenses or mirrors, whether these are laser cavities \cite{Siegman} or
accelerators \cite{ES92}.

The amplitudes and offsets of the betatron oscillation increase in
proportion to the alignment error values\footnote{We note that the
  amplitudes of the betatron oscillation in
  Figs.~\ref{fig:full-simulation-crystal-angle-sys}-\ref{fig:full-simulation-crystal-position}
  are much larger than those in Fig.~\ref{fig:perfect}, which are due
  to the artifact arising from use of a finite number of discrete rays in the beam, as
  discussed in Section~\ref{perfect}. This difference means that the oscillations in  Figs.~\ref{fig:full-simulation-crystal-angle-sys}-\ref{fig:full-simulation-crystal-position} are
  real effects.}.  Only in a few cases, which we will term symmetric,
are the oscillation offsets zero.  One such case is 
systematic angular misalignment, for which the offset of the
angular oscillations is zero at both source and midpoint and is independent
of the alignment error value [see
  Fig.~\ref{fig:full-simulation-crystal-angle-sys}(c)-(d)], while
simultaneously the offset of the spatial oscillations is non-zero [see
  Fig.~\ref{fig:full-simulation-crystal-angle-sys}(a)-(b)]. A second case is a cavity with position
error $\xi$  in crystal C$_{\ind{4}}$ [see Fig.~\ref{fig:full-simulation-crystal-position}(a)], for which
the offset of the spatial oscillations does reach zero, but only at the
midpoint of the return pass.  In all other cases, the oscillation
offsets are non-zero, and we will refer to such cases as asymmetric
betatron oscillations.

Asymmetric betatron oscillations take place only on one side of the
optical axis, with the beam touching or barely crossing
the axis. The side on which oscillations occur depends on the sign of the
  alignment error or on which crystal is misaligned. In the first approximation, the offset of
the betatron oscillations is equal to their amplitude. The smallest deflection
from the optical axis for any error value or type is near zero
(the beam is almost on the optical axis), while the largest deviation
equals two amplitudes.

Another common signature can be observed in the beam dynamics in the
cavities with {\em angular} crystal alignment errors.  In all such
cases (see
Figs.~\ref{fig:full-simulation-crystal-angle-sys}-\ref{fig:full-simulation-crystal-angle-roll}),
the beam characteristics (size and deflections) are very similar, both
at the source center and at the midpoint.  See, for example, the
values in Table~\ref{tab:comparison} for the case of systematically
misaligned yaw angles.  Stated another way, the beam monitors at the
midpoint reflect the beam status at the source.

This similarity of beam characteristics between source and midpoint is
observed in most of the simulations, with one exception. In the case
of spatial error in crystal C$_{\ind{4}}$, the offset values of the
betatron oscillations at the two locations are different, as shown in
Fig.~\ref{fig:full-simulation-crystal-position}. However, the beam
size and the amplitudes of the betatron oscillations are still the
same at both locations.

Table~\ref{tab:angle_misalignment} presents x-ray beam characteristics
at the source location calculated for various crystal misalignment
types from the data in
Figs.~\ref{fig:full-simulation-crystal-angle-sys}-\ref{fig:full-simulation-crystal-position}.
Shown are offsets and amplitudes of the betatron oscillations for the
beam position, angle, and size. If the offsets have different signs at
the two locations, both signs are provided, with the upper sign
corresponding to the source center and lower sign to the midpoint
location.  For the case of position error in crystal C$_{\ind{4}}$, we
provide offset values for both locations.

The {\em smallest} crystal alignment errors shown in the table are
chosen such that the rms amplitude of the betatron oscillations {\em
  for the beam position} is less than 5~$\mu$m, i.e., smaller than one
third of the 16-$\mu$m rms source size (see Table~\ref{tab1}). These
smallest alignment errors could be considered as alignment error
tolerances.

There are several issues to be noted regarding the signs of the
offsets. For all positive crystal alignment errors (angular and
spatial), the offset of the spatial betatron oscillations is positive
at all monitored positions, with one exception, namely, for spatial
error at crystal C$_{\ind{4}}$. For that case, the spatial betatron
oscillations are symmetric with zero offset at the midpoint
location. For the angular betatron oscillations, the offset is zero
for the systematic misalignment case, negative at both source and
midpoint for the case of spatial error in crystal C$_{\ind{4}}$, and
opposite in sign for the cases of yaw and roll error at crystal
C$_{\ind{4}}$ (positive offset at the source, negative offset at the
midpoint). Note that if a different crystal is faulty, the signs may
change relative to those noted in Table~\ref{tab:angle_misalignment}
for errors at crystal C$_{\ind{4}}$. For example, a faulty crystal
C$_{\ind{3}}$ changes the sign of both position and angular
oscillation. For a faulty crystal C$_{\ind{2}}$, the sign of angular
oscillation changes but that of position does not. The reverse is true
for a faulty crystal C$_{\ind{1}}$, where the position sign changes
but the angular sign does not.

The deviations of the beam from the optical axis in the cavity plane
are observed to be more sensitive to systematic yaw angle misalignment
than to the one-crystal yaw angle misalignment, and least sensitive to
roll angle misalignment.

\begin{figure*}[t!]
\centering
\begin{minipage}{0.49\textwidth}
\centering
      \includegraphics[width=0.99\linewidth]{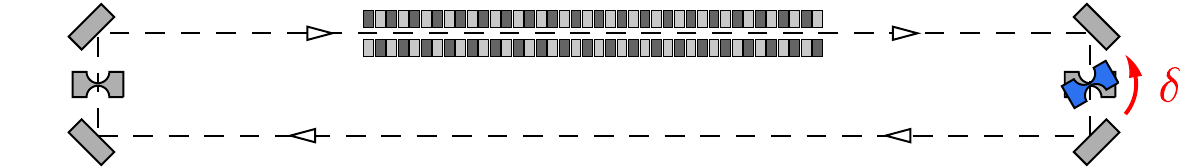}
      \includegraphics[width=0.99\linewidth]{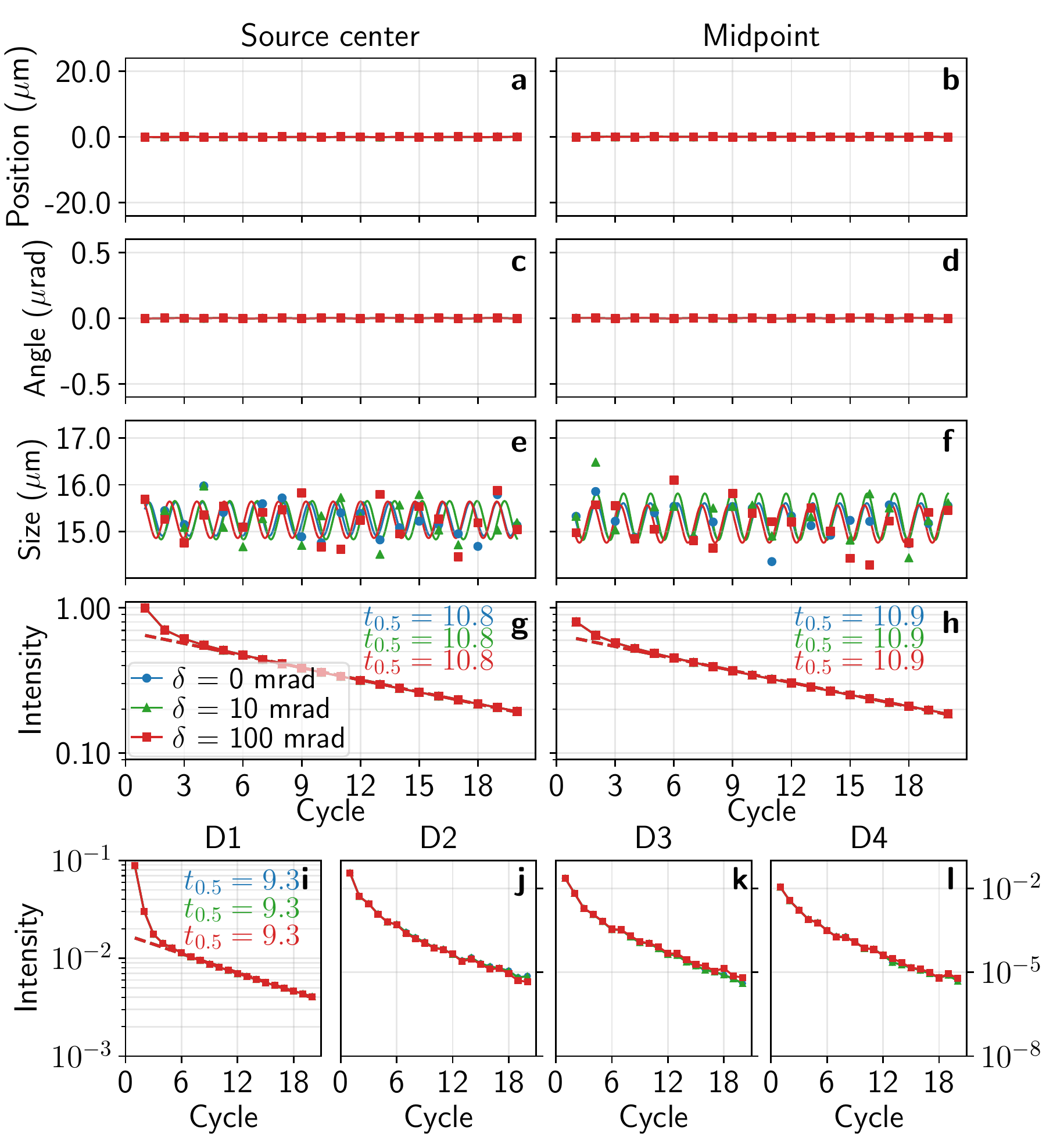}
\caption{Simulation results  of x-ray beam dynamics in the four-crystal generic cavity with yaw angle error $\delta$ at lens L$_{\ind{1}}$.}
    \label{fig:full-simulation-crl-angle}
\end{minipage}
\begin{minipage}{0.02\textwidth}
\end{minipage}  
\begin{minipage}{0.49\textwidth}
  \centering
  \includegraphics[width=0.99\linewidth]{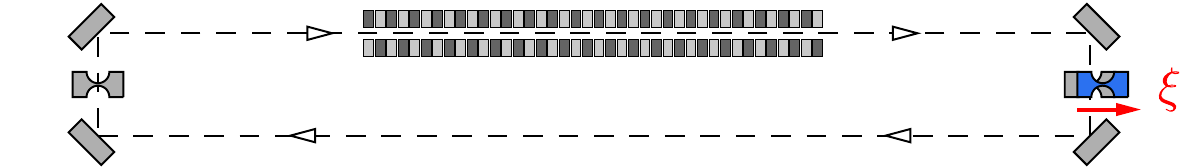}
        \includegraphics[width=0.99\linewidth]{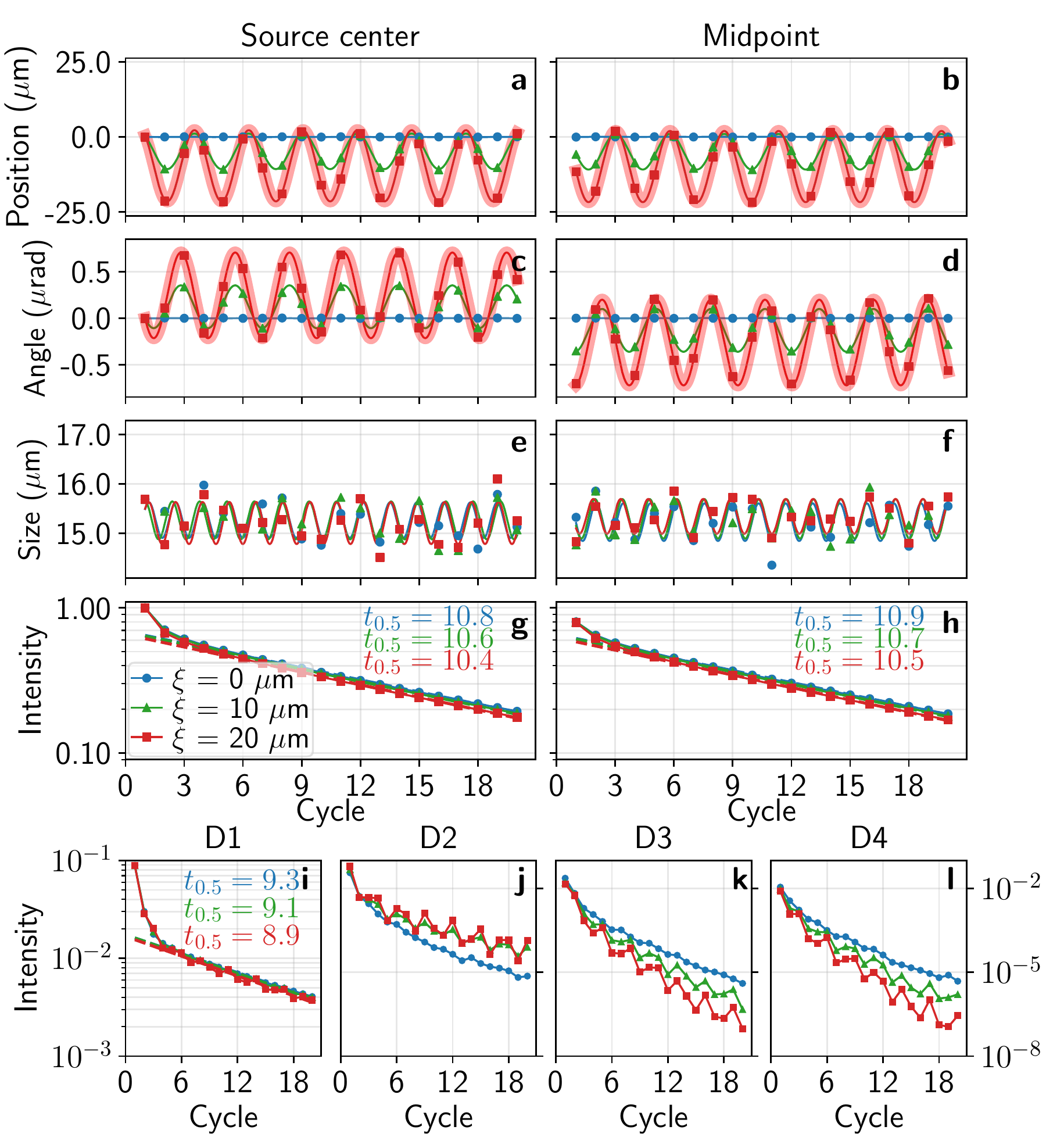}
\caption{Similar to Fig.~\ref{fig:full-simulation-crl-angle}, but with position error $\xi$ at lens L$_{\ind{1}}$.}
    \label{fig:full-simulation-crl-position}
\end{minipage}
\end{figure*}

\begin{table*}[t]
    \centering
    \caption{X-ray beam characteristics at the source position
      simulated for the cavity with yaw angle error $\delta$ or lateral spatial error $\xi$ of one lens derived from the data in
      Figs.~\ref{fig:full-simulation-crl-angle}-\ref{fig:full-simulation-crl-position}. A second sign for offset
      values is provided if the sign differs between the source and the midpoint.  }
    \begin{threeparttable}[t]
      \begin{tabular}{c|c||c|c|c|c|c|c|c}
              \hline
            \multicolumn{2}{c||}{\makecell{\makecell{Alignment\\error}}} &
            \multicolumn{6}{c|}{Betatron oscillation type and parameters} &
            \makecell{Half lifetime \\ (cycles)\\}\\
            \cline{3-8}
            \multicolumn{2}{c||}{} &
            \multicolumn{2}{c|}{\makecell{\makecell{Position \\(\si{\micro\metre})}}} &
            \multicolumn{2}{c|}{\makecell{\makecell{Angle \\(nrad)}}} &
            \multicolumn{2}{c|}{\makecell{\makecell{Size \\(\si{\micro\metre})}}} & \\
            \cline{1-8}
            Type & Value & Offset & \makecell{Amplitude \\(rms)} & Offset & \makecell{Amplitude \\(rms)} & Average & \makecell{Amplitude \\(rms)}\\\hline\hline
             None &0 & 0.00 & 0.02 & 0.00 & 0.00 & 15.47 & 0.26 & 10.8  \\
            \Xhline{1pt}
            \multirow{2}{*}{\makecell{L$_{\ind{1}}$ yaw angle $\delta$\\(mrad)}} 
            & 10 & 0.00 & 0.04 & 0.00 & 0.00 & 15.24 & 0.39 & 10.8\\
            \cline{2-9}
            & 100  & 0.00 & 0.04 & 0.00 & 0.00 & 15.25 & 0.40 & 10.8\\
            \cline{2-9}
            \Xhline{1pt}
            \multirow{ 2}{*}{\makecell{L$_{\ind{1}}$ position $\xi$\\ ($\mu$m)}} 
            & 10 & -4.80 & 4.31 & $\pm$123 & 158 & 15.26 & 0.34 & 10.6\\
            \cline{2-9}
            & 20 & -9.60 & 8.63 & $\pm$246 & 316 & 15.21 & 0.39 & 10.4\\
            \cline{1-9}
        \end{tabular}
    \end{threeparttable}
\label{tab:crl_misalignment}
\end{table*}

\begin{figure*}[t!]
\centering
\begin{minipage}{0.49\textwidth}
  \centering
        \includegraphics[width=0.99\linewidth]{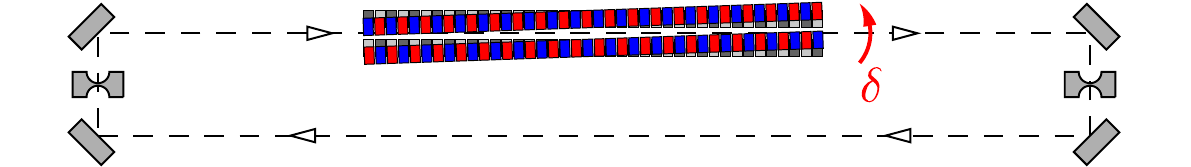}
        \includegraphics[width=0.99\linewidth]{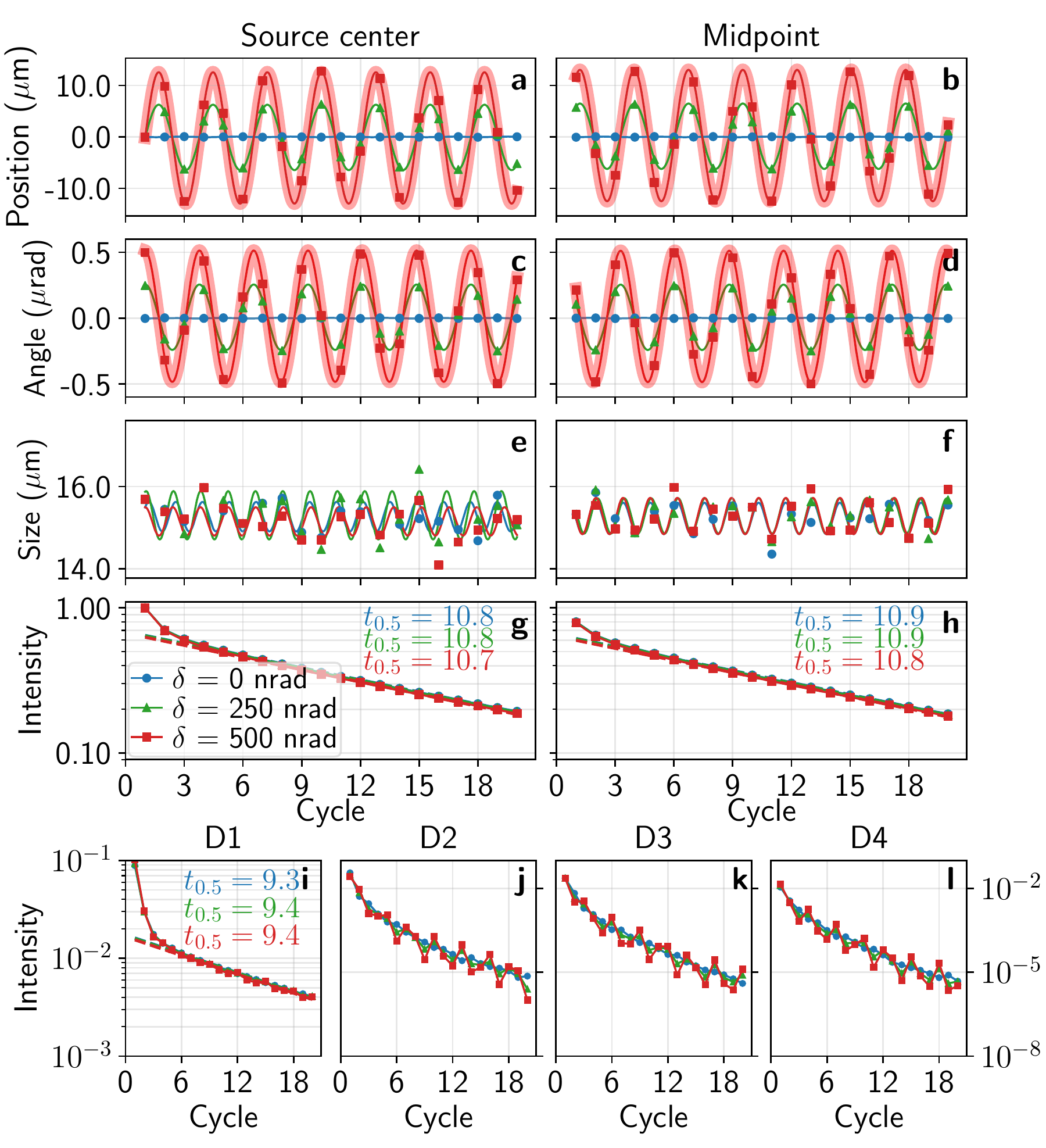}
\caption{Simulation results of x-ray beam dynamics in the
  rectangular four-crystal generic cavity with angle error
  $\delta$ at source.}
    \label{fig:full-simulation-source-angle}
\end{minipage}
\begin{minipage}{0.02\textwidth}
\end{minipage}  
\begin{minipage}{0.49\textwidth}
  \centering
      \includegraphics[width=0.99\linewidth]{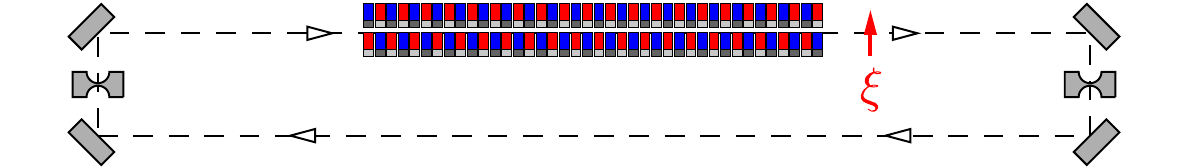}
  \includegraphics[width=0.99\linewidth]{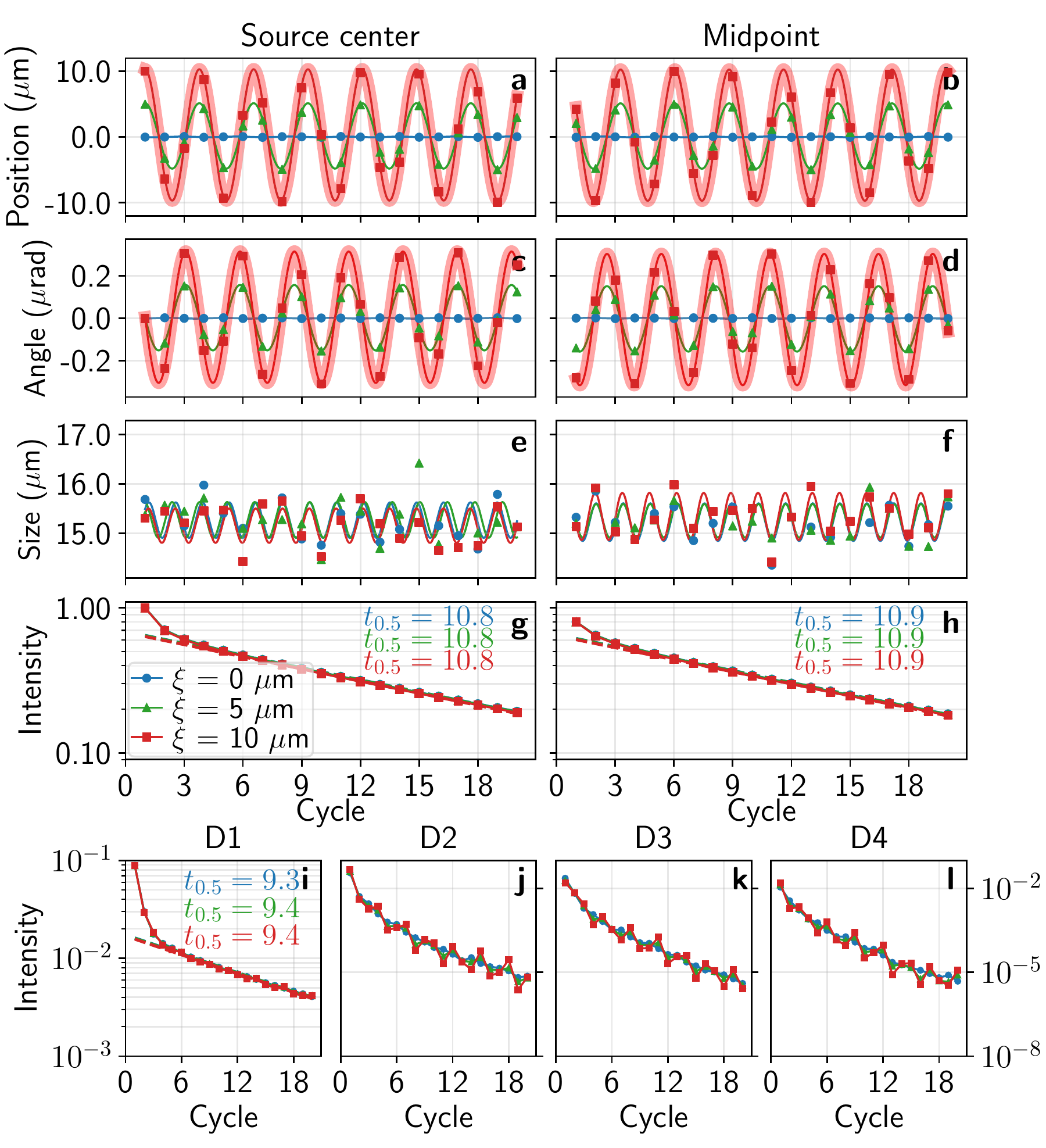}
        \caption{Similar to Fig.~\ref{fig:full-simulation-source-angle}, but with position error $\xi$ at source.}
    \label{fig:full-simulation-source-position}
\end{minipage}
\end{figure*}

\begin{table*}[t]
    \centering
    \caption{X-ray beam characteristics at the source position
      simulated for a cavity with yaw angle alignment error $\delta$
      or lateral spatial error $\xi$ at the source as
      derived from the data in
      Figs.~\ref{fig:full-simulation-source-angle}-\ref{fig:full-simulation-source-position}.} 
    \begin{threeparttable}[t]
      \begin{tabular}{c|c||c|c|c|c|c|c|c}
              \hline
            \multicolumn{2}{c||}{\makecell{\makecell{Alignment\\error}}} &
            \multicolumn{6}{c|}{Betatron oscillation type and parameters} &
            \makecell{Half lifetime \\ (cycles)\\}\\
            \cline{3-8}
            
            \multicolumn{2}{c||}{} &
            \multicolumn{2}{c|}{\makecell{\makecell{Position \\(\si{\micro\metre})}}} &
            \multicolumn{2}{c|}{\makecell{\makecell{Angle \\(nrad)}}} &
            \multicolumn{2}{c|}{\makecell{\makecell{Size \\(\si{\micro\metre})}}} & \\
            
            \cline{1-8}
            Type & Value & Offset & \makecell{Amplitude \\(rms)} & Offset & \makecell{Amplitude \\(rms)} & Average & \makecell{Amplitude \\(rms)} \\\hline\hline
            
            None &0 & 0.00 & 0.02 & 0.00 & 0.00 & 15.47 & 0.60 & 10.7  \\ 
            \Xhline{1pt}
            \multirow{ 2}{*}{\makecell{Source angle $\delta$ \\ (nrad) }   } 
            & 250 & -0.10 & 4.41 & 7 & 178 & 15.30 & 0.51 & 10.7\\
            \cline{2-9}
            & 500 & -0.20 & 8.87 & 15 & 358 & 15.15 & 0.41 & 10.7\\
            \cline{2-9}
            \Xhline{1pt}
            \multirow{ 2}{*}{\makecell{Source position $\xi$\\ ($\mu$m) }} 
            & 5 & 0.15 & 3.58 & 3 & 134 & 15.27 & 0.42 & 10.7\\
            \cline{2-9}
            & 10 & 0.3 & 7.17 & 6 & 269 & 15.16 & 0.38 & 10.7\\
            \cline{1-9}
        \end{tabular}
    \end{threeparttable}
\label{tab:source_misalignment}
\end{table*}

Transverse beam size oscillations observed in panels (e)-(f) of
Figs.~\ref{fig:full-simulation-crystal-angle-sys}-\ref{fig:full-simulation-crystal-position}
do not change either with the type or magnitude of the alignment
errors and have the same origin as those in the perfectly aligned
cavity, see Figs.~\ref{fig:perfect}(e)-(f) and discussion in
Section.~\ref{perfect}.

\subsubsection{Cavity ringdown curves}

In contrast to the lensless cavity, the angular beam variations due to
the crystal alignment errors are much smaller than the width of the
Bragg reflections.  As a result, the intra-cavity ringdown curves in
panels (g)-(h) of Figs.~\ref{fig:full-simulation-crystal-angle-sys},
\ref{fig:full-simulation-crystal-angle-one}, and
\ref{fig:full-simulation-crystal-position}, together with their half
lifetimes, do not change substantially and are close to those of the
perfectly aligned cavity case. For the roll angle misalignment case,
the ringdown curves practically do not change at all; see
Figs.~\ref{fig:full-simulation-crystal-angle-roll}(g)-(h). This lack
of change is due to weak coupling between the variation of the crystal
roll angle and the variation of the angle of incidence of x-rays to
the Bragg reflecting atomic planes in the crystals. Similar behavior
is observed in the analogous lensless cavity case presented in
Fig.~\ref{fig:c4-roll-angle-nolenses}.

The ringdown curves in panels (j)-(l) of
Figs.~\ref{fig:full-simulation-crystal-angle-sys},
\ref{fig:full-simulation-crystal-angle-one}, and
\ref{fig:full-simulation-crystal-position}, reflecting leakage of
x-rays through crystals C$_{\ind{2}}$-C$_{\ind{4}}$, reveal higher
sensitivity to crystal errors. The betatron oscillations can be
detected by XBIMs D$_{\ind{2}}$-D$_{\ind{4}}$, as calculations in
panels (j)-(l) show. They are practically invisible at XBIM
D$_{\ind{1}}$ [panel (i) in each figure], reflecting output coupling
of x-rays through crystal C$_{\ind{1}}$. Additional simulations (not
presented here) show that the visibility of the betatron oscillations
at XBIM D$_{\ind{1}}$ increases with the thickness of crystal
C$_{\ind{1}}$.

Although crystals C$_{\ind{2}}$-C$_{\ind{4}}$ have identical
thicknesses, the ringdown curves in panels (j)-(l) of
Figs.~\ref{fig:full-simulation-crystal-angle-sys},
\ref{fig:full-simulation-crystal-angle-one}, and
\ref{fig:full-simulation-crystal-position} are different. Additional
simulations show that this difference has two causes: first, the
asymmetric shape of the Bragg reflection angular dependence, and
second, the presence of a dispersive crystal setting. As a result,
angular deflection from the Bragg peak always has a different sign
from the preceding reflection and, therefore, a slightly different
reflectivity.

\subsection{Complete generic cavity with misaligned  lenses }

In the next step, we consider the complete four-crystal rectangular
cavity as in Fig.~\ref{fig000}, now including x-ray lenses with
angular and spatial alignment errors.
Figure~\ref{fig:full-simulation-crl-angle} shows numerical simulation
results of the x-ray beam dynamics in the cavity with lens
L$_{\ind{1}}$ having yaw angle error $\delta$, while
Fig.~\ref{fig:full-simulation-crl-position} shows results for lens
L$_{\ind{1}}$ having lateral position error $\xi$. The simulation
results in
Figs.~\ref{fig:full-simulation-crl-angle}-\ref{fig:full-simulation-crl-position}
are organized in the same way as those for the cavity with crystal
alignment errors
(Figs.~\ref{fig:full-simulation-crystal-angle-sys}-\ref{fig:full-simulation-crystal-position}).
Table~\ref{tab:crl_misalignment} presents relevant x-ray beam
characteristics, organized analogously to
Table~\ref{tab:angle_misalignment} for the crystal error cases.

The x-ray beam dynamics are very insensitive to angular errors at the
lens.  This behavior is expected because the x-ray beam (a few tens of
micrometers in cross-section) is at almost normal incidence to the
nearly flat apex of the parabolic surface of the lens (radius of
curvature 200~$\mu$m). The angular errors must be in a radian range to
produce any significant distortion of x-ray beam
trajectory.

In contrast, the lateral spatial errors at the lens are as critical
for x-ray beam dynamics as spatial errors at the crystals. The lateral
spatial error tolerance is $|\xi|\lesssim 10~\mu$m, very similar to
the crystal spatial error case (compare
Tables~\ref{tab:angle_misalignment} and \ref{tab:crl_misalignment}).
The ringdown curves also have similar behavior for both crystals and
lenses.

\subsection{Perfect generic cavity with misaligned x-ray source }

Betatron oscillations of the x-ray beam can be induced even in a
perfectly aligned cavity, provided the x-ray source is misaligned, as
shown in our first analysis of betatron oscillations using analytical
methods, in Section~\ref{selfconsistent-generic}.

The numerical simulations of the beam dynamics for this case are shown
in
Figs.~\ref{fig:full-simulation-source-angle}-\ref{fig:full-simulation-source-position},
and the beam characteristics are presented in
Table~\ref{tab:source_misalignment}.

Both the angular and spatial source misalignment types result in
symmetric betatron oscillations of the x-ray beam position and angle
(see
Figs.~\ref{fig:full-simulation-source-angle}-\ref{fig:full-simulation-source-position},
respectively), in agreement with analytical calculations represented
by Eq.~\eqref{eq0220}. Both approaches---analytical and
numerical---also agree quantitatively. The thick pink lines in
Figs.~\ref{fig:full-simulation-source-angle}-\ref{fig:full-simulation-source-position}
show results calculated with the ray-transfer matrix approach; they
are almost identical to the red lines for the results calculated
numerically with the Shadow3 package \cite{SHADOW3} for the same
magnitudes of the alignment errors.

The sensitivity to the source alignment errors is comparable to the sensitivity to alignment errors of the optical elements. The smallest
alignment errors in Table~\ref{tab:source_misalignment} can be
considered as source alignment tolerance.

\begin{figure*}[t!]
\centering
\begin{minipage}{0.49\textwidth}
\centering
      \includegraphics[width=0.63\linewidth]{systematic.pdf}
\includegraphics[width=0.99\linewidth]{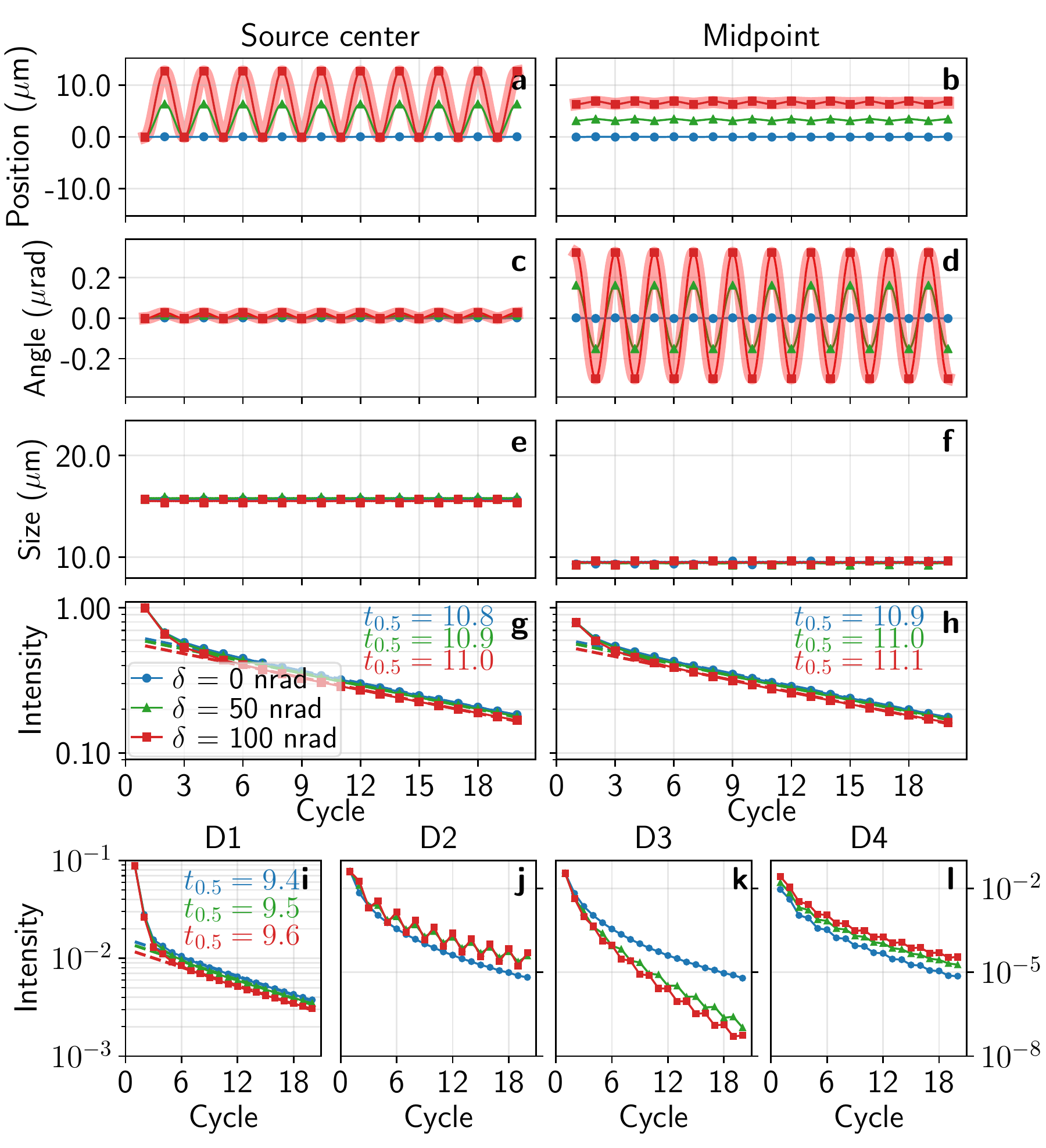}
      \caption{Numerical simulation results  of x-ray beam dynamics in the confocal rectangular four-crystal cavity with systematically misaligned crystal yaw angles.}
    \label{fig:full-conf-simulation-crystal-angle-sys}
\end{minipage}
\begin{minipage}{0.02\textwidth}
\end{minipage}  
\begin{minipage}{0.49\textwidth}
\centering
      \includegraphics[width=0.63\linewidth]{c4-yaw.pdf}
      \includegraphics[width=0.99\linewidth]{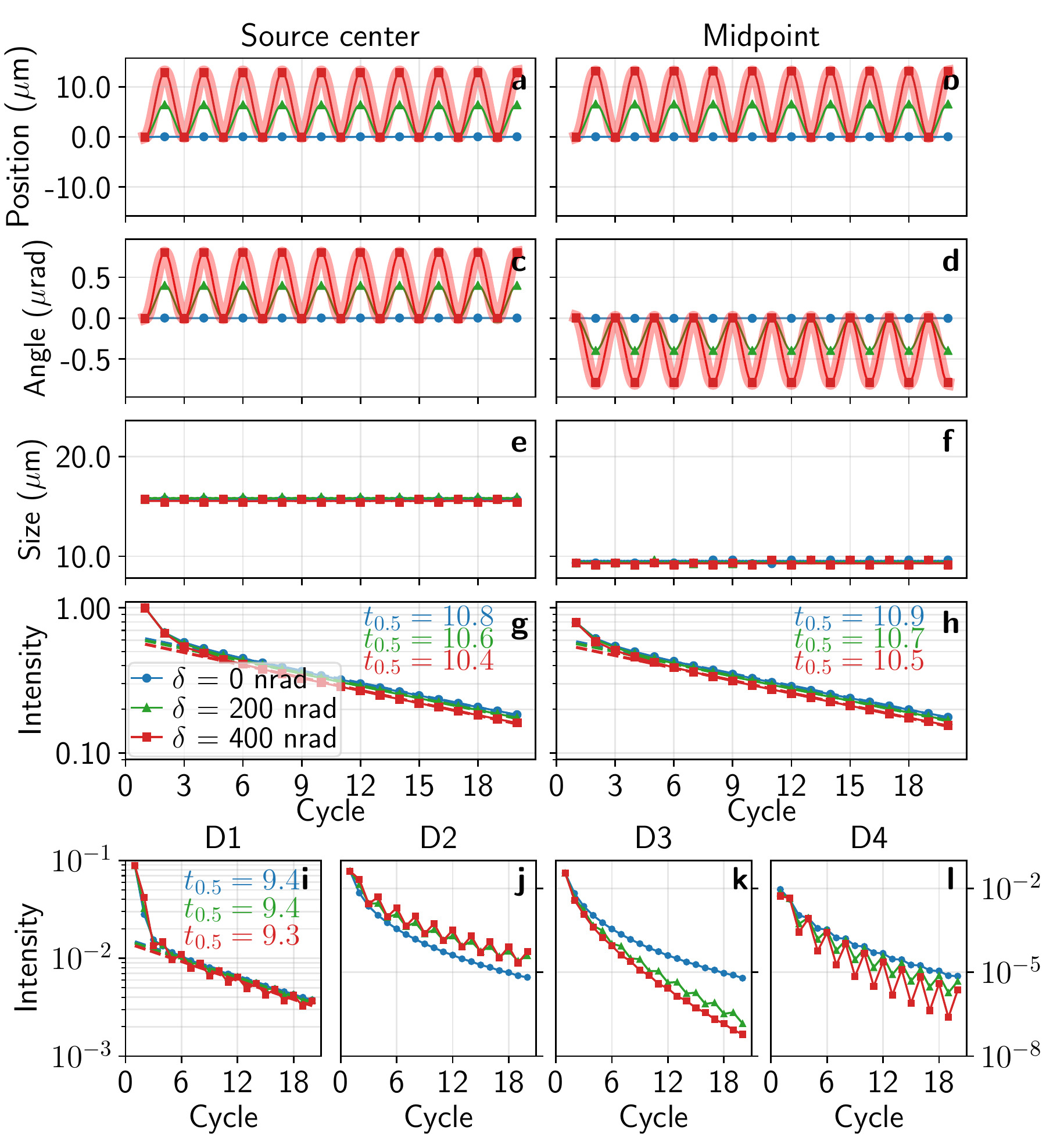}
    \caption{Similar to Fig.~\ref{fig:full-conf-simulation-crystal-angle-sys}, but with yaw angle error $\delta$ at crystal C$_{\ind{4}}$.}
    \label{fig:full-conf-simulation-crystal-angle-one}
\end{minipage}
\centering
\begin{minipage}{0.49\textwidth}
  \centering
      \includegraphics[width=0.63\linewidth]{l1-xi.pdf}
        \includegraphics[width=0.99\linewidth]{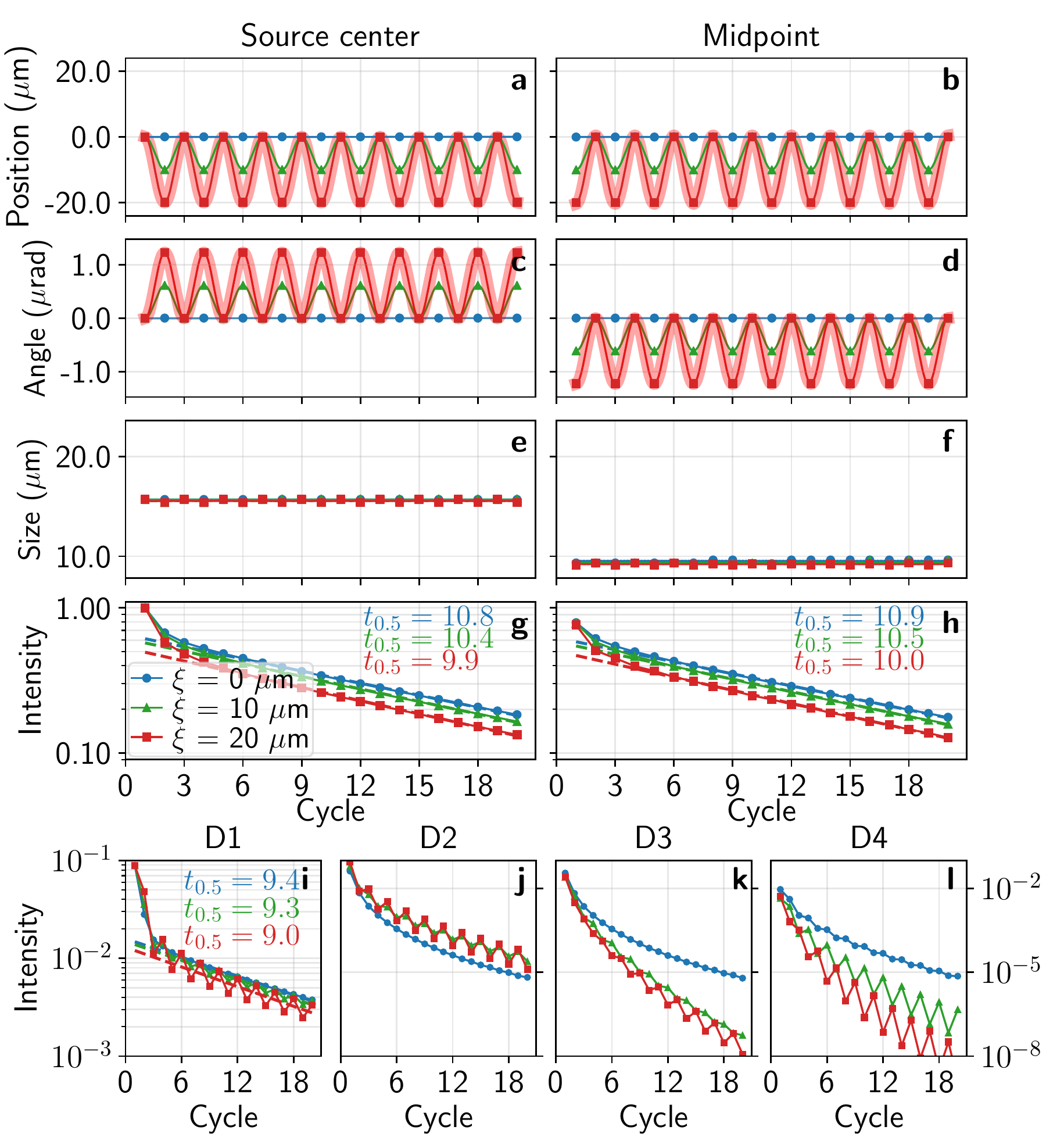}
    \caption{Similar to Fig.~\ref{fig:full-conf-simulation-crystal-angle-sys}, but with position error $\xi$ at lens L$_{\ind{1}}$.}
    \label{fig:full-conf-simulation-crl-position}
\end{minipage}
\begin{minipage}{0.02\textwidth}
\end{minipage}  
\begin{minipage}{0.49\textwidth}
\centering
      \includegraphics[width=0.63\linewidth]{c4-xi.pdf}
      \includegraphics[width=0.99\linewidth]{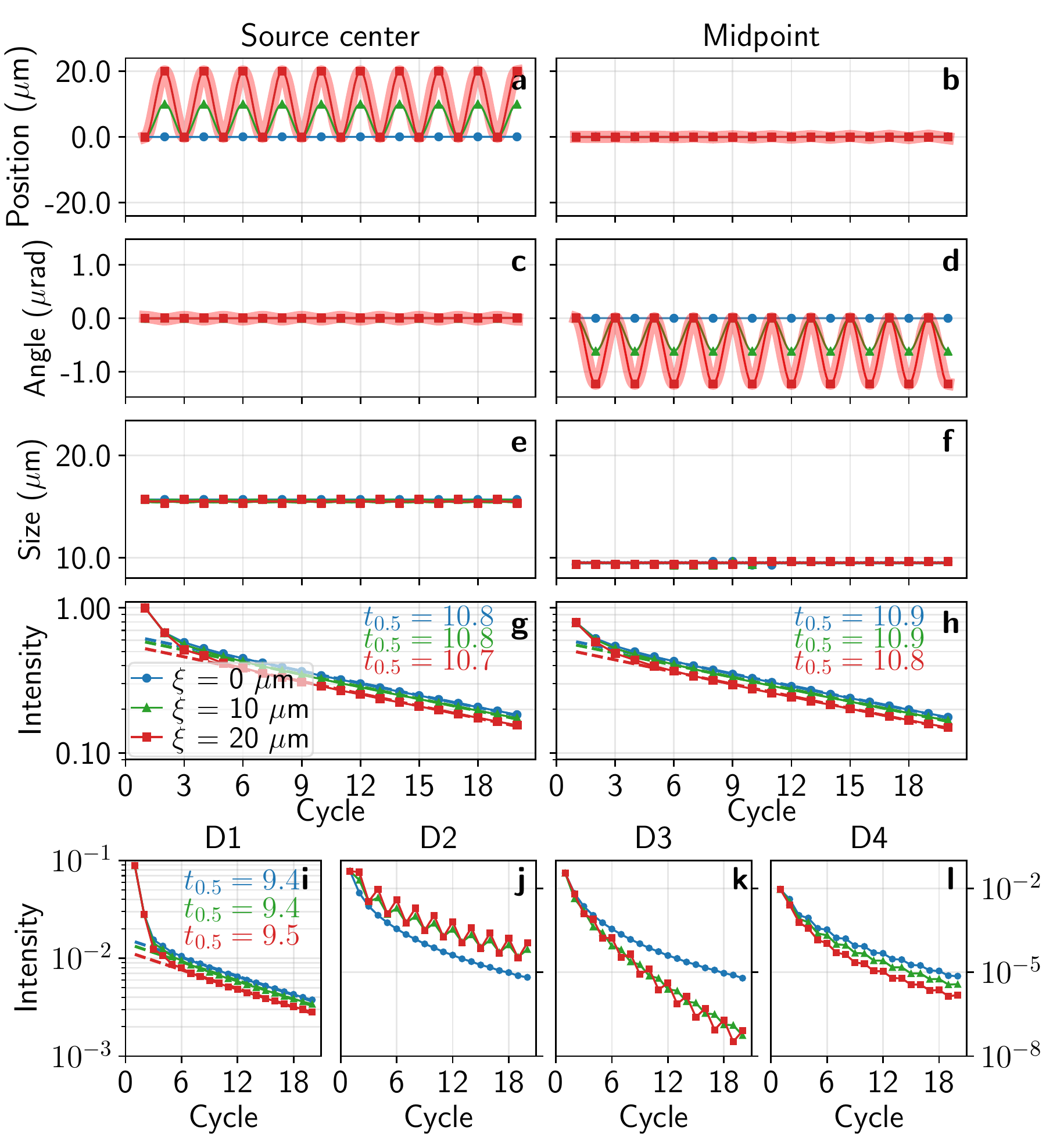}
    \caption{Similar to Fig.~\ref{fig:full-conf-simulation-crystal-angle-sys}, but with position error $\xi$ at crystal C$_{\ind{4}}$.}
    \label{fig:full-conf-simulation-crystal-position}
\end{minipage}
\end{figure*}

\subsection{Misaligned confocal cavity  }
\label{confocal-numeric}

So far in this section we have used numerical methods to study x-ray
beam dynamics in a generic four-crystal rectangular x-ray cavity. Here
we address x-ray beam dynamics in a confocal cavity, which we studied
partially using analytical methods in Section~\ref{confocal}. The only
difference to the generic cavity case is a different choice of the
focal length of both lenses, which is now $f^{(1)}=\ell/4=16.3$~m (see
Table~\ref{tab1}). In the confocal cavity case, Gaussian beams with
any Rayleigh length $\zr$ are stable and self-consistent solutions. In
our calculations, we will use the same parameters for the Gaussian
x-ray source as in the generic cavity case.

Figures~\ref{fig:full-conf-simulation-crystal-angle-sys}-\ref{fig:full-conf-simulation-crystal-position}
show numerical simulation results of the x-ray beam dynamics in the
confocal cavity with different alignment errors: systematically
misaligned yaw angles at crystals C$_{\ind{1}}$-C$_{\ind{4}}$
(Fig.~\ref{fig:full-conf-simulation-crystal-angle-sys}), one-crystal
yaw angle error $\delta$ at C$_{\ind{4}}$
(Fig.~\ref{fig:full-conf-simulation-crystal-angle-one}),  position error $\xi$ at crystal C$_{\ind{4}}$
(Fig.~\ref{fig:full-conf-simulation-crystal-position}), and
position error $\xi$ at lens L$_{\ind{1}}$ 
(Fig.~\ref{fig:full-conf-simulation-crl-position}). The simulation
results in
Figs.~\ref{fig:full-conf-simulation-crystal-angle-sys}-\ref{fig:full-conf-simulation-crystal-position}
are arranged in the same way as those for the
generic cavity in
Figs.~\ref{fig:full-simulation-crystal-angle-sys}-\ref{fig:full-simulation-crystal-position}.

Just as in the generic cavity, alignment errors in the confocal cavity
result in betatron oscillations (both symmetric and asymmetric) of the
x-ray beam position and angle. Unlike the generic cavity case,
however, the beam size at the monitored locations is stable;
there are no beam size oscillations,
in agreement with Eq.~\eqref{eq0340}. Also unlike the
generic cavity, the beam sizes at the source and the midpoint are
different. At the midpoint, the beam size is demagnified to $\simeq
9.8~\mu$m from $\simeq 15.8~\mu$m at the source location due to a
demagnification factor of $M=0.62$, in agreement with $M=f^{(1)}/\zr$
of Eq.~\eqref{eq0125}.

For the same magnitudes of the alignment errors, the magnitudes of the
spatial and angular beam displacements from the optical axis are
close to those observed in the generic cavity discussed earlier.

However, there are some substantial differences.
The period of the betatron oscillation in the confocal cavity is
exactly 2 cycles (compare to 2.77 cycles in the generic cavity), in
agreement with analytical simulations in Section~\ref{confocal} [see
  Eq.~\eqref{eq0126}].

Due to this two-cycle period, for all types of
alignment errors {\em of the optical elements} in a confocal cavity,
the x-ray beam returns to the initial position on the optical axis
every second pass.  The angular beam deviations show similar behavior,
but in some cases the angular deviations are always zero, i.e.,
completely insensitive to some types of alignment errors, such as
systematic angular misalignment and spatial error at crystal
C$_{\ind{4}}$.

This property, however, does not extend to the source alignment
errors. From Eq.~\eqref{eq0126} and results of numerical simulations
that are not shown here, the x-ray source error is reproduced every
second cycle.  The beam size, \deleted{nevertheless, never changes} \added{however, either does not change at the source origin, provided  the source errors are involved, or varies insignificantly by 0.15~$\mu$m (rms) because of the optical elements errors.}.

These very interesting properties of the confocal cavity can be used
to mitigate tolerances on CBXFEL cavity alignment errors (albeit
not on the source errors). For this approach to work, however, the electron bunches
should arrive with a periodicity matching the betatron oscillations, i.e.,
with a period equal to two round-trip times of x-rays in the optical
cavity.  This approach also halves the cavity size for the given electron beam
repetition rate, but at a cost of increased cavity losses,  as every second round trip is idle.

Ringdown curves in the confocal cavity are of course not identical to
those in the generic cavity, showing instead some nuanced differences. However, the decay times are similar.

\section{Cavity alignment}
\label{alignment}

There are tens of degrees of freedom of the optical elements in CBXFEL
cavities (close to 30 in the rectangular four-crystal cavity
considered in this paper)---many more than in a classical laser
oscillator cavity.  All these degrees of freedom have to be aligned
with a precision ensuring that the to-be-amplified x-ray beam a few
tens of microns in cross-section returning into the undulator meets a
fresh electron bunch of a similar size.

The effects of different alignment errors are (in most cases)  entangled.
Where should cavity alignment start, and how should it be performed?  The results
of the numerical simulation presented in 
Section~\ref{numerical} can be used to propose a strategy for cavity alignment.

First, we recall that in the lensless cavity the spatial crystal error
does not affect the angular beam deviations (see Fig.~\ref{fig:c4-position-nolenses}). Similarly, the roll angle
crystal errors in the lensless cavity produce very small effects on the
in-cavity-plane angular beam deviations (see Fig.~\ref{fig:c4-roll-angle-nolenses}).

As a result, the angular and spatial errors can be decoupled by first aligning crystals in the absence of lenses and then moving in and aligning the lenses. The full procedure would be as follows:\\ (1) With lenses removed, align crystal angles (yaw and roll). \\ (2) Align crystal
positions.  \\ (3)
Move in and align lenses.\\ (4) Adjust the round-trip time to the periodicity of the electron bunches by
changing simultaneously the longitudinal positions of two crystals
(e.g., C$_{\ind{2}}$ and C$_{\ind{3}}$) and optimizing the CBXFEL gain.

Time-resolved signals from XBIMs D$_1$, D$_2$, D$_3$, and D$_4$ and from the
XBPM at the midpoint are essential for all steps in the alignment.

As noted earlier, the results of our studies show that the XBPM at the
midpoint reflects the beam status at the source.  This fact is very
useful, as in practice it is much easier to monitor the beam at the
midpoint than in the undulator.

\section{Discussion, conclusions and outlook}
\label{conclusions}

We are studying spatial, angular, and temporal behavior of x-ray beams
generated by a Gaussian source and stored in a {\em stable} x-ray
cavity (i.e., x-ray beam dynamics) using analytical methods (paraxial
ray-transfer matrix, wave-optics Gaussian beam methods) and numerical
ray-tracing computer codes.  An example of a rectangular cavity with a
65-m round-trip length is considered, which is designed for a test
cavity-based XFEL \cite{MAA19}.  The cavity is composed of four
Bragg-reflecting flat crystal mirrors located in the corners of the
rectangular cavity and two lenses in two symmetry points (see
Fig.~\ref{fig000}). The focal length of the lenses is chosen to ensure
a stable, self-consistent solution for the Gaussian beam in the
cavity.

The analytical approaches allow us to determine two stable
self-consistent solutions, which we here term the confocal and generic
cases.  Both cases are studied here.

The analytical approaches are powerful, but they are limited in some
respects because the cavity optical components---the crystals and
lenses---are treated in a simplified fashion that neglects
photoabsorption and the strongly restricted angular and spectral
ranges of Bragg reflections.  Because of these limitations, the
analytical approach does not provide information about beam intensity
that is essential for cavity diagnostics, namely, the intensity of the
x-ray beams stored in the cavity, output-coupled from the cavity, or
leaked through cavity crystals. The numerical approach removes these
limitations.  Both approaches, however, provide similar results
\replaced{for the beam position and angle, if the alignment
  errors are small}{in the areas where they intersect}.

Our studies are focused on the impact on x-ray beam dynamics of
spatial and angular alignment errors of the optical components and of
the x-ray source.

\added{Most of the alignment errors result in reduced lifetime of x-rays stored in the cavity.
But, any} misalignment of an optical component or the x-ray source results
in periodic betatron-like oscillations of the x-ray orbit in the
cavity. The period of the betatron oscillations is a non-integer
number (in a general case) of the passes of the stored x-ray beams in
the cavity.  The period is independent of the type of misalignment and
is defined by the parameters of the cavity and x-ray source. The
betatron oscillation period is different for the confocal and generic
cavities.

The amplitude and offset of the betatron oscillations are signatures
of the misalignment type.  The betatron oscillations are either
symmetric, with zero offset, or asymmetric, with the offset magnitude
equal to the amplitude of the oscillations.  The amplitude and offset
of the betatron oscillations are proportional to the magnitude of the
alignment error.  The offset sign depends on the sign of the alignment
error, on the location of the faulty optical element, and on the
monitored location.  In general, the presence of betatron oscillations
is a signature of cavity misalignment.

In all misalignment cases, the amplitude of the betatron oscillations
are practically identical at the source and midpoint locations.  The
x-ray beam sizes at these locations can differ by a magnification
factor.  These facts are of practical importance for cavity
diagnostics and alignment, as monitoring at the midpoint is easier
than in the undulator.  However, the offset signs of the betatron
oscillations can be different at the source and midpoint locations. In
one case considered in the paper, even the magnitude of the offset can
differ at the source and midpoint locations. This happens due to the
crystal position error (see
Fig.~\ref{fig:full-simulation-crystal-position}).

Transverse beam size at the source and midpoint locations is stable
independent of the type and magnitude of the alignment errors. Beam
size oscillations may appear even in a perfectly aligned cavity
\deleted{(albeit, only in the generic cavity)} if the spatial and
angular source parameters are not equal to those of a Gaussian beam
with Rayleigh length $\zr$ at its waist [See Eq.~\eqref{eq0310} and
  Table~\ref{tab1}].

Tolerances of $\simeq 5-10$ $\mu$m in positional alignment and $\simeq
50-200$ nrad in yaw angle alignment \added{of the crystal mirrors} are required to achieve close to
the perfect half lifetime of 10.5 passes of the stored in the cavity
x-ray beam power and a betatron oscillation amplitude close to
zero. Similar tolerances are required for the lens position errors and
for the source angle and position errors.

The angular and spatial x-ray beam dynamics
results [see panels (a)-(d) of
  Figs.~\ref{fig:full-simulation-crystal-angle-sys}-\ref{fig:full-simulation-source-position}]
obtained with numerical simulations are practically identical to those
obtained with the ray-transfer matrix approach, as long as the angular
deviations are much smaller than the angular width of the Bragg
reflections.

For any misalignment type of the optical elements in the confocal
cavity, the x-ray beam position returns on the optical axis every
second pass. This property may be used to mitigate alignment tolerances
for the CBXFEL cavity, provided electron bunches are arriving every
second pass of x-rays in the cavity. This approach may also halve the cavity
size for the given electron repetition rate, \added{but at a cost of increased cavity losses, because the x-ray pulse is amplified only each second round trip}.

Caution is required in using a confocal cavity, as small variations of
the cavity parameters may result in beam instabilities. For example,
slightly changing the focal length $f^{(1)}=\ell/4$ to
$f^{(1)}(1+\epsilon)$ ($|\epsilon| \ll 1$) may slightly change the
period of the betatron oscillations from exactly 2 to 2+$\delta$
passes, without causing beam instabilities. However, breaking the
cavity symmetry by displacing lenses along the optical axis from the
symmetry points may result in beam instabilities.

Betatron oscillations \added{and reduced lifetime of x-rays stored
  in the cavity} are also expected to be the typical signatures of
misalignments for x-ray cavities of other types.  More studies are
required, including wavefront propagation simulations, to scrutinize
the obtained results and apply them to computer-assisted cavity
alignment procedures, including strategies based on machine learning.

\section{Acknowledgments}

Kwang-Je Kim and Ryan Lindberg (ANL) are acknowledged for pointing out
the importance of the equivalence of the ray optics and the paraxial
wave optics as well as of the betatron oscillation for a closed
optical cavity.  Luca Refuffi, Xianbo Shi (ANL) and Manuel Sanches del
Rio (European Synchrotron, ESRF) are acknowledged for helping to set
up x-ray optics modeling with package Shadow3 in the Oasys
environment.  P.Q. acknowledges L. Dean Chapman (University of
Saskatchewan) for support and useful discussions.  Peifan Liu (ANL) is
acknowledged for reading the manuscript, valuable amendments and
observations.  Deming Shu, Marion White (ANL), Zhirong Huang, Gabriel
Marcus, Diling Zhu, and Tien Tan (SLAC) are acknowledged for useful
discussions. Work at Argonne National Laboratory was supported by the
U.S. Department of Energy, Office of Science, Office of Basic Energy
Sciences, under contract DE-AC02- 06CH11357.


\begin{thebibliography}{36}
\expandafter\ifx\csname natexlab\endcsname\relax\def\natexlab#1{#1}\fi
\expandafter\ifx\csname bibnamefont\endcsname\relax
  \def\bibnamefont#1{#1}\fi
\expandafter\ifx\csname bibfnamefont\endcsname\relax
  \def\bibfnamefont#1{#1}\fi
\expandafter\ifx\csname citenamefont\endcsname\relax
  \def\citenamefont#1{#1}\fi
\expandafter\ifx\csname url\endcsname\relax
  \def\url#1{\texttt{#1}}\fi
\expandafter\ifx\csname urlprefix\endcsname\relax\def\urlprefix{URL }\fi
\providecommand{\bibinfo}[2]{#2}
\providecommand{\eprint}[2][]{\url{#2}}

\bibitem[{\citenamefont{Decking et~al.}(2020)\citenamefont{Decking, Abeghyan,
  Abramian, and et~al.}}]{EXFEL20}
\bibinfo{author}{\bibfnamefont{W.}~\bibnamefont{Decking}},
  \bibinfo{author}{\bibfnamefont{S.}~\bibnamefont{Abeghyan}},
  \bibinfo{author}{\bibfnamefont{P.}~\bibnamefont{Abramian}}, \bibnamefont{and}
  \bibinfo{author}{\bibnamefont{et~al.}}, \bibinfo{journal}{Nature Photonics}
  \textbf{\bibinfo{volume}{14}}, \bibinfo{pages}{391} (\bibinfo{year}{2020}).

\bibitem[{\citenamefont{Raubenheimer}(2018)}]{Raubenheimer18}
\bibinfo{author}{\bibfnamefont{T.}~\bibnamefont{Raubenheimer}}, in
  \emph{\bibinfo{booktitle}{Proc. 60th ICFA Advanced Beam Dynamics Workshop
  (FLS'18), Shanghai, China, 5-9 March 2018}} (\bibinfo{publisher}{JACoW
  Publishing}, \bibinfo{address}{Geneva, Switzerland}, \bibinfo{year}{2018}),
  no.~\bibinfo{number}{60} in \bibinfo{series}{ICFA Advanced Beam Dynamics
  Workshop}, pp. \bibinfo{pages}{6--11}, ISBN
  \bibinfo{isbn}{978-3-95450-206-6},
  \bibinfo{note}{https://doi.org/10.18429/JACoW-FLS2018-MOP1WA02},
  \urlprefix\url{http://jacow.org/fls2018/papers/mop1wa02.pdf}.

\bibitem[{\citenamefont{Siegman}(1986)}]{Siegman}
\bibinfo{author}{\bibfnamefont{A.~E.} \bibnamefont{Siegman}},
  \emph{\bibinfo{title}{Lasers}} (\bibinfo{publisher}{University Science
  Books}, \bibinfo{address}{Sausalito, California}, \bibinfo{year}{1986}).

\bibitem[{\citenamefont{Kondratenko and Saldin}(1980)}]{KS80}
\bibinfo{author}{\bibfnamefont{A.~M.} \bibnamefont{Kondratenko}}
  \bibnamefont{and} \bibinfo{author}{\bibfnamefont{E.~L.}
  \bibnamefont{Saldin}}, \bibinfo{journal}{Part. Accel.}
  \textbf{\bibinfo{volume}{10}}, \bibinfo{pages}{207} (\bibinfo{year}{1980}).

\bibitem[{\citenamefont{Bonifacio et~al.}(1984)\citenamefont{Bonifacio,
  Pellegrini, and Narducci}}]{BPN84}
\bibinfo{author}{\bibfnamefont{R.}~\bibnamefont{Bonifacio}},
  \bibinfo{author}{\bibfnamefont{C.}~\bibnamefont{Pellegrini}},
  \bibnamefont{and} \bibinfo{author}{\bibfnamefont{L.}~\bibnamefont{Narducci}},
  \bibinfo{journal}{Optics Communications} \textbf{\bibinfo{volume}{50}},
  \bibinfo{pages}{373 } (\bibinfo{year}{1984}), ISSN \bibinfo{issn}{0030-4018}.

\bibitem[{\citenamefont{Emma et~al.}(2010)\citenamefont{Emma, Akre, Arthur,
  Bionta, Bostedt, Bozek, Brachmann, Bucksbaum, Coffee, Decker et~al.}}]{EAA10}
\bibinfo{author}{\bibfnamefont{P.}~\bibnamefont{Emma}},
  \bibinfo{author}{\bibfnamefont{R.}~\bibnamefont{Akre}},
  \bibinfo{author}{\bibfnamefont{J.}~\bibnamefont{Arthur}},
  \bibinfo{author}{\bibfnamefont{R.}~\bibnamefont{Bionta}},
  \bibinfo{author}{\bibfnamefont{C.}~\bibnamefont{Bostedt}},
  \bibinfo{author}{\bibfnamefont{J.}~\bibnamefont{Bozek}},
  \bibinfo{author}{\bibfnamefont{A.}~\bibnamefont{Brachmann}},
  \bibinfo{author}{\bibfnamefont{P.}~\bibnamefont{Bucksbaum}},
  \bibinfo{author}{\bibfnamefont{R.}~\bibnamefont{Coffee}},
  \bibinfo{author}{\bibfnamefont{F.-J.} \bibnamefont{Decker}},
  \bibnamefont{et~al.}, \bibinfo{journal}{Nature Photonics}
  \textbf{\bibinfo{volume}{4}}, \bibinfo{pages}{641 } (\bibinfo{year}{2010}).

\bibitem[{\citenamefont{Huang and Ruth}(2006)}]{HR06}
\bibinfo{author}{\bibfnamefont{Z.}~\bibnamefont{Huang}} \bibnamefont{and}
  \bibinfo{author}{\bibfnamefont{R.~D.} \bibnamefont{Ruth}},
  \bibinfo{journal}{Phys. Rev. Lett.} \textbf{\bibinfo{volume}{96}},
  \bibinfo{pages}{144801} (\bibinfo{year}{2006}).

\bibitem[{\citenamefont{Kim et~al.}(2008)\citenamefont{Kim, Shvyd\char39{}ko,
  and Reiche}}]{KSR08}
\bibinfo{author}{\bibfnamefont{K.-J.} \bibnamefont{Kim}},
  \bibinfo{author}{\bibfnamefont{Yu.}~\bibnamefont{Shvyd\char39{}ko}},
  \bibnamefont{and} \bibinfo{author}{\bibfnamefont{S.}~\bibnamefont{Reiche}},
  \bibinfo{journal}{Phys. Rev. Lett.} \textbf{\bibinfo{volume}{100}},
  \bibinfo{pages}{244802} (\bibinfo{year}{2008}).

\bibitem[{\citenamefont{Kim and Shvyd\char39{}ko}(2009)}]{KS09}
\bibinfo{author}{\bibfnamefont{K.-J.} \bibnamefont{Kim}} \bibnamefont{and}
  \bibinfo{author}{\bibfnamefont{Yu.~V.} \bibnamefont{Shvyd\char39{}ko}},
  \bibinfo{journal}{Phys. Rev. ST Accel. Beams} \textbf{\bibinfo{volume}{12}},
  \bibinfo{pages}{030703} (\bibinfo{year}{2009}).

\bibitem[{\citenamefont{Lindberg et~al.}(2011)\citenamefont{Lindberg, Kim,
  Shvyd'ko, and Fawley}}]{LSKF11}
\bibinfo{author}{\bibfnamefont{R.~R.} \bibnamefont{Lindberg}},
  \bibinfo{author}{\bibfnamefont{K.-J.} \bibnamefont{Kim}},
  \bibinfo{author}{\bibfnamefont{Yu.}~\bibnamefont{Shvyd'ko}}, \bibnamefont{and}
  \bibinfo{author}{\bibfnamefont{W.~M.} \bibnamefont{Fawley}},
  \bibinfo{journal}{Phys. Rev. ST Accel. Beams} \textbf{\bibinfo{volume}{14}},
  \bibinfo{pages}{010701} (\bibinfo{year}{2011}).

\bibitem[{\citenamefont{Kim et~al.}(2012)\citenamefont{Kim, Shvyd'ko, and
  Lindberg}}]{KSL12}
\bibinfo{author}{\bibfnamefont{K.-J.} \bibnamefont{Kim}},
  \bibinfo{author}{\bibfnamefont{Yu.}~\bibnamefont{Shvyd'ko}}, \bibnamefont{and}
  \bibinfo{author}{\bibfnamefont{R.~R.} \bibnamefont{Lindberg}},
  \bibinfo{journal}{Synchrotron Radiation News} \textbf{\bibinfo{volume}{25}},
  \bibinfo{pages}{25} (\bibinfo{year}{2012}).

\bibitem[{\citenamefont{Dai et~al.}(2012)\citenamefont{Dai, Deng, and
  Dai}}]{DDD12}
\bibinfo{author}{\bibfnamefont{J.}~\bibnamefont{Dai}},
  \bibinfo{author}{\bibfnamefont{H.}~\bibnamefont{Deng}}, \bibnamefont{and}
  \bibinfo{author}{\bibfnamefont{Z.}~\bibnamefont{Dai}},
  \bibinfo{journal}{Phys. Rev. Lett.} \textbf{\bibinfo{volume}{108}},
  \bibinfo{pages}{034802} (\bibinfo{year}{2012}).

\bibitem[{\citenamefont{Marcus et~al.}(2017)\citenamefont{Marcus, Ding, Duris,
  Feng, Huang, Krzywinski, T.~Maxwell, Kim, Lindberg, Shvyd'ko et~al.}}]{MDD17}
\bibinfo{author}{\bibfnamefont{G.}~\bibnamefont{Marcus}},
  \bibinfo{author}{\bibfnamefont{Y.}~\bibnamefont{Ding}},
  \bibinfo{author}{\bibfnamefont{J.}~\bibnamefont{Duris}},
  \bibinfo{author}{\bibfnamefont{Y.}~\bibnamefont{Feng}},
  \bibinfo{author}{\bibfnamefont{Z.}~\bibnamefont{Huang}},
  \bibinfo{author}{\bibfnamefont{J.}~\bibnamefont{Krzywinski}},
  \bibinfo{author}{\bibfnamefont{T.~R.} \bibnamefont{T.~Maxwell}},
  \bibinfo{author}{\bibfnamefont{K.-J.} \bibnamefont{Kim}},
  \bibinfo{author}{\bibfnamefont{R.}~\bibnamefont{Lindberg}},
  \bibinfo{author}{\bibfnamefont{Yu.}~\bibnamefont{Shvyd'ko}},
  \bibnamefont{et~al.}, in \emph{\bibinfo{booktitle}{Proceedings of 38th
  International Free Electron Laser Conference}} (\bibinfo{address}{Santa Fe,
  NM, USA}, \bibinfo{year}{2017}).

\bibitem[{\citenamefont{Freund et~al.}(2019)\citenamefont{Freund, van~der Slot,
  and Shvyd'ko}}]{FSS19}
\bibinfo{author}{\bibfnamefont{H.~P.} \bibnamefont{Freund}},
  \bibinfo{author}{\bibfnamefont{P.~J.~M.} \bibnamefont{van~der Slot}},
  \bibnamefont{and} \bibinfo{author}{\bibfnamefont{Yu.}~\bibnamefont{Shvyd'ko}},
  \bibinfo{journal}{New Journal of Physics} \textbf{\bibinfo{volume}{21}},
  \bibinfo{pages}{093028} (\bibinfo{year}{2019}).

\bibitem[{\citenamefont{Marcus et~al.}(2020)\citenamefont{Marcus, Halavanau,
  Huang, Krzywinski, MacArthur, Margraf, Raubenheimer, and Zhu}}]{MHH20}
\bibinfo{author}{\bibfnamefont{G.}~\bibnamefont{Marcus}},
  \bibinfo{author}{\bibfnamefont{A.}~\bibnamefont{Halavanau}},
  \bibinfo{author}{\bibfnamefont{Z.}~\bibnamefont{Huang}},
  \bibinfo{author}{\bibfnamefont{J.}~\bibnamefont{Krzywinski}},
  \bibinfo{author}{\bibfnamefont{J.}~\bibnamefont{MacArthur}},
  \bibinfo{author}{\bibfnamefont{R.}~\bibnamefont{Margraf}},
  \bibinfo{author}{\bibfnamefont{T.}~\bibnamefont{Raubenheimer}},
  \bibnamefont{and} \bibinfo{author}{\bibfnamefont{D.}~\bibnamefont{Zhu}},
  \bibinfo{journal}{Phys. Rev. Lett.} \textbf{\bibinfo{volume}{125}},
  \bibinfo{pages}{254801} (\bibinfo{year}{2020}).

\bibitem[{\citenamefont{Shvyd'ko et~al.}(2010)\citenamefont{Shvyd'ko, Stoupin,
  Cunsolo, Said, and Huang}}]{SSC10}
\bibinfo{author}{\bibfnamefont{Yu.~V.} \bibnamefont{Shvyd'ko}},
  \bibinfo{author}{\bibfnamefont{S.}~\bibnamefont{Stoupin}},
  \bibinfo{author}{\bibfnamefont{A.}~\bibnamefont{Cunsolo}},
  \bibinfo{author}{\bibfnamefont{A.}~\bibnamefont{Said}}, \bibnamefont{and}
  \bibinfo{author}{\bibfnamefont{X.}~\bibnamefont{Huang}},
  \bibinfo{journal}{Nature Physics} \textbf{\bibinfo{volume}{6}},
  \bibinfo{pages}{196} (\bibinfo{year}{2010}).

\bibitem[{\citenamefont{Shvyd'ko et~al.}(2011)\citenamefont{Shvyd'ko, Stoupin,
  Blank, and Terentyev}}]{SSB11}
\bibinfo{author}{\bibfnamefont{Yu.~V.} \bibnamefont{Shvyd'ko}},
  \bibinfo{author}{\bibfnamefont{S.}~\bibnamefont{Stoupin}},
  \bibinfo{author}{\bibfnamefont{V.}~\bibnamefont{Blank}}, \bibnamefont{and}
  \bibinfo{author}{\bibfnamefont{S.}~\bibnamefont{Terentyev}},
  \bibinfo{journal}{Nature Photonics} \textbf{\bibinfo{volume}{5}},
  \bibinfo{pages}{539} (\bibinfo{year}{2011}).

\bibitem[{\citenamefont{Lengeler et~al.}(1999)\citenamefont{Lengeler, Schroer,
  T\"ummler, Benner, Richwin, Snigirev, Snigireva, and Drakopoulos}}]{LST99}
\bibinfo{author}{\bibfnamefont{B.}~\bibnamefont{Lengeler}},
  \bibinfo{author}{\bibfnamefont{C.}~\bibnamefont{Schroer}},
  \bibinfo{author}{\bibfnamefont{J.}~\bibnamefont{T\"ummler}},
  \bibinfo{author}{\bibfnamefont{B.}~\bibnamefont{Benner}},
  \bibinfo{author}{\bibfnamefont{M.}~\bibnamefont{Richwin}},
  \bibinfo{author}{\bibfnamefont{A.}~\bibnamefont{Snigirev}},
  \bibinfo{author}{\bibfnamefont{I.}~\bibnamefont{Snigireva}},
  \bibnamefont{and}
  \bibinfo{author}{\bibfnamefont{M.}~\bibnamefont{Drakopoulos}},
  \bibinfo{journal}{J. Synchrotron Radiation} \textbf{\bibinfo{volume}{6}},
  \bibinfo{pages}{1153} (\bibinfo{year}{1999}).

\bibitem[{\citenamefont{Kolodziej et~al.}(2018)\citenamefont{Kolodziej,
  Stoupin, Grizolli, Krzywinski, Shi, Kim, Qian, Assoufid, and
  Shvyd'ko}}]{KSG18}
\bibinfo{author}{\bibfnamefont{T.}~\bibnamefont{Kolodziej}},
  \bibinfo{author}{\bibfnamefont{S.}~\bibnamefont{Stoupin}},
  \bibinfo{author}{\bibfnamefont{W.}~\bibnamefont{Grizolli}},
  \bibinfo{author}{\bibfnamefont{J.}~\bibnamefont{Krzywinski}},
  \bibinfo{author}{\bibfnamefont{X.}~\bibnamefont{Shi}},
  \bibinfo{author}{\bibfnamefont{K.-J.} \bibnamefont{Kim}},
  \bibinfo{author}{\bibfnamefont{J.}~\bibnamefont{Qian}},
  \bibinfo{author}{\bibfnamefont{L.}~\bibnamefont{Assoufid}}, \bibnamefont{and}
  \bibinfo{author}{\bibfnamefont{Yu.}~\bibnamefont{Shvyd'ko}},
  \bibinfo{journal}{Journal of Synchrotron Radiation}
  \textbf{\bibinfo{volume}{25}}, \bibinfo{pages}{354} (\bibinfo{year}{2018}).

\bibitem[{\citenamefont{Shvyd'ko et~al.}(2003)\citenamefont{Shvyd'ko, Lerche,
  Wille, Gerdau, Alp, Lucht, R{\"u}ter, and Khachatryan}}]{SLW02}
\bibinfo{author}{\bibfnamefont{Yu.~V.} \bibnamefont{Shvyd'ko}},
  \bibinfo{author}{\bibfnamefont{M.}~\bibnamefont{Lerche}},
  \bibinfo{author}{\bibfnamefont{H.-C.} \bibnamefont{Wille}},
  \bibinfo{author}{\bibfnamefont{E.}~\bibnamefont{Gerdau}},
  \bibinfo{author}{\bibfnamefont{E.~E.} \bibnamefont{Alp}},
  \bibinfo{author}{\bibfnamefont{M.}~\bibnamefont{Lucht}},
  \bibinfo{author}{\bibfnamefont{H.~D.} \bibnamefont{R{\"u}ter}},
  \bibnamefont{and}
  \bibinfo{author}{\bibfnamefont{R.}~\bibnamefont{Khachatryan}},
  \bibinfo{journal}{Phys. Rev. Lett.} \textbf{\bibinfo{volume}{90}},
  \bibinfo{pages}{013904} (\bibinfo{year}{2003}).

\bibitem[{\citenamefont{Cotterill}(1968)}]{Cotterill68}
\bibinfo{author}{\bibfnamefont{R.~M.~J.} \bibnamefont{Cotterill}},
  \bibinfo{journal}{Appl. Phys. Lett.} \textbf{\bibinfo{volume}{12}},
  \bibinfo{pages}{403} (\bibinfo{year}{1968}).

\bibitem[{\citenamefont{Shvyd'ko}(2013)}]{Shv13}
\bibinfo{author}{\bibfnamefont{Yu.}~\bibnamefont{Shvyd'ko}},
  \bibinfo{journal}{Beam Dynamics Newsletter} \textbf{\bibinfo{volume}{60}},
  \bibinfo{pages}{68} (\bibinfo{year}{2013}), \bibinfo{note}{{I}nternational
  {C}ommittee for {F}uture {A}ccelerators.},
  \urlprefix\url{https://icfa-usa.jlab.org/archive/newsletter/icfa_bd_nl_60.pdf}.

\bibitem[{\citenamefont{Marcus et~al.}(2019)\citenamefont{Marcus, Anton,
  Assoufid, Decker, Gassner, Goetze, Halavanau, Hastings, Huang, Jansma
  et~al.}}]{MAA19}
\bibinfo{author}{\bibfnamefont{G.}~\bibnamefont{Marcus}},
  \bibinfo{author}{\bibfnamefont{J.}~\bibnamefont{Anton}},
  \bibinfo{author}{\bibfnamefont{L.}~\bibnamefont{Assoufid}},
  \bibinfo{author}{\bibfnamefont{F.-J.} \bibnamefont{Decker}},
  \bibinfo{author}{\bibfnamefont{G.}~\bibnamefont{Gassner}},
  \bibinfo{author}{\bibfnamefont{K.}~\bibnamefont{Goetze}},
  \bibinfo{author}{\bibfnamefont{A.}~\bibnamefont{Halavanau}},
  \bibinfo{author}{\bibfnamefont{J.}~\bibnamefont{Hastings}},
  \bibinfo{author}{\bibfnamefont{Z.}~\bibnamefont{Huang}},
  \bibinfo{author}{\bibfnamefont{W.}~\bibnamefont{Jansma}},
  \bibnamefont{et~al.}, in \emph{\bibinfo{booktitle}{Proc. FEL'19}}
  (\bibinfo{publisher}{JACoW Publishing, Geneva, Switzerland},
  \bibinfo{year}{2019}), no.~\bibinfo{number}{39} in \bibinfo{series}{Free
  Electron Laser Conference}, pp. \bibinfo{pages}{282--287}, ISBN
  \bibinfo{isbn}{978-3-95450-210-3},
  \bibinfo{note}{https://doi.org/10.18429/JACoW-FEL2019-TUD04},
  \urlprefix\url{http://jacow.org/fel2019/papers/tud04.pdf}.

\bibitem[{\citenamefont{Kogelnik and Li}(1966)}]{Kl66}
\bibinfo{author}{\bibfnamefont{H.}~\bibnamefont{Kogelnik}} \bibnamefont{and}
  \bibinfo{author}{\bibfnamefont{T.}~\bibnamefont{Li}}, \bibinfo{journal}{Appl.
  Opt.} \textbf{\bibinfo{volume}{5}}, \bibinfo{pages}{1550}
  (\bibinfo{year}{1966}).

\bibitem[{\citenamefont{Kim}(2020)}]{KJK20}
\bibinfo{author}{\bibfnamefont{K.}~\bibnamefont{Kim}}, \bibinfo{type}{Tech.
  Rep.} \bibinfo{number}{Preprint AOP-TN-2020-106},
  \bibinfo{institution}{Argonne National Laboratory} (\bibinfo{year}{2020}).

\bibitem[{\citenamefont{Authier}(2001)}]{Authier}
\bibinfo{author}{\bibfnamefont{A.}~\bibnamefont{Authier}},
  \emph{\bibinfo{title}{Dynamical Theory of X-Ray Diffraction}},
  vol.~\bibinfo{volume}{11} of \emph{\bibinfo{series}{IUCr Monographs on
  Crystallography}} (\bibinfo{publisher}{Oxford University Press},
  \bibinfo{address}{Oxford, New York}, \bibinfo{year}{2001}).

\bibitem[{\citenamefont{Shvyd'ko et~al.}(2017)\citenamefont{Shvyd'ko, Blank,
  and Terentyev}}]{SBT17}
\bibinfo{author}{\bibfnamefont{Yu.}~\bibnamefont{Shvyd'ko}},
  \bibinfo{author}{\bibfnamefont{V.}~\bibnamefont{Blank}}, \bibnamefont{and}
  \bibinfo{author}{\bibfnamefont{S.}~\bibnamefont{Terentyev}},
  \bibinfo{journal}{MRS Bulletin} \textbf{\bibinfo{volume}{62}},
  \bibinfo{pages}{437} (\bibinfo{year}{2017}).

\bibitem[{\citenamefont{Shvyd'ko}(2019)}]{Shvydko19}
\bibinfo{author}{\bibfnamefont{Yu.}~\bibnamefont{Shvyd'ko}},
  \bibinfo{journal}{Phys. Rev. Accel. Beams} \textbf{\bibinfo{volume}{22}},
  \bibinfo{pages}{100703} (\bibinfo{year}{2019}).

\bibitem[{\citenamefont{{RXOPTICS}}(2021)}]{RXOPTICS}
\bibinfo{author}{\bibnamefont{{RXOPTICS}}} (\bibinfo{year}{2021}),
  \bibinfo{note}{http://www.rxoptics.de}.

\bibitem[{\citenamefont{Saleh and Teich}(2007)}]{SalehTeich}
\bibinfo{author}{\bibfnamefont{B.~E.~A.} \bibnamefont{Saleh}} \bibnamefont{and}
  \bibinfo{author}{\bibfnamefont{M.~C.} \bibnamefont{Teich}},
  \emph{\bibinfo{title}{{Fundamentals of photonics; 2nd ed.}}}
  (\bibinfo{publisher}{Wiley}, \bibinfo{address}{New York, NY},
  \bibinfo{year}{2007}).

\bibitem[{\citenamefont{Shvyd'ko}(2015)}]{Shvydko15}
\bibinfo{author}{\bibfnamefont{Yu.}~\bibnamefont{Shvyd'ko}},
  \bibinfo{journal}{Phys. Rev. A} \textbf{\bibinfo{volume}{91}},
  \bibinfo{pages}{053817} (\bibinfo{year}{2015}).

\bibitem[{\citenamefont{Liu}(2021)}]{PeifanLiu21}
\bibinfo{author}{\bibfnamefont{P.}~\bibnamefont{Liu}} (\bibinfo{year}{2021}),
  \bibinfo{note}{{P}eifan Liu is acknowledged for this observation, while
  reading the manuscript}.

\bibitem[{\citenamefont{Edwards and Syphers}(1992)}]{ES92}
\bibinfo{author}{\bibfnamefont{D.~A.} \bibnamefont{Edwards}} \bibnamefont{and}
  \bibinfo{author}{\bibfnamefont{M.~J.} \bibnamefont{Syphers}},
  \emph{\bibinfo{title}{{An Introduction to the Physics of High-Energy
  Accelerators}}}, Wiley Series in Beam Physics and Accelerator Technology
  (\bibinfo{publisher}{Wiley}, \bibinfo{address}{New York},
  \bibinfo{year}{1992}), ISBN \bibinfo{isbn}{978-0-471-55163-8}.

\bibitem[{\citenamefont{Sanchez~del Rio et~al.}(2011)\citenamefont{Sanchez~del
  Rio, Canestrari, Jiang, and Cerrina}}]{SHADOW3}
\bibinfo{author}{\bibfnamefont{M.}~\bibnamefont{Sanchez~del Rio}},
  \bibinfo{author}{\bibfnamefont{N.}~\bibnamefont{Canestrari}},
  \bibinfo{author}{\bibfnamefont{F.}~\bibnamefont{Jiang}}, \bibnamefont{and}
  \bibinfo{author}{\bibfnamefont{F.}~\bibnamefont{Cerrina}},
  \bibinfo{journal}{Journal of Synchrotron Radiation}
  \textbf{\bibinfo{volume}{18}}, \bibinfo{pages}{708} (\bibinfo{year}{2011}).

\bibitem[{\citenamefont{Rebuffi and S{\'{a}}nchez~del R{\'\i}o}(2016)}]{RS16}
\bibinfo{author}{\bibfnamefont{L.}~\bibnamefont{Rebuffi}} \bibnamefont{and}
  \bibinfo{author}{\bibfnamefont{M.}~\bibnamefont{S{\'{a}}nchez~del R{\'\i}o}},
  \bibinfo{journal}{Journal of Synchrotron Radiation}
  \textbf{\bibinfo{volume}{23}}, \bibinfo{pages}{1357} (\bibinfo{year}{2016}).

\bibitem[{\citenamefont{Rebuffi and del Rio}(2017)}]{RS17}
\bibinfo{author}{\bibfnamefont{L.}~\bibnamefont{Rebuffi}} \bibnamefont{and}
  \bibinfo{author}{\bibfnamefont{M.~S.} \bibnamefont{del Rio}}, in
  \emph{\bibinfo{booktitle}{Advances in Computational Methods for X-Ray Optics
  IV}}, edited by \bibinfo{editor}{\bibfnamefont{O.}~\bibnamefont{Chubar}}
  \bibnamefont{and} \bibinfo{editor}{\bibfnamefont{K.}~\bibnamefont{Sawhney}}
  (\bibinfo{organization}{SPIE}, \bibinfo{year}{2017}), vol.
  \bibinfo{volume}{10388}, p.~\bibinfo{pages}{28}.

\end{thebibliography}

\end{document}